\documentclass[12pt, draftclsnofoot, onecolumn]{IEEEtran}

\usepackage{graphicx}
\usepackage{amsmath, epsfig, amssymb, amstext, verbatim, amsopn, cite, subfigure, multirow, multicol, lipsum}
\usepackage{balance}
\usepackage{url}
\usepackage{amsfonts}
\usepackage{epsfig}
\usepackage{algorithmic}
\usepackage{algorithm}
\usepackage{setspace}
\usepackage{stmaryrd}
\usepackage{psfrag}	
\usepackage{multirow}
\usepackage{float}
\usepackage[process=auto]{pstool}
\usepackage{etoolbox}
\usepackage{graphicx}
\usepackage{acronym}
\allowdisplaybreaks

\allowdisplaybreaks
\graphicspath{{images/}}

\newtheorem{theorem}{Theorem}
\newtheorem{lemma}{Lemma}
\newtheorem{remark}{Remark}

\newtheorem{corollary}{Corollary}

\newcommand*{\norm}[1]{\mathopen\| #1 \mathclose\|}

\def\norm#1{\mathopen\| #1 \mathclose\|}

\newcommand{\vv}[1]{\boldsymbol{\mathrm{#1}}}
\newcommand{\mm}[1]{\boldsymbol{\mathrm{#1}}}

\newcommand{\expect}[1]{{\mathbb{E}}\!\left\{ #1 \right\}}

\newcommand{\herm}{{\sf{H}}}
\newcommand{\transp}{{\sf{T}}}

\begin{document}
\IEEEoverridecommandlockouts

\title{I/Q Imbalance Aware Widely-Linear Receiver for Uplink Multi-Cell Massive MIMO Systems\\}
\author{\IEEEauthorblockN{Shahram Zarei, Wolfgang Gerstacker, Jocelyn Aulin, and Robert Schober\vspace*{-10mm}}\\
\IEEEauthorblockA{\thanks{This paper will be presented in part at ISWCS 2015, Brussels, August 2015.}
}}

%
  
\maketitle
%
\begin{abstract}
In-phase/quadrature-phase (I/Q) imbalance is one of the most important hardware impairments in communication systems. It arises in the analogue parts of direct conversion radio frequency (RF) transceivers and can cause severe performance losses. In this paper, I/Q imbalance (IQI) aware widely-linear (WL) channel estimation and data detection schemes for uplink multi-cell massive multiple-input multiple-output (MIMO) systems are proposed. The resulting receiver is a WL extension of the minimum mean square error (MMSE) receiver and jointly mitigates multi-user interference and IQI by processing the real and the imaginary parts of the received signal separately. The IQI arising at both the base station (BS) and the user terminals (UTs) is then taken into account. The considered channel state information (CSI) acquisition model includes the effects of both estimation errors and pilot contamination, which is caused by the reuse of the same training sequences in neighboring cells. We apply results from random matrix theory to derive analytical expressions for the achievable sum rates of the proposed IQI aware and conventional IQI unaware receivers. Our simulation results show that the performance of the proposed IQI aware WLMMSE receiver in a system with IQI is close to that of the MMSE receiver in an ideal system without IQI. Moreover, our results for the sum rate of the IQI unaware MMSE receiver reveal that the performance loss due to IQI can be large and, if left unattended, does not vanish for large numbers of BS antennas. 
\end{abstract}
\vspace*{-3mm}
\IEEEpeerreviewmaketitle
\vspace*{-3mm}
%
\section{Introduction}\label{Sec_Intro}
\IEEEPARstart{M}{ultiple}-input multiple-output (MIMO) techniques have become a central part of modern communication systems such as Long Term Evolution (LTE) and WiMAX. With MIMO technology, high throughput and transmission reliability can be achieved. An emerging research field in wireless communications are so-called massive MIMO systems \cite{Marzetta2010}, \cite{Rusek2013}. Massive MIMO systems employ a large number of antennas, e.g. one hundred antennas or more at the base station (BS), and can achieve very high spectral and energy efficiencies \cite{Ngo2012}. Moreover, in massive MIMO systems, the transmit power of the BS and the user terminals (UTs) can be decreased by increasing the number of antennas at the BS \cite{Ngo2012}. These and other desirable features render massive MIMO a promising technology for future wireless communication systems. In this paper, we consider the uplink of a multi-cell massive MIMO system. In an uplink single-cell massive MIMO system, which embodies a multiple access channel (MAC), successive interference cancellation (SIC) is the optimal detection scheme \cite{Verdu_MU_Detection}. Nevertheless, since SIC is not feasible in most practical systems due to its high computational complexity, linear detectors such as matched filter (MF) and minimum mean square error (MMSE) detectors are often preferred as they provide a good trade-off between performance and complexity \cite{Rusek2013}, \cite{Ngo2012}. 

However, hardware (H/W) impairments, which exist in all practical systems, can severely degrade the performance of linear detectors. One of the most important H/W impairments in digital communication systems is in-phase/quadrature-phase (I/Q) imbalance, which arises in direct conversion transceivers \cite{Razavi1998}. In such systems, the real and imaginary parts of the received RF signal are mixed with the high-frequency carrier signal and its $90^\circ$ phase-shifted version, respectively, to produce the baseband signal. Ideally, the phase difference between the carrier signal and its phase-shifted version is exactly $90^\circ$ and the mixers for the real and the imaginary part have the same amplitude gain. However, in practical systems, both amplitude and phase mismatches between the real and imaginary parts occur, which leads to an I/Q imbalance (IQI) in each antenna branch at the base station and in the RF chain of each UT, and impair the received data vectors \cite{Tarighat2005}, \cite{Gottumukkala2012}. Thus, for reliable detection, besides the mitigation of multi-user interference, IQI compensation is necessary as well. One approach to overcome the negative effects of IQI is to measure and compensate the individual IQIs in each antenna branch. However, this solution becomes very costly in massive MIMO systems, where the BS may be equipped with several hundred RF chains. Another approach, which we consider in this paper, is joint data detection and IQI mitigation. In this case, the equivalent channel, which comprises the actual channel and the IQI, is estimated at the BS. For this estimation, we exploit the received training sequence and the channel statistics which also include the effect of IQI. The detection matrix is then constructed based on this equivalent channel estimate and can be used for joint data detection and IQI mitigation. To the best of the authors' knowledge, the problem of joint channel estimation, data detection, and IQI mitigation for uplink multi-cell massive MIMO systems has not been investigated yet. 

Recently, a WLMMSE beamformer for systems with IQI has been proposed in \cite{Hakkarainen2013}. Here, the authors propose a beamforming scheme, where a multi-antenna receiver suffering from IQI detects signals coming from a specific direction while suppressing signals arriving from other directions. Another related work is \cite{Hakkarainen_TWC_Arx2014}, where the authors propose a detection scheme for uplink single-cell multi-user-MIMO (MU-MIMO) systems impaired by IQI. In \cite{Bjoernson_TIT_2014}, the authors model the residual H/W impairments, which remain after compensation, in massive MIMO systems as an additive Gaussian impairment and derive a capacity bound. Moreover, in our recent work \cite{Zarei_GC2014}, we have proposed an IQI aware precoder for downlink single-cell massive MIMO systems assuming perfect CSI and IQI present only at the BS. Another recent work is \cite{Kolomvakis_ICC2015}, where the authors consider the uplink of a single-cell massive MIMO system with IQI present only at the BS. In this paper, we propose a widely-linear MMSE (WLMMSE) receiver for CSI acquisition and data detection in an uplink multi-cell massive MIMO system, where both the BS and the UTs are impaired by IQI and both the received data and training signals for channel estimation are affected by IQI. WL filtering was introduced in \cite{Picinbono1995} and is used to estimate complex signals by filtering the real and imaginary parts separately. Several works have employed WL filtering in single-user MIMO systems, cf. \cite{Darsena2012}, \cite{Darsena2013}. WL processing results in a higher performance than strictly linear processing, if rotationally variant signals are involved, which is the case when IQI is present. This motivates the use of WL filtering for channel estimation and data detection at the BS of uplink multi-cell massive MIMO systems suffering from IQI. 

In contrast to \cite{Hakkarainen2013}, \cite{Hakkarainen_TWC_Arx2014}, and \cite{Zarei_GC2014}, where a single-cell system with perfect CSI was considered, in this paper, a more sophisticated system model is adopted, which includes the effects of multi-cell interference and CSI imperfection originating from  pilot contamination, channel estimation errors, and IQI. Furthermore, contrary to \cite{Kolomvakis_ICC2015}, where a single-cell system with ideal UTs was considered, our system model includes multi-cell interference and takes IQI at both the UTs and the BSs into account. Moreover, as opposed to \cite{Bjoernson_TIT_2014}, where the residual H/W impairments were modelled by an additional additive Gaussian noise term and a non-augmented system model was employed, we adopt an augmented system model, which is essential for the analysis and mitigation of IQI. In addition, we use results from random matrix theory and provide analytical results for the sum rate performance of the proposed IQI aware (IQA) and conventional IQI unaware (IQU) receivers.

This paper is organized as follows. In Section \ref{Sec_SystemModel}, the system model is presented. In Section \ref{Sec_IQU_Detector}, the conventional IQU-MMSE receiver is investigated and an analytical expression for its sum rate in the presence of IQI is derived. In Section \ref{Sec_IQA_Detector}, we introduce the proposed IQA-WLMMSE receiver, and present an analytical expression for the corresponding sum rate. Numerical results are provided in Section \ref{Sec_NumResults}, and conclusions are drawn in Section \ref{Sec_Conclusion}.

\emph{Notation:} Boldface lower and upper case letters represent column vectors and matrices, respectively. $\mathrm{blockdiag}\left(\mm{Q}_1,\ldots,\mm{Q}_N\right)$ is a block diagonal matrix with matrices $\mm{Q}_1,\ldots,\mm{Q}_N$ on its main diagonal. $\mm{I}_K$ denotes the $K \times K$ identity matrix and ${\left[\mm{A}\right]}_{k,:}$, ${\left[\mm{A}\right]}_{:,l}$, and ${\left[\mm{A}\right]}_{k, l}$ stand for the $k$th row, the $l$th column, and the element in the $k$th row and the $l$th column of matrix $\mm{A}$, respectively. $(\cdot)^*$ denotes the complex conjugate and $\mathrm{det}(\cdot)$, $\norm{\cdot}^2_\mathrm{F}$, $\mathrm{tr}(\cdot)$, $(\cdot)^{\transp}$, and $(\cdot)^{\herm}$ are the determinant, Frobenius norm, trace, transpose, and Hermitian transpose of a matrix, respectively. $\Re\lbrace \cdot \rbrace$ and $\Im\lbrace \cdot \rbrace$ denote the real and imaginary parts of a complex number, respectively. $\expect{\cdot}$ stands for the expectation operator and $\mathcal{C} \mathcal{N} \left(\vv{u}, \mm{\Phi} \right)$ denotes a circular symmetric complex Gaussian distribution with mean vector $\vv{u}$ and covariance matrix $\mm{\Phi}$. Moreover, $\mathrm{a.s.}$ stands for almost sure convergence.
%
\vspace*{-4mm}
\section{System Model}\label{Sec_SystemModel}
In this paper, we consider the uplink of a multi-cell massive MIMO system with universal frequency reuse. The number of cells is denoted by $L$, and in each cell, $K$ single-antenna UTs simultaneously transmit data to a BS with $N$ antennas. $K$ and $N$ are assumed to be very large with their ratio $\beta=K/N$ being constant. Furthermore, we assume a block fading channel. The channel matrix between the UTs in the $l$th cell and the BS in the $i$th cell is denoted by $\mm{H}_{i,l}=\left[ \vv{h}_{i,l,1} \ldots \vv{h}_{i,l,K} \right] \in \mathbb{C}^{N \times K}$. Here, $\vv{h}_{i,l,k}=\check{\mm{R}}_{i,l,k} \vv{\nu}_{i,l,k} \in \mathbb{C}^{N\times1}$ is the channel vector between the $k$th UT in the $l$th cell and the BS in the $i$th cell, where $\vv{\nu}_{i,l,k} \sim \mathcal{C} \mathcal{N} \left(\mm{0}, \mm{I}_N \right)$ and $\mm{R}_{i,l,k} = \mathbb{E}\lbrace \vv{h}_{i,l,k} \vv{h}^\herm_{i,l,k} \rbrace = \check{\mm{R}}_{i,l,k} \check{\mm{R}}^\herm_{i,l,k}$ represents the channel covariance matrix.
Since the detection schemes considered in this paper, i.e., IQU-MMSE and IQA-WLMMSE detection, have fundamentally different structures, we adopt two different representations for the system model, namely a complex-valued and a real-valued representation, which are presented in the following subsections.
%
\vspace*{-3mm}
\subsection{Complex-Valued Representation} \label{Subsec_SystemModel_IQU}
In this subsection, the complex representation of the system model, which is used for the IQU-MMSE detector, is introduced. The transmitted data symbols of the $K$ UTs in the $l$th cell are stacked into a vector, which is denoted by $\vv{d}_l=\left[ d_{l,1},\ldots, d_{l,K} \right]^\transp \sim \mathcal{C}\mathcal{N}\left(\mm{0},\mm{I}_K\right)$, where $d_{l,k}$ is the data symbol transmitted by the $k$th UT in the $l$th cell. The received signal at the $i$th BS can be modeled as
\vspace*{-2mm}
\begin{align}
\vv{r}_{i}=\mm{\Psi}_{A,i} \left( \sum_{l=1}^{L} \sqrt{\rho_{\mathrm{UL}}} \vv{H}_{i,l}  \left( \mm{\Xi}_{A,l} \vv{d}_{l} 
\hspace*{-1mm}+\hspace*{-1mm}  \mm{\Xi}_{B,l} \vv{d}^*_{l} \right) \hspace*{-1mm}+\hspace*{-1mm} \vv{z}_{i} \right) 
\hspace*{-1mm}+\hspace*{-1mm} \mm{\Psi}_{B,i} \left( \sum_{l=1}^{L} \sqrt{\rho_{\mathrm{UL}}} \vv{H}^*_{i,l}  \left( \mm{\Xi}^*_{A,l} \vv{d}^*_{l} \hspace*{-1mm}+\hspace*{-1mm}  \mm{\Xi}^*_{B,l} \vv{d}_{l} \right) \hspace*{-1mm}+\hspace*{-1mm} \vv{z}^*_{i} \right),
\vspace*{-4mm}
\label{Eqn_SysMod_IQU}
\end{align}
where $\rho_\mathrm{UL}$ denotes the uplink transmit signal-to-noise ratio (SNR), and $\mm{\Xi}_{A,l} = \mathrm{diag} (\xi_{A,l,1}, \ldots, \\ \xi_{A,l,K} )$ and $\mm{\Xi}_{B,l} \hspace*{-1mm}=\hspace*{-1mm} \mathrm{diag} \left(\xi_{B,l,1}, \ldots, \xi_{B,l,K} \right)$ with $\xi_{A,l,k} = \cos \left( \check{\theta}_{l,k} / 2 \right) + j \check{\epsilon}_{l,k} \sin \left( \check{\theta}_{l,k} / 2 \right)$ and $\xi_{B,l,k} \hspace*{-1mm}=\hspace*{-1mm} \check{\epsilon}_{l,k} \cos \left( \\
 \check{\theta}_{l,k} / 2 \right) - j \sin \left( \check{\theta}_{l,k} / 2 \right)$ representing the IQI at the $k$th UT in the $l$th cell. $\check{\theta}_{l,k}$ and $\check{\epsilon}_{l,k}$ denote the phase and amplitude imbalances at the corresponding UT, respectively. The IQI at the $i$th BS is modelled by diagonal matrices $\mm{\Psi}_{A,i}=\mathrm{diag}\left(  \psi_{A,i,1}, \ldots, \psi_{A,i,N} \right)$ and $\mm{\Psi}_{B,i}=\mathrm{diag}\left(  \psi_{B,i,1}, \ldots, \psi_{B,i,N} \right)$, where $\psi_{A,i,n}=\cos \left( \theta_{i,n} / 2 \right) + j \epsilon_{i,n} \sin \left( \theta_{i,n} / 2 \right)$ and $\psi_{B,i,n}=\epsilon_{i,n} \cos \big( \theta_{i,n} / $ $2 \big) - j \sin \big( \theta_{i,n} / 2 \big)$ with $\theta_{i,n}$ and $\epsilon_{i,n}$ being the phase and amplitude imbalances at the $n$th antenna branch of the $i$th BS, respectively. If IQI is absent, we have $\epsilon_{i,n}=0$, $\theta_{i,n}=0, \forall i \in \left\lbrace1,\ldots,L\right\rbrace, \forall n \in \left\lbrace1,\ldots,N\right\rbrace$ and $\check{\epsilon}_{l,k}=0$, $\check{\theta}_{l,k}=0, \forall l \in \left\lbrace1,\ldots,L\right\rbrace, \forall k \in \left\lbrace1,\ldots,K\right\rbrace$. $\vv{z}_i \sim \mathcal{C} \mathcal{N} \left( \mm{0}, \mm{I}_N \right) $ represents the complex additive white Gaussian noise (AWGN) at the $i$th BS.

%
\vspace*{-3mm}
\subsection{Augmented Real-Valued Representation} \label{Subsec_SystemModel_IQA}
Since IQI affects the real and imaginary parts of a signal differently, IQA channel estimation and data detection should allow for processing the real and imaginary parts of received signals individually. Hence, we use an augmented representation for the IQA system model, where the real and imaginary parts of the signals are stacked together. More precisely, the real and imaginary parts of the independent and identically distributed (i.i.d.) zero-mean complex Gaussian transmit data symbols of the $K$ UTs in the $l$th cell are stacked into the augmented vector $\tilde{\vv{d}}_l=\left[ \tilde{d}_{l,1}, \ldots, \tilde{d}_{l,2K} \right]^\transp=\left[\vv{d}_{\mathrm{R}_l}^\transp \ \vv{d}_{\mathrm{I}_l}^\transp\right]^\transp \in \mathbb{R}^{2K}$, where $\vv{d}_{\mathrm{R}_l}=\left[ d_{\mathrm{R}_{l,1}},\ldots, d_{\mathrm{R}_{l,K}} \right]^\transp$ and $\vv{d}_{\mathrm{I}_l}=\left[ d_{\mathrm{I}_{l,1}},\ldots, d_{\mathrm{I}_{l,K}} \right]^\transp$ contain the real and imaginary parts of the $K$ complex transmit data symbols in the $l$th cell, respectively, and $ \mathbb{E} \left\lbrace \tilde{\vv{d}}_l \tilde{\vv{d}}^\transp_l \right\rbrace = 0.5 \ \mm{I}_{2K}$. The augmented data vector received at the $i$th BS, $\tilde{\vv{r}}_{i} \in \mathbb{R}^{2K}$, from all UTs in the $L$ cells can be expressed as
\vspace*{-3mm}
\begin{align}
\tilde{\vv{r}}_{i} = \left[\vv{r}_{\mathrm{R}_i}^\transp \ \vv{r}_{\mathrm{I}_i}^\transp\right]^\transp = \tilde{\mm{\Psi}}_{i} \left( \sum_{l=1}^{L} \sqrt{\rho_{\mathrm{UL}}} \tilde{\mm{H}}_{i,l}  \tilde{\vv{\Xi}}_{l} \tilde{\vv{d}}_{l} + \tilde{\vv{z}}_{i} \right),
\label{Eqn_SysMod_IQA}
\end{align}
where $\vv{r}_{\mathrm{R}_i}$ and $\vv{r}_{\mathrm{I}_i}$ are the real and the imaginary parts of the received signal $\vv{r}_{i}$ in (\ref{Eqn_SysMod_IQU}), respectively. Here, the augmented real-valued channel matrix $\tilde{\mm{H}}_{i,l}$ is given by $ \begin{bmatrix} \mm{H}_{\mathrm{R}_{i,l}} & -\mm{H}_{\mathrm{I}_{i,l}}\\ \mm{H}_{\mathrm{I}_{i,l}} & \mm{H}_{\mathrm{R}_{i,l}} \end{bmatrix} \in \mathbb{R}^{2N \times 2K}$ with $\mm{H}_{\mathrm{R}_{i,l}}$ and $\mm{H}_{\mathrm{I}_{i,l}}$ being the real and imaginary parts of the complex-valued channel matrix $\mm{H}_{i,l} = \mm{H}_{\mathrm{R}_{i,l}} + j \mm{H}_{\mathrm{I}_{i,l}}$, respectively. The augmented real-valued vector $\tilde{\vv{z}}_i = \left[\vv{z}_{\mathrm{R}_i}^\transp \vv{z}_{\mathrm{I}_i}^\transp\right]^\transp \in \mathbb{R}^{2N}$ contains the real and imaginary parts of the complex-valued AWGN vector $ \vv{z}_i = \left[z_{i,1},\ldots,z_{i,N}\right]^\transp = \vv{z}_{\mathrm{R}_i} + j \vv{z}_{\mathrm{I}_i} \sim \mathcal{C} \mathcal{N} \left(\vv{0}, {\vv{I}}_N\right)$ at the $i$th BS, respectively. $\tilde{\mm{\Psi}}_{i} = \mm{\Pi} \check{\mm{\Psi}}_i \mm{\Pi}^{-1} \in \mathbb{R}^{2N\times2N}$ models the IQI at the $i$th BS. Here, $\tilde{\mm{\Psi}}_{i}$ is a permuted version of $\check{\mm{\Psi}}_{i}=\mathrm{blockdiag}\left(\mm{A}_{i,1},\ldots,\mm{A}_{i,N}\right)$, 
where $\mm{A}_{i,n}, \ n \in \left\lbrace 1, \ldots, N \right\rbrace$, represents the IQI of the $n$th RF branch of the $i$th BS and is given by \cite{Tarighat2005}
\vspace*{-6mm}
\begin{align}
\mm{A}_{i,n} = \begin{bmatrix}
\Re\left\lbrace \psi_{A,i,n}+\psi_{B,i,n} \right\rbrace  & \Im\left\lbrace \psi_{B,i,n}-\psi_{A,i,n} \right\rbrace\\
\Im\left\lbrace \psi_{A,i,n}+\psi_{B,i,n} \right\rbrace & \Re\left\lbrace \psi_{A,i,n}-\psi_{B,i,n} \right\rbrace\\
\end{bmatrix}.
\label{Eqn_IQA_MatrixA}
\end{align}
The elements of permutation matrix $\mm{\Pi}\in\mathbb{R}^{2N\times2N}$ are defined as
\vspace*{-4mm}
\begin{align}
{\left[\mm{\Pi}\right]}_{p, q}=\begin{cases}
  1  & \mathrm{if} \ p=2n-1, q=n, \forall n \in \left\lbrace 1,\ldots,N \right\rbrace, \\
  1  & \mathrm{if} \ p=2n, q=n+N, \forall n \in \left\lbrace 1,\ldots,N \right\rbrace, \\
  0  & \mathrm{otherwise.}
\end{cases}
\label{Equ_PermutMatrix}
\end{align}
This permutation is required, since the stacked received data vector $\tilde{\vv{r}}_i$ contains the in-phase and quadrature-phase components in its upper and lower parts, respectively. Moreover, in (\ref{Eqn_SysMod_IQA}), $\tilde{\vv{\Xi}}_{l}=\check{\mm{\Pi}} \check{\mm{\Xi}}_l \check{\mm{\Pi}}^{-1} \in \mathbb{R}^{2K\times2K}$ denotes the stacked IQIs of the $K$ UTs in the $l$th cell, where the permutation matrix $\check{\mm{\Pi}}$ is similarly defined as $\mm{\Pi}$ and is obtained by replacing $N$ with $K$ and $n$ with $k$ in (\ref{Equ_PermutMatrix}). Furthermore, $\check{\vv{\Xi}}_{l}=\mathrm{blockdiag}\left(\vv{\Xi}_{l,1},\ldots,\vv{\Xi}_{l,K}\right)$, 
where $\vv{\Xi}_{l,k} \in \mathbb{R}^2$ is the IQI matrix of the $k$th UT in the $l$th cell, which can be expressed as
\vspace*{-4mm}
\begin{align}
\vv{\Xi}_{l,k} = \begin{bmatrix}
\Re\left\lbrace \xi_{A,l,k}+\xi_{B,l,k} \right\rbrace  & \Im\left\lbrace \xi_{B,l,k}-\xi_{A,l,k} \right\rbrace\\
\Im\left\lbrace \xi_{A,l,k}+\xi_{B,l,k} \right\rbrace & \Re\left\lbrace \xi_{A,l,k}-\xi_{B,l,k} \right\rbrace\\
\end{bmatrix}. 
\label{Eqn_IQ_UT_Def}
\end{align}
%
\vspace*{-5mm}
\section{IQI Unaware MMSE Receiver} \label{Sec_IQU_Detector}
In this section, as a performance benchmark, the sum rate of a conventional IQU-MMSE receiver comprising an IQU-MMSE channel estimator and an IQU-MMSE data detector is investigated in the presence of IQI. For IQU estimation and detection, we adopt the conventional MMSE estimator and detector, respectively, which are not designed for IQI mitigation. 
%
\vspace*{-2mm}
\subsection{Channel Estimation}{\label{Subsec_ChEst}}
In this subsection, channel estimation for an IQU system in the presence of IQI is presented. For channel estimation, at the beginning of every coherence interval, training sequences are transmitted by all UTs to their serving BS. Due to the limited length of the coherence interval, there are not enough orthogonal training sequences for all UTs in all cells. Hence, UTs with the same index in different cells use the same training sequence \cite{Jose_ISIT2009}. This leads to a corrupted channel estimate and this effect is known as pilot contamination in the massive MIMO literature \cite{Jose_ISIT2009}. Since we consider full pilot reuse, when pilot contamination is present, UTs having the same index $k$ in different cells employ the same training sequence $\vv{x}_k \in \mathbb{R}^{T\times1}$, where $T$ is the length of the training sequence. The received training signal at each BS is  multiplied by the original transmitted training sequence to eliminate the interference caused by other UTs. Considering (\ref{Eqn_SysMod_IQU}), and the orthonormality of the training sequences, i.e., $\vv{x}_k^\transp\vv{x}_k=1$, $\vv{x}_k^\transp\vv{x}_j=0, k \neq j$, we have the following expression for the received training sequence of the $k$th UT in the $i$th cell
\vspace*{-4mm}
\begin{align}
{\vv{y}}_{i,k}=&\mm{Y}_{i} \vv{x}_k=\mm{\Psi}_{A,i} \Big( \sum_{l=1}^{L} \sqrt{\rho_\mathrm{TR}} \vv{h}_{i,l,k}  \big( \xi_{A,l,k} +  \xi_{B,l,k}  \big) + \check{\vv{w}}_i \big) + \mm{\Psi}_{B,i} \Big( \sum_{l=1}^{L} \sqrt{\rho_\mathrm{TR}} \vv{h}^*_{i,l,k}  \big( \xi^*_{A,l,k} +  \nonumber \\
& \xi^*_{B,l,k} \big) + \check{\vv{w}}^*_i \Big),
\label{Eqn_ChEstIQU_Y}
\end{align}
where $\mm{Y}_{i} \in \mathbb{C}^{N \times T}$ is the received training signal at the $i$th BS. Here, $\rho_\mathrm{TR}$ is the transmit training signal-to-noise ratio (SNR) and $\check{\vv{w}}_i = \vv{W}_i \vv{x}_k \sim \mathcal{C} \mathcal{N} \left(\mm{0}, \mm{I}_N\right)$, where $\vv{W}_i \in \mathbb{C}^{N \times T}$ is the AWGN at the $i$th BS during the training period. In this paper, for IQU channel estimation, we assume that the estimator tries to estimate ${\vv{g}}_{i,i,k} = \xi_{A,i,k} \mm{\Psi}_{A,i} \vv{h}_{i,i,k}$ as the desired channel between the $k$th UT in the $i$th cell and the $i$th BS. Since the IQU estimator does not process the real and imaginary parts of the received training sequence separately, it can consider only one component, i.e., $\xi_{A,i,k} \mm{\Psi}_{A,i} \vv{h}_{i,i,k}$ of the equivalent channel vector. We note that with the conventional complex-valued system model, which is assumed for the IQU-MMSE estimator, it is not possible to fully model the equivalent channel vector, which comprises both the actual channel and IQI. Taking this into account and considering ${\vv{y}}_{i,k}$ as the observation, the MMSE channel estimate can be expressed as \cite{Kay_StSiProDet2013}
\vspace*{-4mm}
\begin{align}
\hat{\vv{g}}_{i,i,k} = \mm{\Phi}_{\vv{g}_{i,i,k} {\vv{y}}_{i,k}} \left(\mm{\Phi}_{{\vv{y}}_{i,k} {\vv{y}}_{i,k}}\right)^{-1} {\vv{y}}_{i,k},
\label{Eqn_ChEstIQU1}
\end{align}
where $\mm{\Phi}_{\vv{g}_{i,i,k} {\vv{y}}_{i,k}}$ is the cross-correlation matrix of the desired channel estimate and the observation, and given by
\vspace*{-4mm}
\begin{align}
\mm{\Phi}_{\vv{g}_{i,i,k} {\vv{y}}_{i,k}} = \mathbb{E} \left\lbrace {\vv{g}}_{i,i,k} {\vv{y}}_{i,k}^\herm \right\rbrace = \sqrt{\rho_\mathrm{TR}} \xi_{A,i,k} \left(\xi_{A,i,k}^* + \xi_{B,i,k}^* \right) \mm{\Psi}_{A,i} \mm{R}_{i,i,k} \mm{\Psi}^\herm_{A,i}.
\label{Eqn_ChEstIQU_GY}
\end{align}
The auto-correlation matrix of the received signal in (\ref{Eqn_ChEstIQU_Y}) can be expressed as
\vspace*{-4mm}
\begin{align}
\mm{\Phi}_{{\vv{y}}_{i,k} {\vv{y}}_{i,k}}&=\mathbb{E} \left\lbrace{\vv{y}}_{i,k} {\vv{y}}_{i,k}^\herm \right\rbrace = \sum_{l=1}^{L} \rho_\mathrm{TR} \big( | \xi_{A,l,k} +  \xi_{B,l,k}  |^2 \mm{\Psi}_{A,i} \mm{R}_{i,l,k} \mm{\Psi}^\herm_{A,i} + | \xi_{A,l,k} +  \xi_{B,l,k}  |^2 \mm{\Psi}_{B,i} \mm{R}^*_{i,l,k} \nonumber \\
& \times \mm{\Psi}^\herm_{B,i} \big) + \mm{\Psi}_{A,i} \mm{\Psi}^\herm_{A,i} + \mm{\Psi}_{B,i} \mm{\Psi}^\herm_{B,i}. 
\label{Eqn_ChEstIQU_YY}
\end{align}
Now, we substitute (\ref{Eqn_ChEstIQU_GY}) and (\ref{Eqn_ChEstIQU_YY}) into (\ref{Eqn_ChEstIQU1}) and obtain the following expression for IQU-MMSE estimation of the $k$th UT's channel vector
\vspace*{-3mm}
\begin{align}
\hat{\vv{g}}_{i,i,k} =& \mm{\Omega}_{i,k} \bigg( \mm{\Psi}_{A,i} \Big( \sum_{l=1}^{L} \vv{h}_{i,l,k}  \big( \xi_{A,l,k} +  \xi_{B,l,k}  \big) + \frac{1}{\sqrt{\rho_\mathrm{TR}}}\check{\vv{w}}_i \Big) + \mm{\Psi}_{B,i} \Big( \sum_{l=1}^{L} \vv{h}^*_{i,l,k}  \big( \xi^*_{A,l,k} +  \xi^*_{B,l,k} \big) + \nonumber \\
& \frac{1}{\sqrt{\rho_\mathrm{TR}}}\check{\vv{w}}^*_i \Big) \bigg),
\label{Eqn_ChEstIQU_PilCo_ghat}
\end{align}
where deterministic matrix $\mm{\Omega}_{i,k}$ is given by
\vspace*{-4mm}
\begin{align}
\mm{\Omega}_{i,k} \triangleq & \xi_{A,i,k} \left(\xi_{A,i,k}^* + \xi_{B,i,k}^* \right) \mm{\Psi}_{A,i} \mm{R}_{i,i,k} \mm{\Psi}^\herm_{A,i} \Bigg( \sum_{l=1}^{L} \Big( \big( | \xi_{A,l,k} +  \xi_{B,l,k}  |^2 \mm{\Psi}_{A,i} \mm{R}_{i,l,k} \mm{\Psi}^\herm_{A,i} \nonumber \\
&+ | \xi_{A,l,k} +  \xi_{B,l,k}  |^2 \mm{\Psi}_{B,i} \mm{R}^*_{i,l,k} \mm{\Psi}^\herm_{B,i} \big) + \frac{1}{\rho_\mathrm{TR}} \big( \mm{\Psi}_{A,i} \mm{\Psi}^\herm_{A,i} + \mm{\Psi}_{B,i} \mm{\Psi}^\herm_{B,i} \big) \Big) \Bigg)^{-1}.
\label{Eqn_ChEstIQU_Omega}
\end{align}
If pilot contamination is absent, (\ref{Eqn_ChEstIQU_PilCo_ghat}) reduces to 
\vspace*{-4mm}
\begin{align}
\hat{\vv{g}}_{i,i,k} =& \mm{\Omega}^\prime_{i,k} \bigg( \mm{\Psi}_{A,i} \Big( \vv{h}_{i,i,k}  \big( \xi_{A,i,k} \hspace*{-1mm}+\hspace*{-1mm}  \xi_{B,i,k}  \big) \hspace*{-1mm}+\hspace*{-1mm} \frac{1}{\sqrt{\rho_\mathrm{TR}}} \check{\vv{w}}_i \Big) \hspace*{-1mm}+\hspace*{-1mm} \mm{\Psi}_{B,i} \Big( \vv{h}^*_{i,i,k}  \big( \xi^*_{A,i,k} +  \xi^*_{B, i,k} \big) \hspace*{-1mm}+\hspace*{-1mm} \frac{1}{\sqrt{\rho_\mathrm{TR}}} \check{\vv{w}}^*_i \Big) \Bigg), 
\label{Eqn_ChEstIQU_NoPilCo_ghat}
\vspace*{-3mm}
\end{align}
where the deterministic matrix $\mm{\Omega}^\prime_{i,k}$ is equal to $\mm{\Omega}_{i,k}$ if we set $L=1$ and $l=i$ in (\ref{Eqn_ChEstIQU_Omega}).
%
\vspace*{-4mm}
\subsection{Data Detection}{\label{Subsec_DataDet}}
In this subsection, we investigate IQU-MMSE data detection. The IQU-MMSE detector adopted here is the conventional single-cell MMSE detector. The IQU-MMSE detection vector for the $k$th UT at the $i$th BS is given by
\vspace*{-4mm}
\begin{align}
\vv{u}_{i,k} = \hat{\vv{g}}^\herm_{i,i,k} \left( \hat{\mm{G}}_{i,i} \hat{\mm{G}}^\herm_{i,i} + \frac{1}{\rho_\mathrm{UL}} \mm{I}_N \right)^{-1},
\end{align}
where the $k$th column of the estimated channel matrix $\hat{\mm{G}}_{i,i}$ is $\hat{\vv{g}}_{i,i,k}$ and given in (\ref{Eqn_ChEstIQU_PilCo_ghat}) and (\ref{Eqn_ChEstIQU_NoPilCo_ghat}) for the cases with and without pilot contamination, respectively. Thus, the detected signal corresponding to the $k$th UT in the $i$th cell at the output of the IQU-MMSE detector of the $i$th BS can be expressed as
\vspace*{-4mm}
\begin{align}
\check{d}^\mathrm{IQU}_{i, k} =& \vv{u}_{i,k} \vv{r}_i = \vv{u}_{i,k} \Bigg( \sum_{l=1}^{L}  \mm{\Psi}_{A,i} \Big( \sqrt{\rho_{\mathrm{UL}}} \vv{H}_{i,l} \left( \mm{\Xi}_{A,l} \vv{d}_{l} + \mm{\Xi}_{B,l} \vv{d}^*_{l} \right) + \vv{z}_{i} \Big) + \mm{\Psi}_{B,i} \Big( \sqrt{\rho_{\mathrm{UL}}} \vv{H}^*_{i,l}  \big( \mm{\Xi}_{A,l}^* \vv{d}^*_{l} +  \nonumber \\
& {\Xi}^*_{B,l} \vv{d}_{l} \big) + \vv{z}^*_{i} \Big) \Bigg).
\label{Eqn_IQU_EstDataWithDet}
\end{align}
%
\subsection{Asymptotic Sum Rate Analysis}\label{Subsec_IQU_Asy}
The performance metric considered in this paper is the ergodic sum rate, which is a commonly used metric for performance evaluation of wireless communication systems. For the $i$th cell, the ergodic sum rate is given by  
\vspace*{-6mm}
\begin{align}
\bar{R}^\mathrm{IQU}_i = \sum_{k=1}^{K} \mathbb{E} \left\lbrace \mathrm{log}_2 \left(1+ \mathrm{SINR}^\mathrm{IQU}_{i,k} \right) \right\rbrace,
\end{align}
where the expectation is taken with respect to the channel realizations. $\mathrm{SINR}^\mathrm{IQU}_{i,k}$ is the signal-to-noise-plus-interference ratio (SINR) for the $k$th UT in the $i$th cell at the $i$th BS and given by
\vspace*{-3mm}
\begin{align}
\mathrm{SINR}^\mathrm{IQU}_{i,k} = \frac{ S^\mathrm{IQU}_{i,k}}{I^\mathrm{IQU}_{i,k} + Z^\mathrm{IQU}_{i,k}},
\end{align}
where $S^\mathrm{IQU}_{i,k}$, $I^\mathrm{IQU}_{i,k}$, and $Z^\mathrm{IQU}_{i,k}$ are the useful signal power, interference power, and noise power for the $k$th UT in the $i$th cell, respectively. In this paper, using results from random matrix theory, first, an analytical expression for $\mathrm{SINR}^{\mathrm{IQU}^\circ}_{i,k}$, the asymptotic value of $\mathrm{SINR}^\mathrm{IQU}_{i,k}$ for large numbers of antennas $N$, is derived. Then, using $\mathrm{SINR}^{\mathrm{IQU}^\circ}_{i,k}$, the asymptotic sum rate is calculated as
\vspace*{-4mm}
\begin{align}
R^{\mathrm{IQU}^\circ}_i = \sum_{k=1}^{K} \mathrm{log}_2 \left(1+ \mathrm{SINR}^{\mathrm{IQU}^\circ}_{i,k} \right).
\vspace*{-4mm}
\label{Eqn_AsySumRate_IQU}
\end{align}
In the following Theorem, we provide an analytical expression for the asymptotic SINR of the IQU-MMSE detector.
\begin{theorem} \label{Theorem_IQU_SINR}
In an uplink multi-cell massive MIMO system employing an IQU-MMSE receiver at the BS in the presence of IQI, the asymptotic SINR corresponding to the $k$th UT in the $i$th cell for $N \rightarrow \infty$ is given by
\vspace*{-6mm}
\begin{align}
\mathrm{SINR}^{\mathrm{IQU}^\circ}_{i,k} = \frac{ S^{\mathrm{IQU}^\circ}_{i,k}}{I^{\mathrm{IQU}^\circ}_{i,k} + Z^{\mathrm{IQU}^\circ}_{i,k}},
\end{align}
where the asymptotic useful signal power, $ S^{\mathrm{IQU}^\circ}_{i,k}$, the asymptotic interference power, $ I^{\mathrm{IQU}^\circ}_{i,k}$, and the asymptotic noise power, $ Z^{\mathrm{IQU}^\circ}_{i,k}$, are given by
\vspace*{-2mm}
\begin{align}
S^{\mathrm{IQU}^\circ}_{i,k} =& \lim_{N \rightarrow \infty} S^{\mathrm{IQU}}_{i,k}=\frac{\rho_\mathrm{UL}}{\left(1+\delta_{i,k}\right)^2} \left( \left| \xi_{A,i,k} \lambda^{(A)}_{i,i,k} \right|^2 + \left| \xi^*_{B,i,k} \lambda^{(B)}_{i,i,k} \right|^2 \right) \label{Eqn_IQU_SigPow} \\
I^{\mathrm{IQU}^\circ}_{i,k} =& \lim_{N \rightarrow \infty} I^{\mathrm{IQU}}_{i,k}= \frac{\rho_\mathrm{UL}}{\left(1+\delta_{i,k}\right)^2} \Bigg(\left| \xi_{B,i,k} {\lambda}^{(A)}_{i,i,k} \right|^2 + \left| \xi^*_{A,i,k} {\lambda}^{(B)}_{i,i,k} \right|^2 + \mathop{\sum_{q=1}^{K}\sum_{l=1}^{L}}_{(q, l) \neq (k, i)} \left( \left|\xi_{A,l,q}\right|^2 + \left|\xi_{B,l,q}\right|^2\right) \nonumber \\
& \times \left(\varrho^{(A)}_{i,l,q} + \varrho^{(B)}_{i,l,q} \right)+ \mathop{\sum_{l=1}^{L}}_{l \neq i} \left(\left|\xi_{A,l,k} \right|^2 + \left| \xi_{B,l,k} \right|^2\right)\left(\lambda^{(A)}_{i,l,k} +\lambda^{(B)}_{i,l,k} \right) \Bigg) \label{Eqn_IQU_Theo2InterfPow} \\
Z^{\mathrm{IQU}^\circ}_{i,k} =& \lim_{N \rightarrow \infty} Z^{\mathrm{IQU}}_{i,k}=\frac{1}{ N^2\left(1+\delta_{i,k}\right)^2} \left( \mathrm{tr}\left( \mm{\Theta}_{i,i,k} \left( \mm{\Gamma}^{\prime^{(A)}}_{i,k}  +  \mm{\Gamma}^{\prime^{(B)}}_{i,k} \right) \right)\right). \label{Eqn_IQU_Theo1_Noise}
\end{align}
Here, $\delta_{i,k}$, $\lambda^{(A)}_{i,l,k}$, $\lambda^{(B)}_{i,l,k}$, and $\varrho^{(A)}_{i,l,q}$ are defined as
\vspace*{-3mm}
\begin{align}
\delta_{i,k} &\triangleq \frac{1}{N} \mathrm{tr} \left( \mm{\Theta}_{i,i,k} \mm{\Gamma}_{i} \right) \label{Eqn_IQU_delta} \\
\lambda^{(A)}_{i,l,k} &\triangleq \frac{1}{N} \left( \xi^*_{A,l,k} + \xi^*_{B,l,k} \right) \mathrm{tr} \left( \mm{\Psi}_{A,i} \mm{R}_{i,l,k} \mm{\Psi}_{A,i}^\herm \mm{\Omega}_{i,k} \mm{\Gamma}_{i} \right) \label{Eqn_IQU_lambda_A} \\
\lambda^{(B)}_{i,l,k} &\triangleq \frac{1}{N} \left( \xi_{A,l,k} + \xi_{B,l,k} \right) \mathrm{tr} \left( \mm{\Psi}_{B,i} \mm{R}^*_{i,l,k} \mm{\Psi}_{B,i}^\herm \mm{\Omega}_{i,k} \mm{\Gamma}_{i} \right) \label{Eqn_IQU_lambda_B} \\
\varrho^{(A)}_{i,l,q} &\triangleq \frac{1}{N^2} \mathrm{tr} \left( \mm{R}_{i,l,q} \mm{\Psi}_{A,i}^\herm \mm{\Gamma}_{i,k}^\prime \mm{\Psi}_{A,i} \right) + \frac{\left|\lambda^{(A)}_{i,l,q}\right|^2 \mathrm{tr} \left( \mm{\Theta}_{i,i,q}  \mm{\Gamma}_{i,k}^\prime \right) }{N^2\left(1+\delta_{i,q}\right)^2} \nonumber \\
& -2\Re{\left\lbrace \frac{\lambda^{(A)}_{i,l,q} \xi_{A,l,q} \mathrm{tr} \left( \mm{R}_{i,l,q} \mm{\Psi}_{A, i}^\herm \mm{\Gamma}^\prime_{i,k} \mm{\Omega}_{i,k} \mm{\Psi}_{A, i} \right) }{N^2\left(1+\delta_{i,q}\right)} \right\rbrace }, \label{Eqn_IQU_Theo1_RhoA}
\end{align}
and $\varrho^{(B)}_{i,l,q}$ is identical to $\varrho^{(A)}_{i,l,q}$ after replacing in (\ref{Eqn_IQU_Theo1_RhoA}) superscript $(A)$ by $(B)$ and $\mm{R}_{i,l,q}$ by $\mm{R}^*_{i,l,q}$. Furthermore, for the case of pilot contamination, $\mm{\Theta}_{i,i,k}=\mathbb{E}\left\lbrace \hat{\vv{g}}_{i,i,k} \hat{\vv{g}}^\herm_{i,i,k} \right\rbrace$ is obtained as
\vspace*{-4mm}
\begin{align}
\mm{\Theta}_{i,i,k} = & \sum_{l=1}^{L} \left( \left| \xi_{A,i,k} + \xi_{B,i,k} \right|^2 \mm{\Omega}_{i,k} \mm{\Psi}_{A,i} \mm{R}_{i,l,k} \mm{\Psi}_{A,i}^\herm \mm{\Omega}_{i,k}^\herm + \left| \xi_{A,i,k} + \xi_{B,i,k} \right|^2 \mm{\Omega}_{i,k} \mm{\Psi}_{B,i} \mm{R}^*_{i,l,k} \mm{\Psi}_{B,i}^\herm \mm{\Omega}_{i,k}^\herm \right) + \nonumber \\
& \frac{1}{\rho_\mathrm{TR}} \left( \mm{\Omega}_{i,k} \mm{\Psi}_{A,i} \mm{\Psi}_{A,i}^\herm \mm{\Omega}_{i,k}^\herm + \mm{\Omega}_{i,k} \mm{\Psi}_{B,i} \mm{\Psi}_{B,i}^\herm \mm{\Omega}_{i,k}^\herm \right),
\label{Eqn_IQU_Theta}
\end{align}
and for the case without pilot contamination, we have
\begin{align}
\mm{\Theta}_{i,i,k} = & \left| \xi_{A,i,k} + \xi_{B,i,k} \right|^2 {\mm{\Omega}^\prime}_{i,k} \mm{\Psi}_{A,i} \mm{R}_{i,i,k} \mm{\Psi}_{A,i}^\herm {\mm{\Omega}^\prime}_{i,k}^\herm + \left| \xi_{A,i,k} + \xi_{B,i,k} \right|^2 {\mm{\Omega}^\prime}_{i,k} \mm{\Psi}_{B,i} \mm{R}^*_{i,i,k} \mm{\Psi}_{B,i}^\herm {\mm{\Omega}^\prime}_{i,k}^\herm + \nonumber \\
& \frac{1}{\rho_\mathrm{TR}} \left( {\mm{\Omega}^\prime}_{i,k} \mm{\Psi}_{A,i} \mm{\Psi}_{A,i}^\herm {\mm{\Omega}^\prime}_{i,k}^\herm + {\mm{\Omega}^\prime}_{i,k} \mm{\Psi}_{B,i} \mm{\Psi}_{B,i}^\herm {\mm{\Omega}^\prime}_{i,k}^\herm \right).
\label{Eqn_IQUnoPilCo_Theta}
\end{align}
Moreover, $\mm{\Gamma}_{i}$ is given by $\mm{T}$ in Lemma \ref{Lemma_RMT2} in Appendix A for $\mm{\Delta}_k = \mm{\Theta}_{i,i,k} / N$, $\alpha=1/(N\rho_\mathrm{UL})$, and $\mm{B}=\mm{0}_N$. Furthermore, $\mm{\Gamma}^\prime_{i, k}$, $\mm{\Gamma}^{\prime^{(A)}}_{i}$, and $\mm{\Gamma}^{\prime^{(B)}}_{i}$ are given by $\mm{T}^\prime$ in Lemma \ref{Lemma_RMT3} in Appendix A after setting $\mm{\Delta}_k = \mm{\Theta}_{i,i,k}/N$ and replacing $\mm{C}$ with $\mm{\Theta}_{i,i,k}$, $\mm{\Psi}_{A,i}\mm{\Psi}_{A,i}^\herm$, and $\mm{\Psi}_{B,i}\mm{\Psi}_{B,i}^\herm$, respectively.
\end{theorem}
\begin{IEEEproof}
Please refer to Appendix B.
\vspace*{0mm}
\end{IEEEproof}
%
\subsection{Asymptotic Sum Rate Analysis for Single-Cell Case}\label{Subsec_IQU_Asy_SingleCell}
Due to the very general setting considered in Theorem \ref{Theorem_IQU_SINR}, the obtained analytical expression for the asymptotic SINR of the IQU-MMSE detector is quite involved. Nevertheless, using these analytical results for performance evaluation is still much more convenient than performing lengthy Monte-Carlo simulations. However, to get some insight for system design, in this subsection, we provide analytical results for the simplified single-cell case with i.i.d. channel vectors and perfect CSI. In particular, we investigate the impact of the IQI at the BS and at the UTs separately to determine whether the IQI at the BS or at UTs is more harmful.
\vspace*{-2mm}
\begin{corollary} \label{Corollary1}
In an uplink single-cell massive MIMO system with i.i.d. channel vectors, perfect CSI, and IQI only at the BS, the asymptotic SINR of the $k$th UT for the IQU-MMSE detector for $K, N \rightarrow \infty, \ K \ll N$, and $\epsilon_n, \theta_n \ll 1$, is given by
\vspace*{-2mm}
\begin{align}
\mathrm{SINR}_\mathrm{IQU-BS}^\circ = \frac{ N^2 \rho_\mathrm{UL} \left( 1 + \sum_{n=1}^{N} \left( \epsilon^2_{n} -1 \right) \sin^2 \frac{\theta_{n}}{2} \right)^2 }{ N \rho_\mathrm{UL} \beta \sum_{n=1}^{N} \left( \epsilon^2_{n}  + \frac{1}{4} \sin^2 \theta_{n} \right) \hspace*{-1mm} + \sum_{n=1}^{N} \left( 1 + \left(\epsilon^2_{n} -1\right) \sin^2 \frac{\theta_{n}}{2} \right)^2},
\label{Eqn_IQU_SINR_Cor1}
\end{align}
where $\epsilon_n$ and $\theta_n$ are the amplitude and phase imbalances of the RF chain of the $n$th antenna at the BS.
\end{corollary}
\begin{IEEEproof}
Please refer to Appendix C.
\end{IEEEproof}
\vspace*{-2mm}
\begin{remark} \label{Remark2_IQU_Coro} For identical IQI at all BS antenna branches, i.e., $\epsilon_n = \epsilon, \theta_n=\theta, \forall n$, the asymptotic SINR of the IQU-MMSE receiver for $K, N \rightarrow \infty, \ K \ll N,$ and $\epsilon_n, \theta_n \ll 1$, is given by
\vspace*{-4mm}
\begin{align}
\mathrm{SINR}_\mathrm{IQU-BS}^\circ = \frac{N \rho_\mathrm{UL}}{ K \rho_\mathrm{UL} \left( \epsilon^2 + \frac{1}{4} \theta^2 \right) + 1}.
\label{Eqn_IQU_SINR_Rem2}
\end{align}
\end{remark}
\begin{IEEEproof}
Substituting $\epsilon_n = \epsilon, \theta_n=\theta, \forall n$ into (\ref{Eqn_IQU_SINR_Cor1}) and considering $\epsilon, \theta \ll 1$ leads to (\ref{Eqn_IQU_SINR_Rem2}).
\end{IEEEproof}
Remark \ref{Remark2_IQU_Coro} reveals that for a fixed number of users $K$, the SINR increases with increasing number of BS antennas $N$.
\vspace*{-2mm}
\begin{remark} \label{Remark1_IQU_Coro} In the absence of IQI, the asymptotic SINR of the IQU-MMSE receiver for $K, N \rightarrow \infty, \ K \ll N$, is given by $\mathrm{SINR}_\mathrm{No-IQI}^\circ = N \rho_\mathrm{UL} $, which is obtained by setting $\epsilon=\theta=0$ in (\ref{Eqn_IQU_SINR_Rem2}).
\end{remark}
\vspace*{-2mm}
\begin{corollary} The asymptotic SINR loss of an IQU-MMSE receiver due to IQI in the uplink massive MIMO system defined in Corollary \ref{Corollary1}, and for $\epsilon_n=\epsilon \ll 1, \theta_n=\theta \ll 1, \forall n \in\{1,\ldots,N\}$, is given by
\vspace*{-4mm}
\begin{align}
\lim_{\beta \rightarrow 0} \Delta^\circ_\mathrm{SINR} = \lim_{\beta \rightarrow 0} \frac{\mathrm{SINR}_\mathrm{No-IQI}^\circ}{\mathrm{SINR}_\mathrm{IQU}^\circ} = K \rho_\mathrm{UL} \left(\epsilon^2 + \frac{1}{4} \theta^2 \right) + 1.
\label{Eqn_Coro2}
\end{align}
\end{corollary}
\begin{IEEEproof}
Considering (\ref{Eqn_IQU_SINR_Cor1}), Remark \ref{Remark1_IQU_Coro}, and performing simple mathematical manipulations yields (\ref{Eqn_Coro2}).
\end{IEEEproof}
From (\ref{Eqn_Coro2}), it can be seen that the SINR loss of the IQU-MMSE detector compared to the ideal case without IQI does not vanish even
in the asymptotic scenario where the number of the BS antennas is much larger than the number of the UTs. This motivates the need for a receiver, which mitigates both multi-user interference and IQI, cf. Section \ref{Sec_IQA_Detector}. 
\begin{corollary} \label{CoCorollary3}
In an uplink single-cell massive MIMO system with i.i.d. channel vectors, perfect CSI, IQI only at the UTs, and equal amplitude and phase mismatches, i.e., $\check{\epsilon}_k = \check{\epsilon}, \check{\theta}_k=\check{\theta}, \forall k$, the asymptotic SINR of the $k$th UT for the IQU-MMSE detector for $K, N \rightarrow \infty, \ K \ll N$, and $\check{\epsilon}, \check{\theta} \ll 1$, is given by
\vspace*{-4mm}
\begin{align}
\mathrm{SINR}_\mathrm{IQU-UT}^\circ = \frac{N \rho_\mathrm{UL}}{N \rho_\mathrm{UL}\left(\check{\epsilon}^2 + \frac{1}{4} {\check{\theta}}^2 \right)+1}.
\label{Eqn_Coro3_IQU}
\end{align}
\end{corollary}
\begin{IEEEproof}
Please refer to appendix D.
\end{IEEEproof}
From Corollary \ref{CoCorollary3}, we observe that, in an uplink massive MIMO system with IQI only at the UTs, for $N \rightarrow \infty,$ the SINR does not increase with increasing number of BS antennas. Moreover, by comparing (\ref{Eqn_Coro3_IQU}) and (\ref{Eqn_IQU_SINR_Rem2}) we conclude that assuming identical values for the amplitude and phase imbalances at the BS and the UTs, and $K \ll N$, the SINR of a massive MIMO system with IQI only at the BS is higher than that of a massive MIMO system with IQI only at the UTs. Thus, if not compensated, IQI at the UTs is more harmful than IQI at the BS.

%
\vspace*{-2mm}
\section{I/Q Imbalance Aware WLMMSE Receiver} \label{Sec_IQA_Detector}
In this section, the proposed IQA-WLMMSE receiver, which comprises an IQA-WLMMSE channel estimator and an IQA-WLMMSE data detector, is presented. The IQA-WLMMSE channel estimator and detector process the real and imaginary parts of the received signals separately. The widely-linear filtering is necessary for mitigation of the IQI, since the real and imaginary parts of the signals are affected differently by the IQI. 
%
\vspace*{-4mm}
\subsection{Channel Estimation} \label{SubSection_IQA_ChEst}
The proposed IQA-WLMMSE channel estimation scheme is the widely-linear extension of the strictly-linear IQU-MMSE channel estimator introduced in Section \ref{Subsec_ChEst} and performs joint IQI compensation and MMSE channel estimation. 
Since we assume full pilot reuse, UTs in different cells with the same UT index $k$ use the same training sequence $\vv{x}_k$ resulting in the same augmented training sequence $\tilde{\mm{X}}_k = \begin{bmatrix} \Re \left\lbrace \vv{x}_k \right\rbrace \ \ &-\Im \left\lbrace \vv{x}_k \right\rbrace \\ \Im \left\lbrace \vv{x}_k \right\rbrace \ \ & \Re \left\lbrace \vv{x}_k \right\rbrace \end{bmatrix} \in \mathbb{R}^{2T \times 2}$ in the augmented system model.
In order to mitigate the interference from other UTs, the received training signal, $\tilde{\mm{Y}}_{i}$, is multiplied by the augmented training sequence $\tilde{\mm{X}}_k$. Considering (\ref{Eqn_SysMod_IQA}), and taking into account the orthonormality of the pilot sequences, the training signal of the $k$th UT received at the $i$th BS is given by
\vspace*{-4mm}
\begin{align}
{\tilde{\check{\mm{Y}}}}_{i,k} = {\tilde{\mm{Y}}}_{i} \tilde{\mm{X}}_k = \sum_{l=1}^{L} \sqrt{\rho_\mathrm{TR}} \tilde{\mm{\Psi}}_{i} \tilde{\vv{H}}_{i,l,k} \mm{\Xi}_{l,k}  + \tilde{\mm{\Psi}}_{i} \tilde{\check{\mm{W}}}_i
\label{Eqn_ChEstIQA_Y_Mod}
\end{align}
with $\tilde{\check{\mm{W}}}_i  = \tilde{\mm{W}}_i \tilde{\mm{X}}_k $, where $\tilde{\mm{W}}_i \in \mathbb{C}^{2N\times2T}$ is the augmented AWGN matrix at the $i$th BS. In (\ref{Eqn_ChEstIQA_Y_Mod}), $\tilde{\vv{H}}_{i,l,k} \in \mathbb{R}^{2N \times 2}$ represents the augmented channel between the $k$th UT in the $l$th cell and the $i$th BS and is defined as
\vspace*{-3mm}
\begin{align}
\tilde{\vv{H}}_{i,l,k} = \begin{bmatrix}
\Re \left\lbrace \vv{h}_{i,l,k} \right\rbrace  & -\Im \left\lbrace \vv{h}_{i,l,k} \right\rbrace\\
\Im \left\lbrace \vv{h}_{i,l,k} \right\rbrace & \Re \left\lbrace \vv{h}_{i,l,k} \right\rbrace\\
\end{bmatrix} = \check{\tilde{\mm{R}}}_{i,l,k} \begin{bmatrix}
\Re \left\lbrace \vv{\nu}_{i,l,k} \right\rbrace  & -\Im \left\lbrace \vv{\nu}_{i,l,k} \right\rbrace\\
\Im \left\lbrace \vv{\nu}_{i,l,k} \right\rbrace & \Re \left\lbrace \vv{\nu}_{i,l,k} \right\rbrace\\
\end{bmatrix},
\end{align}
where $\vv{\nu}_{i,l,k} \sim \mathcal{C} \mathcal{N}\left( \vv{0}, \mm{I}_N \right)$ and $\check{\tilde{\mm{R}}}_{i,l,k}$ is the square root of the augmented channel covariance matrix between the $k$th UT in the $l$th cell and the $i$th BS. Here, $\check{\tilde{\mm{R}}}_{i,l,k} \in \mathbb{R}^{2N \times 2N}$ is given by
\vspace*{-3mm}
\begin{align}
\check{\tilde{\mm{R}}}_{i,l,k} = \begin{bmatrix}
\Re \left\lbrace \check{\mm{R}}_{i,l,k} \right\rbrace  & -\Im \left\lbrace \check{\mm{R}}_{i,l,k} \right\rbrace\\
\Im \left\lbrace \check{\mm{R}}_{i,l,k} \right\rbrace & \Re \left\lbrace \check{\mm{R}}_{i,l,k} \right\rbrace\\
\end{bmatrix}.
\end{align}
The proposed WLMMSE channel estimator estimates the equivalent augmented channel matrix of the $k$th UT, which is given by $\tilde{\tilde{\vv{g}}}_{i,i,k}=[ \tilde{\vv{g}}_{i,i,k} $ $\tilde{\vv{g}}_{i,i,k+K} ] = \tilde{\mm{\Psi}}_{i} \tilde{\vv{H}}_{i,i,k} \mm{\Xi}_{i,k} $, where $\tilde{\mm{\Psi}}_{i}$, $\tilde{\vv{H}}_{i,i,k}$, and $\mm{\Xi}_{i,k}$ represent the IQI at the $i$th BS, the actual augmented channel of the $k$th UT, and the IQI at the $k$th UT, respectively. Note that for the IQA-WLMMSE detector, both $\tilde{\vv{g}}_{i,i,k}$ and $\tilde{\vv{g}}_{i,i,k+K}$ are required and are used for detection of the real and imaginary parts of the data signal, respectively, c.f. Section \ref{SubSection_IQA_DataDet}. The proposed WLMMSE channel estimate is given by
\vspace*{-2mm}
\begin{align}
\left[ \hat{\tilde{{\vv{g}}}}_{i,i,k} \ \ \hat{\tilde{{\vv{g}}}}_{i,i,k+K} \right] = \mm{\Phi}_{\tilde{{\mm{G}}}_{i,i,k} {\tilde{\check{\mm{Y}}}}_{i,k} } \left( \mm{\Phi}_{ \tilde{\check{\mm{Y}}}_{i,k} {\tilde{\check{\mm{Y}}}}_{i,k} } \right)^{-1} {\tilde{\check{\mm{Y}}}}_{i,k},
\label{Eqn_ChEstIQU}
\end{align}
where $\mm{\Phi}_{ {\tilde{\check{\vv{Y}}}}_{i,k} {\tilde{\check{\vv{Y}}}}_{i,k} }$ is the auto-correlation matrix of the received training sequence ${\tilde{\check{\mm{Y}}}}_{i,k}$ and is given by
\vspace*{-5mm}
\begin{align}
\mm{\Phi}_{ {\tilde{\check{\mm{Y}}}}_{i,k} {\tilde{\check{\mm{Y}}}}_{i,k} } & \triangleq \mathbb{E} \left\lbrace {\tilde{\check{\mm{Y}}}}_{i,k} {\tilde{\check{\mm{Y}}}}_{i,k}^\transp \right\rbrace = \rho_\mathrm{TR} \sum_{l=1}^{L} \tilde{\mm{\Psi}}_{i} \mathbb{E} \left\lbrace \tilde{\vv{H}}_{i,l,k} \mm{\Xi}_{l,k} \mm{\Xi}_{l,k}^\transp \tilde{\vv{H}}_{i,l,k}^\transp \right\rbrace \tilde{\mm{\Psi}}_{i}^\transp +  0.5 \tilde{\mm{\Psi}}_{i} \tilde{\mm{\Psi}}_{i}^\transp \xrightarrow[N\rightarrow\infty]{\mathrm{a.s.}} \nonumber \\
& \rho_\mathrm{TR} \sum_{l=1}^{L} \tilde{\mm{\Psi}}_{i} \mm{\Upsilon}_{i,l,k} \tilde{\mm{\Psi}}_{i}^\transp + 0.5 \tilde{\mm{\Psi}}_{i} \tilde{\mm{\Psi}}_{i}^\transp.
\label{Eqn_IQA_ChEst_AutoCorr}
\end{align}
Here, $\mm{\Upsilon}_{i,l,k} \triangleq \mathbb{E} \left\lbrace \tilde{\vv{H}}_{i,l,k} \mm{\Xi}_{l,k} \mm{\Xi}_{l,k}^\transp \tilde{\vv{H}}_{i,l,k}^\transp \right\rbrace$ is obtained as
\vspace*{-4mm}
\begin{align}
\mm{\Upsilon}_{i,l,k} = \check{ \tilde{ \mm{R}}}_{i,l,k} \mathbb{E} \left\lbrace  \tilde{\mm{V}}_{i,l,k} \mm{\Xi}_{l,k} \mm{\Xi}_{l,k}^\transp \tilde{\mm{V}}_{i,l,k}^\transp \right\rbrace \check{ \tilde{ \mm{R}}}_{i,l,k}^\transp \xrightarrow[N\rightarrow\infty]{\mathrm{a.s.}} 0.5 \norm {\mm{\Xi}_{{l,k}}}^2_\mathrm{F} \tilde{\mm{R}}_{i,l,k},
\end{align}
where $\tilde{\mm{V}}_{i,l,k}=\begin{bmatrix}
\Re \left\lbrace \vv{\nu}_{i,l,k} \right\rbrace  & -\Im \left\lbrace \vv{\nu}_{i,l,k} \right\rbrace\\
\Im \left\lbrace \vv{\nu}_{i,l,k} \right\rbrace & \Re \left\lbrace \vv{\nu}_{i,l,k} \right\rbrace\\
\end{bmatrix}$ and the last step involved Lemma \ref{Lemma_RMT1} and straightforward mathematical operations. In (\ref{Eqn_ChEstIQU}), $\mm{\Phi}_{\tilde{{\mm{G}}}_{i,i,k} {\tilde{\check{\vv{Y}}}}_{i,k} }$ is the cross-correlation between the received training sequence and the desired channel estimate, and is given by
\vspace*{-4mm}
\begin{align}
\mm{\Phi}_{ \tilde{{\mm{G}}}_{i,i,k} {\tilde{\check{\vv{Y}}}}_{i,k} } \hspace*{-3mm} \triangleq \mathbb{E} \left\lbrace \tilde{\tilde{\mm{g}}}_{i,i,k} {\tilde{\check{\mm{Y}}}}_{i,k}^\transp \right\rbrace \hspace*{-1mm} = \hspace*{-1mm} \rho_\mathrm{TR} \tilde{\mm{\Psi}}_{i} \mathbb{E} \left\lbrace \tilde{\vv{H}}_{i,i,k} \mm{\Xi}_{i,k}  \mm{\Xi}_{i,k}^\transp \tilde{\vv{H}}_{i,i,k}^\transp \right\rbrace \tilde{\mm{\Psi}}_{i}^\transp \xrightarrow[N\rightarrow\infty]{\mathrm{a.s.}} 0.5 \rho_\mathrm{TR} \norm {\mm{\Xi}_{{i,k}}}^2_\mathrm{F} \tilde{\mm{\Psi}}_{i}  \tilde{\mm{R}}_{i,i,k} \tilde{\mm{\Psi}}_{i}^\transp.
\label{Eqn_IQA_ChEst_CrossCorr}
\end{align}
Substituting (\ref{Eqn_IQA_ChEst_CrossCorr}) and (\ref{Eqn_IQA_ChEst_AutoCorr}) into (\ref{Eqn_ChEstIQU}) leads to the following expression for the IQA-WLMMSE estimate of the channel between the $k$th UT in the $l$th cell and the $i$th BS
\vspace*{-3mm}
\begin{align}
\left[ \hat{\tilde{{\vv{g}}}}_{i,i,k} \ \ \hat{\tilde{{\vv{g}}}}_{i,i,k+K} \right] =  {\tilde{\mm{\Omega}}}_{i,i,k} \left( \sum_{l=1}^{L} \sqrt{\rho_\mathrm{TR}} \tilde{\mm{\Psi}}_{i} \tilde{\vv{H}}_{i,l,k} \mm{\Xi}_{l,k}  +  \tilde{\mm{\Psi}}_{i} \tilde{\check{\mm{W}}}_i  \right),
\label{Eqn_IQA_ChEst_Omega}
\end{align}
where the IQA-WLMMSE channel estimator ${\tilde{\mm{\Omega}}}_{i,i,k}=\mm{\Phi}_{\tilde{{\mm{G}}}_{i,i,k} {\tilde{\check{\mm{Y}}}}_{i,k} } \left( \mm{\Phi}_{ \tilde{\check{\mm{Y}}}_{i,k} {\tilde{\check{\mm{Y}}}}_{i,k} } \right)^{-1}$ can be expressed as
\vspace*{-4mm}
\begin{align}
{\tilde{\mm{\Omega}}}_{i,i,k} = \norm {\mm{\Xi}_{{i,k}}}^2_\mathrm{F} \tilde{\mm{\Psi}}_{i}  \tilde{\mm{R}}_{i,i,k} \tilde{\mm{\Psi}}_{i}^\transp \left(  \sum_{l^\prime=1}^{L} \norm {\mm{\Xi}_{{l^\prime,k}}}^2_\mathrm{F} \tilde{\mm{\Psi}}_{i} \tilde{\mm{R}}_{i,l^\prime,k} \tilde{\mm{\Psi}}_{i}^\transp + \frac{1}{ \rho_\mathrm{TR}} \tilde{\mm{\Psi}}_{i} \tilde{\mm{\Psi}}_{i}^\transp \right)^{-1}.
\label{Eqn_IQA_OmegaTilde}
\end{align}
\begin{remark} One of the main features of the proposed IQA-WLMMSE channel estimator is that it does not require the explicit knowledge of the IQI at the BS and UTs; indeed, only the equivalent channel covariance matrix and the autocorrelation matrix of the received training signal are needed. This can be observed from (\ref{Eqn_IQA_ChEst_Omega}), where the augmented channel estimate is obtained by filtering the received training signal, i.e., the term in parenthesis, with the IQA-WLMMSE estimator ${\tilde{\mm{\Omega}}}_{i,i,k}$. The IQA-WLMMSE estimator comprises two components: $\mm{\Phi}_{\tilde{{\mm{G}}}_{i,i,k} {\tilde{\check{\mm{Y}}}}_{i,k} }$ and $\mm{\Phi}_{ \tilde{\check{\mm{Y}}}_{i,k} {\tilde{\check{\mm{Y}}}}_{i,k} }$. The first component $\mm{\Phi}_{\tilde{{\mm{G}}}_{i,i,k} {\tilde{\check{\mm{Y}}}}_{i,k} }$, which is given in (\ref{Eqn_IQA_ChEst_CrossCorr}), can be rewritten as $\rho_\mathrm{TR} \mathbb{E} \left\lbrace \tilde{\tilde{\mm{g}}}_{i,i,k} \tilde{\tilde{\mm{g}}}_{i,i,k}^\transp \right\rbrace$, which is the product of the training SNR and the covariance matrix of the augmented equivalent channel $\tilde{\tilde{\mm{g}}}_{i,i,k}$, and both can be estimated using SNR estimation techniques and channel statistics estimation techniques, respectively. The second component in the IQA-WLMMSE channel estimator is $\mm{\Phi}_{ \tilde{\check{\mm{Y}}}_{i,k} {\tilde{\check{\mm{Y}}}}_{i,k} }$, which is the autocorrelation of the received training signal. 
\end{remark}
For the case without pilot contamination, the IQA-WLMMSE channel estimates are given in (\ref{Eqn_IQA_ChEst_Omega}) after setting $L=1$ and $l=l^\prime=i$ by (\ref{Eqn_IQA_ChEst_Omega}) and (\ref{Eqn_IQA_OmegaTilde}).
%
\vspace*{-2mm}
\subsection{Data Detection} \label{SubSection_IQA_DataDet}
The proposed IQA-WLMMSE data detector is the widely-linear extension of the conventional MMSE detector considered in Section {\ref{Subsec_DataDet}}, and employs the estimate of the equivalent channel, $\hat{\tilde{\mm{G}}}_{i,i}=\left[ \hat{\tilde{\vv{g}}}_{i,i,1},\ldots,\hat{\tilde{\vv{g}}}_{i,i,2K} \right]$, which comprises the actual channel, the IQI at the BS, and the IQI at the UT. The IQA-WLMMSE detector includes the filter vectors for the real and imaginary parts of the signal of the $k$th UT at the $i$th BS and is given by
\vspace*{-3mm}
\begin{align}
\begin{bmatrix} \tilde{\mm{u}}_{i,k} \\ \tilde{\mm{u}}_{i,k+K} \end{bmatrix}  &= \begin{bmatrix} \hat{\tilde{\vv{g}}}_{i,i,k}^\transp \\ \hat{\tilde{\vv{g}}}_{i,i,k+K}^\transp \end{bmatrix} \left( \hat{\tilde{\mm{G}}}_{i,i} \hat{\tilde{\mm{G}}}^\transp_{i,i} + \frac{1}{\rho_\mathrm{UL}} \mm{I}_{2N} \right)^{-1}, \label{Eqn_IQA_u_k}
\end{align}
where $\hat{\tilde{\vv{g}}}_{i,i,k}$ and $\hat{\tilde{\vv{g}}}_{i,i,k+K}$ are the $k$th and the $(k+K)$th columns of the estimated augmented channel matrix $\hat{\tilde{\mm{G}}}_{i,i}$, respectively, and are given in (\ref{Eqn_IQA_ChEst_Omega}). Hence, the decision variable at the output of the IQA-WLMMSE detector at the $i$th BS corresponding to the $k$th UT can be expressed as
\vspace*{-2mm}
\begin{align}
\check{d}^\mathrm{IQA}_{i, k} =& \tilde{\mm{u}}_{i,k} \left( \sum_{l=1}^{L} \sqrt{\rho_{\mathrm{UL}}} \tilde{\mm{\Psi}}_{i} \tilde{\mm{H}}_{i,l}  \tilde{\vv{\Xi}}_{l} \tilde{\vv{d}}_{l} + \tilde{\mm{\Psi}}_{i} \tilde{\vv{z}}_{i} \right) + j \tilde{\mm{u}}_{i,k+K} \left( \sum_{l=1}^{L} \sqrt{\rho_{\mathrm{UL}}} \tilde{\mm{\Psi}}_{i} \tilde{\mm{H}}_{i,l}  \tilde{\vv{\Xi}}_{l} \tilde{\vv{d}}_{l} + \tilde{\mm{\Psi}}_{i} \tilde{\vv{z}}_{i} \right).
\label{IQA_DetectedData}
\end{align}
%
\vspace*{-2mm}
\subsection{Asymptotic Sum Rate Analysis} \label{SubSection_IQA_DE}
The ergodic sum rate of the IQA-WLMMSE receiver is given by 
\vspace*{-4mm}
\begin{align}
\bar{R}^\mathrm{IQA}_i = \sum_{k=1}^{K} \mathbb{E} \left\lbrace \mathrm{log}_2 \left(1+ \mathrm{SINR}^\mathrm{IQA}_{i,k} \right) \right\rbrace,
\end{align}
where the expectation is taken with respect to channel realizations. Here, $\mathrm{SINR}^\mathrm{IQA}_{i,k}$ is the SINR of the $k$th UT in the $i$th cell at the $i$th BS and is defined as
\vspace*{-4mm}
\begin{align}
\mathrm{SINR}^{\mathrm{IQA}}_{i,k} = \frac{ S^{\mathrm{IQA}}_{i,k} }{ I^{\mathrm{IQA}}_{i,k} + Z^{\mathrm{IQA}}_{i,k} },
\end{align}
where $S^{\mathrm{IQA}}_{i,k}$, $I^{\mathrm{IQA}}_{i,k}$, and $Z^{\mathrm{IQA}}_{i,k}$ are the useful signal power, interference power, and noise power of the $k$th UT at the $i$th BS, respectively. Using again results from random matrix theory, we will show that the asymptotic sum rate of the IQA-WLMMSE detector can be expressed as
\vspace*{-4mm}
\begin{align}
R^{\mathrm{IQA}^\circ}_i = \sum_{k=1}^{K} \mathrm{log}_2 \left(1+ \mathrm{SINR}^{\mathrm{IQA}^\circ}_{i,k} \right),
\label{Eqn_AsySumRate_IQA}
\end{align}
where the asymptotic SINR expression $\mathrm{SINR}^{\mathrm{IQA}^\circ}_{i,k}$ is provided in the following theorem. In Section \ref{Sec_NumResults}, we show that the derived asymptotic sum rate accurately predicts the ergodic sum rate, which is obtained through lengthy Monte-Carlo simulations.
%
\begin{theorem} \label{Theorem_IQA}
In an uplink multi-cell massive MIMO system employing an IQA-WLMMSE receiver, for $N \rightarrow \infty$, the asymptotic SINR of the $k$th UT in the $i$th cell is given by
\vspace*{-3mm}
\begin{align}
\mathrm{SINR}^{\mathrm{IQA}^\circ}_{i,k} = \frac{S^{\mathrm{IQA}^\circ}_{i,k}}{I^{\mathrm{IQA}^\circ}_{i,k} + Z^{\mathrm{IQA}^\circ}_{i,k}},
\end{align}
where the asymptotic useful signal power, $S^{\mathrm{IQA}^\circ}_{i,k}$, the asymptotic interference power, $I^{\mathrm{IQA}^\circ}_{i,k}$, and the asymptotic noise power, $Z^{\mathrm{IQA}^\circ}_{i,k}$, are given by
\vspace*{-3mm}
\begin{align}
S^{\mathrm{IQA}^\circ}_{i,k} &= \lim_{N \rightarrow \infty} S^{\mathrm{IQA}}_{i,k} = 0.5 \rho_\mathrm{UL}  \left( \frac{  \left|\tilde{\chi}_{i,k} \right|^2 }{\left( 1+ {\tilde{\delta}}_{i,k} \right)^2} +\frac{  \left|\tilde{\chi}_{i,k+K} \right|^2 }{\left( 1+ {\tilde{\delta}}_{i,k+K} \right)^2}\right) \label{Eqn_IQA_AsySiggPow}\\
I^{\mathrm{IQA}^\circ}_{i,k} &=  \lim_{N \rightarrow \infty} I^{\mathrm{IQA}}_{i,k} = 0.5 \rho_\mathrm{UL} \Bigg( \hspace*{-2mm} \mathop{\sum_{l=1, l \neq i}^{L}}_{{q \in \left\lbrace k, k+K \right\rbrace}} \hspace*{-2mm} \frac{\left|\tilde{\lambda}_{i,l,q}\right|^2}{\left(1+\tilde{\delta}_{i,q}\right)^2} + \mathop{\sum_{l=1}^{L}\sum_{q^\prime=1}^{2K}}_{q^\prime \neq k, 2k} \frac{1}{\left(1+\tilde{\delta}_{i,k}\right)^2} \Bigg( \tilde{\zeta}_{i,l,q^\prime} + \frac{\left|\tilde{\lambda}_{i,l,q^\prime}\right|^2 \tilde{\kappa}_{i,i,q^\prime}}{\left(1+\tilde{\delta}_{i,k}\right)^2} \nonumber \\
& -2\Re{\left\lbrace \frac{\tilde{\lambda}_{i,l,q^\prime} \tilde{\phi}_{i,i,q^\prime}}{1+\tilde{\delta}_{i,k}} \right\rbrace } \Bigg) + \frac{1}{\left(1+\tilde{\delta}_{i,k+K}\right)^2} \Bigg( \tilde{\zeta}^\prime_{i,l,q^\prime} + \frac{\left|\tilde{\lambda}_{i,l,q^\prime}\right|^2 \tilde{\kappa}^\prime_{i,i,q^\prime}}{\left(1+\tilde{\delta}_{i,k+K}\right)^2} -2\Re{\left\lbrace \frac{\tilde{\lambda}_{i,l,q^\prime} \tilde{\phi}^\prime_{i,i,q^\prime}}{1+\tilde{\delta}_{i,k+K}} \right\rbrace } \Bigg) \Bigg) \label{Eqn_IQA_AsySigPow}
\\
Z^{\mathrm{IQA}^\circ}_{i,k} &= \lim_{N \rightarrow \infty} Z^{\mathrm{IQA}}_{i,k} = 0.5 \left( \frac{\mathrm{tr}\left( \tilde{\mm{\Phi}}^{(1)}_{i, i, k} \tilde{\mm{\Gamma}}^{\prime\prime\prime}_{i,q} \right) }{4N^2\left(1+\tilde{\delta}_{i,k}\right)^2}    +   \frac{ \mathrm{tr} \left( \tilde{\mm{\Phi}}^{(1)}_{i, i, k+K} \tilde{\mm{\Gamma}}^{\prime\prime\prime}_{i,q} \right) }{4N^2\left(1+\tilde{\delta}_{i,k+K}\right)^2}  \right).
\label{Eqn_IQA_AsyNoisePow}
\end{align}
Here, we use $\tilde{\chi}_{i,k}=\mathrm{tr} \left( \tilde{\mm{\Phi}}^{(2)}_{i,i,k} \tilde{\mm{\Gamma}}_{i} \right)/(2N)$,  $\tilde{\lambda}_{i,l,q}=\mathrm{tr} \left( \tilde{\mm{\Phi}}^{(2)}_{i, l, k} \tilde{\mm{\Gamma}}_{i} \right)/(2N)$, and 
\vspace*{-4mm}
\begin{align}
\tilde{\mm{\Gamma}}_{i} \triangleq \left( \frac{1}{2N} \sum_{k=1}^{K} \left( \frac{\tilde{\mm{\Phi}}^{(1)}_{i,i,k}}{1+\tilde{\delta}_{i,k}} + \frac{ \tilde{\mm{\Phi}}^{(1)}_{i,i,k+K}}{1+\tilde{\delta}_{i,k+K} }\right) + \frac{1}{2N\rho_\mathrm{UL}} \mm{I}_{2N} \right)^{-1},
\label{Eqn_IQA_Gamma}
\end{align}
where $\tilde{\mm{\Phi}}^{(1)}_{i,i,k}$ and $\tilde{\mm{\Phi}}^{(2)}_{i,i,k}$ are defined as
\vspace*{-3mm}
\begin{align}
\tilde{\mm{\Phi}}^{(1)}_{i,i,k} &\triangleq  \tilde{\mm{\Omega}}_{i,i,k} \left( 0.5 \sum_{l=1}^{L} \left( {\left[ \mm{\Xi}_{l,k} \right]}_{1,1}^2 +{\left[ \mm{\Xi}_{l,k} \right]}_{2,1}^2 \right)\tilde{\mm{\Psi}}_{i} \tilde{\mm{R}}_{i,l,k} \tilde{\mm{\Psi}}_{i}^\herm + \frac{1}{2\rho_\mathrm{TR}} \tilde{\mm{\Psi}}_{i} \tilde{\mm{\Psi}}_{i}^\herm \ \right)  \tilde{\mm{\Omega}}_{i,i,k}^\herm  \label{Theo2_Phi1}\\
\tilde{\mm{\Phi}}^{(2)}_{i,i,k} &\triangleq 0.5 \left({\left[ \mm{\Xi}_{i,k} \right]}_{1,1}^2 +{\left[ \mm{\Xi}_{i,k} \right]}_{2,1}^2 \right) \tilde{\mm{\Omega}}_{i,i,k} \tilde{\mm{\Psi}}_{i} \tilde{\mm{R}}_{i,i,k} \tilde{\mm{\Psi}}_{i}^\herm. \label{Theo2_Phi2}
\end{align}
Furthermore, $\tilde{\mm{\Phi}}^{(1)}_{i,i,k+K}$ and $\tilde{\mm{\Phi}}^{(2)}_{i,i,k+K}$ are obtained by replacing ${\left[ \mm{\Xi}_{l,k} \right]}_{1,1}^2 +{\left[ \mm{\Xi}_{l,k} \right]}_{2,1}^2$ with ${\left[ \mm{\Xi}_{l,k} \right]}_{1,2}^2 +{\left[ \mm{\Xi}_{l,k} \right]}_{2,2}^2$ and ${\left[ \mm{\Xi}_{i,k} \right]}_{1,1}^2 +{\left[ \mm{\Xi}_{i,k} \right]}_{2,1}^2$ with ${\left[ \mm{\Xi}_{i,k} \right]}_{1,2}^2 +{\left[ \mm{\Xi}_{i,k} \right]}_{2,2}^2$ in (\ref{Theo2_Phi1}) and (\ref{Theo2_Phi2}), respectively. In (\ref{Eqn_IQA_Gamma}), $\tilde{\delta}_{i,k}$ and $\tilde{\delta}_{i,k+K}$ are the solutions to the following fixed-point equations
\vspace*{-2mm}
\begin{align}
\tilde{\delta}_{i,k} &\triangleq \frac{1}{2N} \mathrm{tr} \left( \tilde{\mm{\Phi}}^{(1)}_{i,i,k} \left( \frac{1}{2N} \sum_{k=1}^{K} \left( \frac{\tilde{\mm{\Phi}}^{(1)}_{i,i,k}}{1+\tilde{\delta}_{i,k}} + \frac{ \tilde{\mm{\Phi}}^{(1)}_{i,i,k+K}}{1+\tilde{\delta}_{i,k+K} }\right) + \frac{1}{2N\rho_\mathrm{UL}} \mm{I}_{2N} \right)^{-1} \right) \label{Eqn_IQA_delta0} \\
\tilde{\delta}_{i,k+K} &\triangleq \frac{1}{2N} \mathrm{tr} \left( \tilde{\mm{\Phi}}^{(1)}_{i,i,k+K} \left( \frac{1}{2N} \sum_{k=1}^{K} \left( \frac{\tilde{\mm{\Phi}}^{(1)}_{i,i,k}}{1+\tilde{\delta}_{i,k}} + \frac{ \tilde{\mm{\Phi}}^{(1)}_{i,i,k+K}}{1+\tilde{\delta}_{i,k+K} }\right) + \frac{1}{2N\rho_\mathrm{UL}} \mm{I}_{2N} \right)^{-1} \right).
\end{align}
Moreover, $\tilde{\zeta}_{i,l,q}$, $\tilde{\kappa}_{i,i,q}$, and $\tilde{\phi}_{i,i,q}$ in (\ref{Eqn_IQA_AsySigPow}) are given by $\mathrm{tr} \left( \mm{B} \tilde{\mm{\Gamma}}^\prime_{i,q} \right) / (4N^2)$, where $\mm{B}$ is equal to $\tilde{\mm{R}}_{i,l,q}$, $\tilde{\mm{\Phi}}^{(1)}_{i,i,k}$, and $\tilde{\mm{\Omega}}_{i,k} \tilde{\mm{R}}_{i,l,q}$, respectively, and $\tilde{\mm{\Gamma}}^\prime_{i,q}$ is given by $\mm{T}^\prime$ in Lemma \ref{Lemma_RMT3} after replacing $N$ by $2N$ and setting $\mm{T}=\tilde{\mm{\Gamma}}_i$, $\mm{C}=\tilde{\mm{\Phi}}^{(1)}_{i,i,k}$, and $\mm{\Delta}_k=\tilde{\mm{\Phi}}^{(1)}_{i,i,k}/(2N)$. Similarly, $\tilde{\zeta}^\prime_{i,l,q}$, $\tilde{\kappa}^\prime_{i,i,q}$, and $\tilde{\phi}^\prime_{i,i,q}$ are given by $\mathrm{tr} \left( \mm{B} \tilde{\mm{\Gamma}}^{\prime\prime}_{i,q} \right) / (4N^2)$, where $\mm{B}$ is equal to $\tilde{\mm{R}}_{i,l,q}$, $\tilde{\mm{\Phi}}^{(1)}_{i,i,k}$, and $\tilde{\mm{\Omega}}_{i,k} \tilde{\mm{R}}_{i,l,q}$, respectively, and $\tilde{\mm{\Gamma}}^{\prime\prime}_{i,q}$ is equal to $\mm{T}^\prime$ in Lemma \ref{Lemma_RMT3} after replacing $N$ by $2N$ and setting $\mm{T}=\tilde{\mm{\Gamma}}_i$, $\mm{C}=\tilde{\mm{\Phi}}^{(1)}_{i,i,k+K}$, and $\mm{\Delta}_k=\tilde{\mm{\Phi}}^{(1)}_{i,i,k+K}/(2N)$. Furthermore, $\tilde{\mm{\Gamma}}^{\prime\prime\prime}_{i,q}$ in (\ref{Eqn_IQA_AsyNoisePow}) is equal to $\mm{T}^\prime$ in Lemma \ref{Lemma_RMT3} after replacing $N$ by $2N$ and setting $\mm{T}=\tilde{\mm{\Gamma}}_i$, $\mm{C}=\tilde{\mm{\Psi}}_i$, and $\mm{\Delta}_k=\tilde{\mm{\Phi}}^{(1)}_{i,i,k}/(2N)$. We note that if pilot contamination is not present, $l$ and $L$ in (\ref{Theo2_Phi1}) are set to $i$ and 1, respectively, and $\tilde{\mm{\Omega}}_{i,i,k}$ is obtained from (\ref{Eqn_IQA_OmegaTilde}) after setting $l^\prime=i$ and $L=1$.
\end{theorem}
\vspace*{-2mm}
\begin{IEEEproof}
Please refer to Appendix E.
\vspace*{0mm}
\end{IEEEproof}
%
\subsection{Asymptotic Sum Rate Analysis for the Single-Cell Case} \label{SubSection_IQA_DE_SingleCell}
Although the provided asymptotic sum rate expression is easy to evaluate numerically, since Theorem \ref{Theorem_IQA} considers a very general case, it does not offer much insight for system design. Hence, in order to get some insight regarding the influence of IQI on the performance of uplink massive MIMO systems employing the IQA-WLMMSE receiver, similar to the analysis for the conventional IQU-MMSE receiver, we consider the simpler single-cell case with perfect CSI, and i.i.d. channels for all UTs. In particular, in order to investigate the influence of the IQI on the performance, the cases with IQI present only at the BS and only at UTs are analyzed separately and their asymptotic SINRs and the corresponding improvements compared to the conventional IQU-MMSE receivers are evaluated in the following Corollaries.
%
\vspace*{-2mm}
\begin{corollary} \label{Corollary_IQA}
In an uplink single-cell massive MIMO system with i.i.d. channel vectors, perfect CSI, and IQI only at the BS, the asymptotic SINR of the $k$th UT for the proposed IQA-WLMMSE receiver for $K, N \rightarrow \infty, \ K \ll N$, and $\epsilon_n, \theta_n \ll 1$, is given by
\vspace*{-4mm}
\begin{align}
\mathrm{SINR}^{\mathrm{IQA-BS}^\circ}_{k} = \frac{N \rho_\mathrm{UL} \left(1+\frac{1}{N}\sum_{n=1}^N\epsilon_{n}^2\right)^2 }{1+\frac{1}{N}\sum_{n=1}^{N} \left( \left( 6 - 2\sin^2 \theta_n \right)\epsilon_n^2 + \sin^2 \theta_n \right)},
\label{Eqn_IQA_SINR_Cor2}
\end{align}
where $\epsilon_n$ and $\theta_n$ are the amplitude and phase imbalances of the RF chain of the $n$th antenna at the BS.
\end{corollary}
\begin{IEEEproof}
Please refer to Appendix F. 
\end{IEEEproof}
\vspace*{-2mm}
\remark For the system described in Corollary \ref{Corollary_IQA} and identical IQI at all BS antenna branches, i.e., $\epsilon_n=\epsilon, \theta_n=\theta, \forall n$, the asymptotic SINR is given by
\begin{align}
\mathrm{SINR}^{\mathrm{IQA-BS}^\circ}_{k} = \frac{N \rho_\mathrm{UL} \left(1+ \epsilon^2\right)^2 }{1+ \left( 6 - 2\sin^2 \theta \right)\epsilon^2 + \sin^2 \theta}.
\label{Eqn_IQA_SINR_Cor2}
\end{align}
Substituting $\epsilon=\theta=0$ into (\ref{Eqn_IQA_SINR_Cor2}) leads to $\mathrm{SINR}^{\mathrm{IQA}^\circ}_{k} = N \rho_\mathrm{UL}$ for an ideal system without IQI. From (\ref{Eqn_IQA_SINR_Cor2}), we observe that with increasing number of BS antennas, the sum rate of the proposed IQA-WLMMSE receiver increases too. In particular, it can be shown that for $\epsilon_{n}, \theta_n \ll 1$, which is valid for typical IQI values, the asymptotic SINR in (\ref{Eqn_IQA_SINR_Cor2}) is smaller than but very close to $N \rho_\mathrm{UL}$, i.e., the sum rate of the ideal system without IQI. In addition, comparing the SINR of the IQA-WLMMSE receiver with IQI present only at the BS given in (\ref{Eqn_IQA_SINR_Cor2}) and the corresponding SINR of the conventional IQU-MMSE receiver given in (\ref{Eqn_IQU_SINR_Rem2}), the following asymptotic SINR loss can be obtained
\vspace*{-4mm}
\begin{align}
\lim_{\beta \rightarrow 0} \Delta^\circ_{\mathrm{SINR-BS}} = \lim_{\beta \rightarrow 0} \frac{\mathrm{SINR}^{\mathrm{IQA-BS}^\circ}_{k}}{\mathrm{SINR}^{\mathrm{IQU-BS}^\circ}_{k}} = \frac{\left( 1+ \epsilon^2 \right)^2 \left( K \rho_\mathrm{UL} \left(\epsilon^2 + \frac{1}{4} \theta^2\right) + 1 \right) }{1+\left(6-2 \sin^2 \theta\right) \epsilon^2 + \sin^2 \theta}.
\label{Eqn_SINR_Loss_BS}
\end{align}
From (\ref{Eqn_SINR_Loss_BS}), it can be observed that for systems with IQI only at the BS, the SINR loss increases with increasing SNR and increasing number of UTs.
\vspace*{-2mm}
\begin{corollary} \label{Corollary_IQA_UT}
In an uplink single-cell massive MIMO system with i.i.d. channel vectors, perfect CSI, IQI only at the UTs, and equal amplitude and phase mismatches, i.e., $\check{\epsilon}_k=\check{\epsilon}, \check{\theta}_k=\check{\theta}, \forall k$, the asymptotic SINR of the $k$th UT for the proposed IQA-WLMMSE receiver for $K, N \rightarrow \infty, \ K \ll N$, and $\check{\epsilon}, \check{\theta} \ll 1$, is given by
\vspace*{-2mm}
\begin{align}
\mathrm{SINR}^{\mathrm{IQA-UT}^\circ}_{k} \hspace*{-3mm} = \hspace*{-1mm} \frac{N \rho_\mathrm{UL} \left(1 + 2\check{\epsilon}^2 \left(1-2\cos^2 \check{\theta} \right) \right)}{2\left( 1+\check{\epsilon}^2 \right)} \hspace*{-1mm} \left( \hspace*{-1mm} \left( \frac{N \rho_\mathrm{UL}\left(1+2\check{\epsilon}\right)}{1+N \rho_\mathrm{UL}\left(1+2\check{\epsilon}\right)} \hspace*{-1mm} \right)^2  \hspace*{-3mm}+\hspace*{-1mm}    \left( \frac{N \rho_\mathrm{UL}\left(1-2\check{\epsilon}\right)}{1+N\rho_\mathrm{UL}\left(1-2\check{\epsilon}\right)} \hspace*{-0.5mm} \right)^2   \right).
\label{Eqn_IQA_Coro_UT}
\end{align}
\end{corollary}
\begin{IEEEproof}
Please refer to appendix G.
\end{IEEEproof}
Substituting typical values for $\check{\epsilon}$ and $\check{\theta}$ into (\ref{Eqn_IQA_Coro_UT}), it can be observed that similar to the system with IQI only at the BS, the asymptotic sum rate of the system with IQI only at the UTs increases with increasing number of BS antennas, and is smaller than but almost identical to the sum rate of an ideal system without IQI for $\check{\epsilon}, \check{\theta} \ll 1$ and $\rho_\mathrm{UL} \gg 1$. Moreover, considering (\ref{Eqn_IQA_Coro_UT}) and (\ref{Eqn_Coro3_IQU}), we obtain the following asymptotic SINR loss of the conventional IQU-MMSE receiver compared to the IQA-WLMMSE receiver in a system, where the IQI is only present at the UTs
\vspace*{-4mm}
\begin{align}
\lim_{\beta \rightarrow 0} \frac{\mathrm{SINR}^{\mathrm{IQA-UT}^\circ}_{k}}{\mathrm{SINR}^{\mathrm{IQU-UT}^\circ}_{k}} \hspace*{-1mm} = \hspace*{-1mm} \frac{ N \rho_\mathrm{UL} \left( \check{\epsilon}^2 + \frac{1}{4} \check{\theta}^2  \right) }{2\left( 1+\check{\epsilon}^2 \right)} \hspace*{-1mm} \left( \hspace*{-1mm} \left( \frac{N \rho_\mathrm{UL}\left(1+2\check{\epsilon}\right)}{1+N\rho_\mathrm{UL}\left(1+2\check{\epsilon}\right)} \right)^2 \hspace*{-3mm} + \hspace*{-1mm}   \left( \frac{N\rho_\mathrm{UL}\left(1-2\check{\epsilon}\right)}{1+N\rho_\mathrm{UL}\left(1-2\check{\epsilon}\right)} \right)^2   \right).
\label{Eqn_SINR_Loss_UT}
\end{align}
From (\ref{Eqn_SINR_Loss_UT}), we observe that, if IQI is present only at the UTs, the SINR loss of the conventional IQU-MMSE receiver compared to the IQA-WLMMSE receiver increases with increasing number of BS antennas, $N$. Substituting typical values for $\check{\epsilon}$ and $\check{\theta}$, it can also be observed that the SINR loss of the conventional IQU-MMSE receiver compared to IQA-WLMMSE receiver in systems with IQI only at the UTs is slightly larger than the corresponding loss in systems, where the IQI is present only at the BS. We validate this observation in Section \ref{Sec_NumResults} for multi-cell systems and more general settings.
%
\vspace*{-2mm}
\section{Numerical Results}\label{Sec_NumResults}
In order to evaluate the performance of the proposed IQA-WLMMSE receiver and to validate our analytical results, Monte-Carlo simulations have been performed. Here, we assume a system consisting of seven hexagonal cells with a normalized cell radius of one. Without loss of generality, we further assume that the central cell is the target cell. In each cell, there is a BS in the cell center and there are $K$ UTs, which are uniformly distributed on a circle with a radius of 2/3. The channel model used here comprises path-loss, antenna correlation, and Rayleigh fading. Moreover, we assume that the BS employs a uniform linear array (ULA) and adopt the ULA channel correlation model used in \cite{Hoydis2013}. In particular, we have $\check{\mm{R}}_{i,l,k}=c_{i,l,k}^{-3/2} [\mm{B} \ \mm{0}_{ N \times \left( N-M \right) }]$, where $c_{i,l,k}$ is the distance between the $k$th UT in the $l$th cell and the $i$th BS, and $\mm{0}_{N \times (N-M)}$ and $M$ are an $N \times (N-M)$ all-zero matrix and the number of dimensions of the antenna's physical model, respectively. We adopt $\mm{B}=[\vv{b}\left(\phi_1\right), \ldots, \vv{b}\left(\phi_M\right) ]$, where the steering vector $\vv{b}\left(\phi_m\right)$ is defined as $\vv{b}\left(\phi_m\right) \hspace*{-1mm} = \hspace*{-1mm} \left[1, \ldots, e^{ -2 \pi j \gamma \left( N-1 \right) \sin\left(\phi_m\right) / \lambda} \right]^\transp / {\sqrt{M}}$
with $\phi_m = -\pi/2+\left(m-1\right)\pi/M$ being the $m$th angle of arrival (AoA), and $\gamma$ and $\lambda$ being the antenna spacing and the wavelength, respectively \cite{Hoydis2013}.
%
%
\begin{figure}[t]
\begin{center}
\includegraphics[width=\linewidth, clip=true]{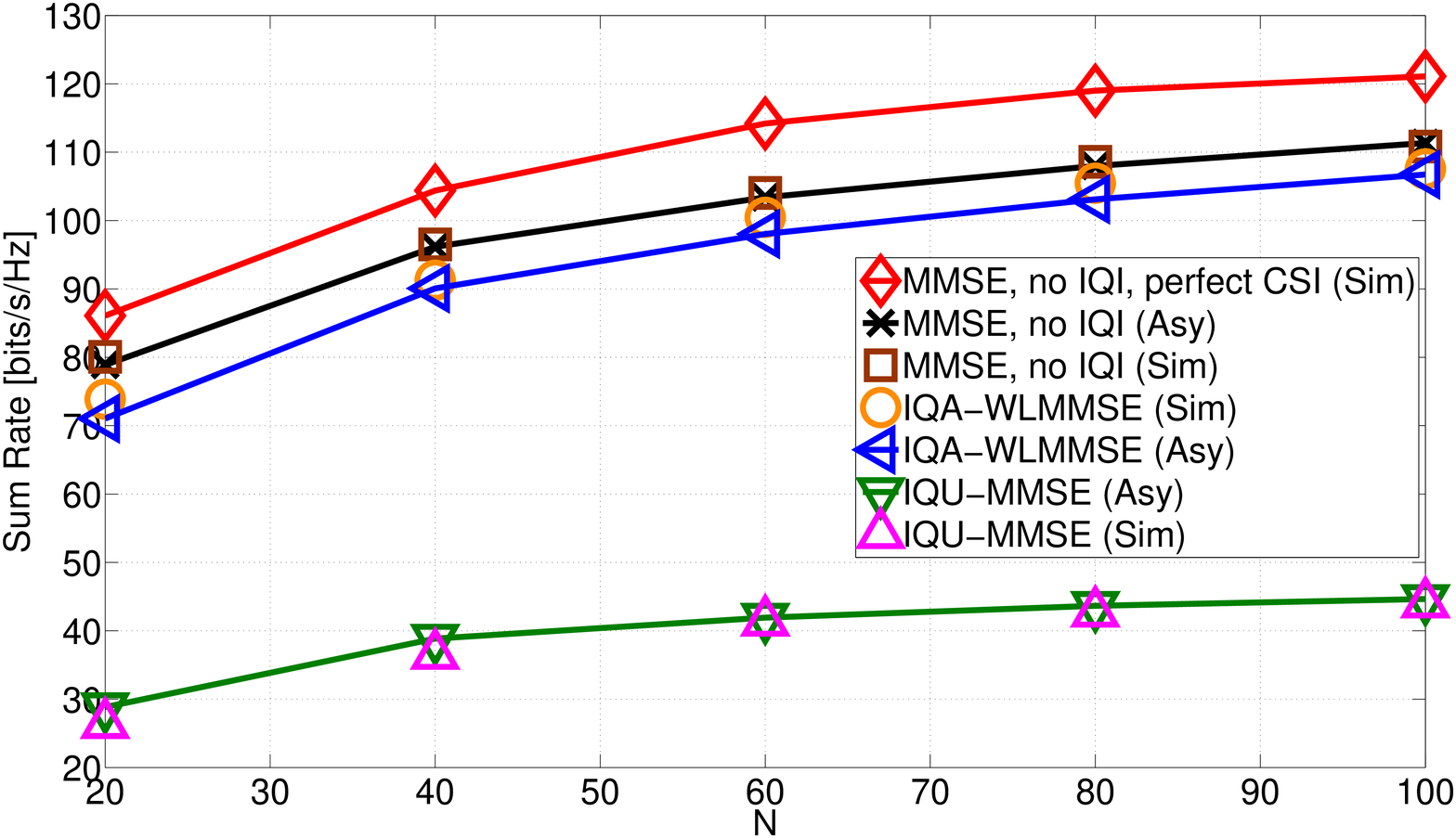}
\vspace*{-10mm}
\caption{\linespread{0.9} \small Sum rate vs. number of BS antennas $N$ for $K=10$, $\rho_\mathrm{UL}$ = 15 dB, $\rho_\mathrm{TR}$ = 10 dB, and pilot contamination with IQI present at the BSs and UTs.}
\label{Fig1}
\end{center}
\vspace*{-7mm}
\end{figure}

In Fig. \ref{Fig1}, the ergodic sum rates of the IQA-WLMMSE receiver, the IQU-MMSE receiver, the MMSE receiver in the absence of IQI, and the MMSE receiver in an ideal system with perfect CSI are depicted. Here, we assume $\rho_\mathrm{UL}=15 \ \mathrm{dB}$ and $\rho_\mathrm{TR}=10 \ \mathrm{dB}$, and we further assume that the IQI is present at both the BS and the UTs. Moreover, except for the perfect CSI case, full pilot contamination is assumed. The number of UTs is set to $K=10$ and the amplitude and phase mismatches at the UTs and the different antenna branches of the BSs are randomly and uniformly distributed in the range of $ 0.15 \leq \check{\epsilon}_{l,k}, \epsilon_{i,n} \leq 0.2, \ 1^\circ \leq \check{\theta}_{l,k}, \theta_{i,n} \leq 2^\circ, \ \forall k \in \left\lbrace 1,\cdots,K\right\rbrace, i \in \left\lbrace1,\cdots,N \right\rbrace, l \in \left\lbrace 1,\cdots,L \right\rbrace$, respectively. As can be observed from Fig. \ref{Fig1}, even small amplitude and phase mismatches lead to a high sum rate loss of the IQU-MMSE receiver compared to the ideal system without IQI. As expected from the analysis of the simplified single-cell channel model in Section \ref{Subsec_IQU_Asy_SingleCell}, the rate loss associated with IQI does not vanish even if the number of BS antennas is much larger than the number of UTs. For example, for $N=80$, the rate loss of the IQU-MMSE receiver compared to the system without IQI is approximately $60\%$. Furthermore, as expected from the analysis of the simplified single-cell channel model in Section \ref{SubSection_IQA_DE_SingleCell}, the proposed IQA-WLMMSE receiver achieves a substantially higher sum rate than the IQU-MMSE receiver and closely approaches the sum rate of the MMSE receiver in an ideal system without IQI. In Fig. \ref{Fig1}, we also present analytical results for the asymptotic sum rates of the IQU-MMSE receiver and the IQA-WLMMSE receiver given in (\ref{Eqn_AsySumRate_IQU}) and (\ref{Eqn_AsySumRate_IQA}), respectively. For large $N$, a perfect match between analytical and simulation results is observed for all receivers. Nevertheless, even for small numbers of BS antennas, the match between asymptotic and simulation results is good.
\begin{figure}[t]
\begin{center}
\includegraphics[width=\linewidth, clip=true]{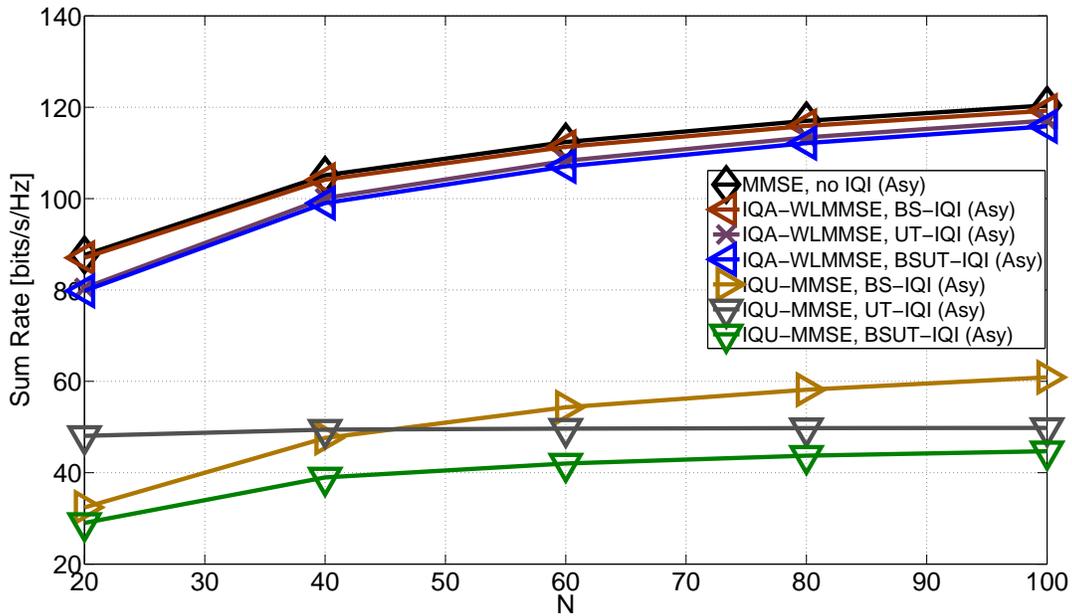}
\vspace*{-10mm}
\caption{\linespread{0.9} \small Sum rate vs. number of BS antennas $N$ for $K=10$, $\rho_\mathrm{UL}$ = 15 dB, $\rho_\mathrm{TR}$ = 10 dB, and no pilot contamination.}
\label{Fig2}
\end{center}
\vspace*{-7mm}
\end{figure}

In Fig. \ref{Fig2}, we investigate whether IQI at the UTs or IQI at the BSs is more harmful. To do so, we compare the sum rate performance of systems with IQI only at the BS (BS-IQI), IQI only at the UTs (UT-IQI), and IQI at both BSs and UTs (BSUT-IQI). For clarity of presentations, only analytical results are shown in Fig. \ref{Fig2}. However, all results were verified by simulations. Here, we consider a system without pilot contamination but with channel estimation errors, and $\rho_\mathrm{UL}=15 \ \mathrm{dB}$ and $\rho_\mathrm{TR}=10 \ \mathrm{dB}$. The amplitude and phase mismatches are generated in the same manner as for Fig. \ref{Fig1}. From Fig. \ref{Fig2}, we observe that if the IQU-MMSE receiver is employed, the system with IQI at both UTs and BSs yields the lowest sum rate, as expected. Furthermore, the system with IQI only at the BSs achieves a higher sum rate than the system with IQI only at the UTs. We note that this effect could also be observed in Section \ref{Subsec_IQU_Asy_SingleCell}, where analytical expressions for the asymptotic SINRs in the simplified single-cell system were derived. In fact, the sum rate of the system with IQI both at the BSs and the UTs approaches the sum rate of the system with IQI only the UTs for large numbers of BS antennas. A similar behavior can be observed for the sum rate performance of the IQA-WLMMSE receiver. Again, the system with IQI only at the BSs achieves the highest sum rate followed by the system with IQI only at the UTs and the system with IQI both at the UTs and the BSs. We note that this behavior supports the results in \cite{Bjoernson_TIT_2014}, where the authors claim that in the asymptotic regime, where the number of BS antennas is very large, H/W imperfections at the UTs are more harmful than those at the BS.
%
%
\begin{figure}[t]
\begin{center}
\psfrag{eps}[cc][cc][0.9]{$\epsilon $}
\includegraphics[width=\linewidth, clip=true]{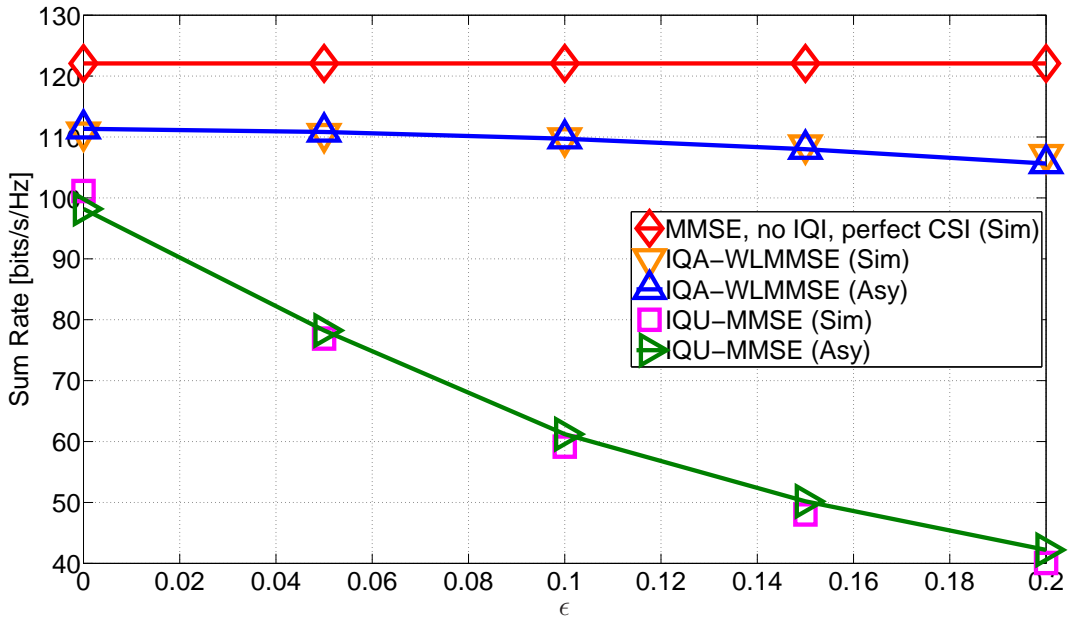}
\vspace*{-10mm}
\caption{\linespread{0.9} \small Sum rate vs. $\epsilon$ for $N=100$, $K=10$, $\rho_\mathrm{UL}$ = 15 dB, $\rho_\mathrm{TR}$ = 10 dB, and pilot contamination.}
\label{Fig3}
\end{center}
\vspace*{-7mm}
\end{figure}
\begin{figure}[t]
\begin{center}
\psfrag{theta}[cc][cc][0.8]{$\Theta $ [degree]}
\includegraphics[width=\linewidth, clip=true]{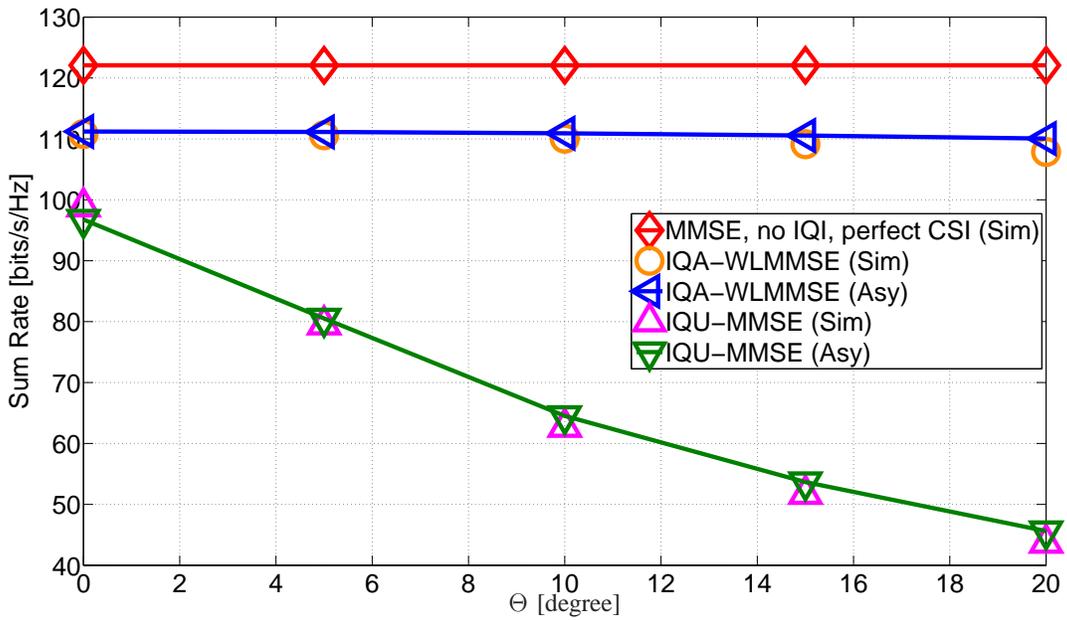}
\vspace*{-10mm}
\caption{\linespread{0.9} \small Sum rate vs. $\theta$ for $N=100$, $K=10$, $\rho_\mathrm{UL}$ = 15 dB, $\rho_\mathrm{TR}$ = 10 dB, and pilot contamination.}
\label{Fig4}
\end{center}
\vspace*{-10mm}
\end{figure}

In Fig. \ref{Fig3}, the influence of the amplitude mismatch $\epsilon$ on the sum rate of the conventional IQU-MMSE and the proposed IQA-WLMMSE receivers is investigated. For convenience, we assume that all UTs and BS antenna branches have the same phase and amplitude mismatches, i.e., $\check{\epsilon}_{l,k} = \epsilon_{i,n} =\epsilon, \check{\theta}_{l,k}= \theta_{i,n} = \theta = 2^\circ, \ \forall k \in \left\lbrace 1,\cdots,K\right\rbrace, i \in \left\lbrace1,\cdots,N \right\rbrace, l \in \left\lbrace 1,\cdots,L \right\rbrace$. The number of BS antennas and the number of UTs are set to $N=100$ and $K=10$, respectively. Moreover, we assume full pilot contamination and we further assume that the transmit data and training SNRs are $\rho_\mathrm{UL}=15 \ \mathrm{dB}$ and $\rho_\mathrm{TR}=10 \ \mathrm{dB}$, respectively. From Fig. \ref{Fig3}, it can be seen that the sum rate of the IQU-MMSE receiver rapidly decreases with increasing amplitude mismatch $\epsilon$. The proposed IQA-WLMMSE receiver performs significantly better than the IQU-MMSE receiver and its performance loss compared to the ideal system without IQI is small even for large amplitude mismatches.

The impact of phase mismatch on the sum rate performance is depicted in Fig. \ref{Fig4}. The same simulation parameters as for Fig. \ref{Fig3} are adopted. However, now the amplitude mismatch at all UTs and BS antenna branches is set to $\epsilon=0.02$, and the performance is evaluated for different values of phase mismatch $\theta$. Similar to Fig. \ref{Fig3}, it can be observed that the sum rate of the IQU-MMSE receiver decreases rapidly with increasing mismatch. Moreover, the proposed IQA-WLMMSE receiver has a substantially higher performance than the IQU-MMSE receiver and its performance loss compared to the ideal system without IQI is negligible even for large phase mismatches.
%
\vspace*{-2mm}
\section{Conclusion}\label{Sec_Conclusion}
We have proposed an IQA-WLMMSE receiver for uplink multi-cell massive MIMO systems suffering from IQI at both the BS and the UTs. In the considered system, CSI acquisition and data detection are affected by IQI, pilot contamination, and multi-cell interference. The proposed receiver comprises a channel estimator and a data detector and processes the real and imaginary parts of the received signal separately. Our simulation and analytical results show that, if left unattended, IQI causes severe performance losses even if the number of BS antennas is much larger than the number of UTs. The proposed IQA-WLMMSE receiver yields a substantially higher sum rate than the IQU-MMSE receiver and approaches the performance of the MMSE receiver in an ideal system without IQI. Furthermore, we observed that for the conventional IQU-MMSE receiver, IQI at the UTs is more harmful than IQI at the BS. We validated our simulation results by providing analytical expressions for the asymptotic sum rate using tools from random matrix theory.
%
\vspace*{-2mm}
\appendices
\section*{Appendix A - Some Useful Lemmas}
\begin{lemma} [{\cite[Theorem 7]{Evans2000}}] \label{Lemma_RMT1}
Let $\vv{p}, \vv{q} \in \mathbb{C}^{N \times 1}$ have mutually independent, i.i.d. zero-mean unit variance Gaussian distributed elements and $\mm{A} \in \mathbb{C}^{N \times N}$ be a Hermitian matrix with bounded spectral norm, whose elements are independent of $\vv{p}$ and $\vv{q}$. Then,
\vspace*{-3mm}
\begin{align}
& \frac{1}{N} \vv{q}^\herm \mm{A} \vv{q}  \xrightarrow[N\rightarrow\infty]{\mathrm{a.s.}} \frac{1}{N} \mathrm{tr} \left( \mm{A} \right) \\
& \frac{1}{N} \vv{p}^\herm \mm{A} \vv{q}  \xrightarrow[N\rightarrow\infty]{\mathrm{a.s.}} 0.
\end{align}
\end{lemma}
%
\begin{lemma} [{\cite[Rank-1 Perturbation Lemma]{Couillet2011}}] \label{Lemma_Perturb}
Let $\mm{A} \in \mathbb{C}^{N \times N}$, $\vv{q} \in \mathbb{C}^{N \times 1}$, and $\mm{B} \in \mathbb{C}^{N \times N}$ be a Hermitian nonnegative definite matrix. Moreover, let $\alpha$ and $\alpha^\prime$ be positive real numbers. Then,
\vspace*{-3mm}
\begin{align}
\left\lvert \mathrm{tr} \left( \mm{A} \left( \mm{B} + \alpha \mm{I}_N \right)^{-1} - \mm{A} \left( \mm{B} + \alpha^\prime \vv{q} \vv{q}^\herm + \alpha \mm{I}_N \right)^{-1} \right) \right\rvert \leq \frac{\norm{\mm{A}}}{\alpha}.
\end{align}
\end{lemma}
%
\begin{lemma} [{\cite[Theorem 1]{Wagner2012}}] \label{Lemma_RMT2}
Let $\mm{A} \in \mathbb{C}^{N \times N}$ and $\mm{B} \in \mathbb{C}^{N \times N}$ be Hermitian nonnegative definite and $\mm{G} \in \mathbb{C}^{N \times K}$ have random column vectors $\vv{g}_k \sim \mathcal{C}\mathcal{N}\left(\mm{0}, \mm{\Delta}_k\right)$. Moreover, $\mm{A}$, $\mm{B}$, and $\mm{\Delta}_k, \forall k = 1, \ldots,K$, have bounded spectral norms. Then, for positive $\alpha$,
\vspace*{-3mm}
\begin{align}
\frac{1}{N} \mathrm{tr} \left( \mm{A} \left( \mm{G} \mm{G}^\herm + \mm{B} + \alpha \mm{I}_N \right)^{-1} \right) \xrightarrow[N\rightarrow\infty]{\mathrm{a.s.}} \frac{1}{N} \mathrm{tr} \left( \mm{A} \mm{T} \right),
\end{align}
where $\mm{T}$ is given by
\vspace*{-6mm}
\begin{align}
\mm{T} = \left( \frac{1}{N} \sum_{k=1}^{K} \frac{\mm{\Delta}_k}{1+\delta_k} + \mm{B} + \alpha \mm{I}_N \right)^{-1},
\label{Lemma3_T}
\end{align}
with $\delta_k$ being the solution to the fixed-point equation
\vspace*{-2mm}
\begin{align}
\delta_k = \frac{1}{N} \mathrm{tr} \left( \mm{\Delta}_k \left( \frac{1}{N} \sum_{j=1}^{K} \frac{\mm{\Delta}_j}{1+\delta_j} + \mm{B} + \alpha \mm{I}_N \right)^{-1} \right).
\label{Lemma3_delta}
\end{align}
\end{lemma}
%
\vspace*{-2mm}
\begin{lemma} [{\cite{Wagner2012}}] \label{Lemma_RMT3}
Let $\mm{A} \in \mathbb{C}^{N \times N}$, $\mm{B} \in \mathbb{C}^{N \times N}$, and $\mm{G} \in \mathbb{C}^{N \times K}$ be defined as in Lemma 3. Then, for Hermitian nonnegative definite $\mm{C} \in \mathbb{C}^{N \times N}$ with bounded spectral norm,
\vspace*{-2mm}
\begin{align}
\frac{1}{N} \mathrm{tr} \left( \mm{A} \left( \mm{G} \mm{G}^\herm + \mm{B} + \alpha \mm{I}_N \right)^{-1} \mm{C} \left( \mm{G} \mm{G}^\herm + \mm{B} + \alpha \mm{I}_N \right)^{-1} \right) \xrightarrow[N\rightarrow\infty]{\mathrm{a.s.}} \frac{1}{N} \mathrm{tr} \left( \mm{A} \mm{T}^\prime \right),
\end{align}
where $\mm{T}^\prime$ is defined as
\vspace*{-4mm}
\begin{align}
\mm{T}^\prime = \mm{T} \mm{C} \mm{T} + \frac{1}{N} \mm{T} \sum_{k=1}^{K} \frac{\mm{\Delta}_k \delta^\prime_k}{\left(1+\delta_k\right)^2} \mm{T},
\end{align}
where $\mm{T}$ and $\delta_k$ are given by (\ref{Lemma3_T}) and (\ref{Lemma3_delta}), respectively and $\vv{\delta}^\prime = \left[\delta^\prime_1, \ldots, \delta^\prime_K\right]^\transp = \left( \mm{I}_K - \mm{Y} \right)^{-1} \vv{x}$, where the elements of $\mm{Y} \in \mathbb{C}^{K \times K}$ and $\mm{x}\in \mathbb{C}^{K \times 1}$ are given by
\vspace*{-4mm}
\begin{align}
{\left[\mm{Y}\right]}_{k, q} &= \frac{\frac{1}{N}\mathrm{tr}\left( \mm{\Delta}_k \mm{T} \mm{\Delta}_q \mm{T} \right)}{N \left(1+\delta_q\right)^2} \ \ \mathrm{and} \ \
{\left[\vv{x}\right]}_k = \frac{1}{N} \mathrm{tr} \left( \mm{\Delta}_k \mm{T} \mm{C} \mm{T} \right).
\end{align}
\end{lemma}
%
\vspace*{-3mm}
\section*{Appendix B - Proof of Theorem 1}\label{Theo1proof}
According to (\ref{Eqn_IQU_EstDataWithDet}), for the IQU-MMSE receiver, the useful signal of the $k$th UT received at the $i$th BS can be expressed as
\vspace*{-3mm}
\begin{align}
\hat{d}_{i,k} =& \Big( \sqrt{\rho_\mathrm{UL}}\xi_{A,i,k} \hat{\vv{g}}_{i,i,k}^\herm \big( \hat{\mm{G}}_{i,i} \hat{\mm{G}}_{i,i}^\herm + \frac{1}{\rho_\mathrm{UL}} \mm{I}_N \big)^{-1} \mm{\Psi}_{A,i} {\vv{h}}_{i,i,k} + \sqrt{\rho_\mathrm{UL}} \xi^*_{B,i,k} \hat{\vv{g}}_{i,i,k}^\herm \big( \hat{\mm{G}}_{i,i} \hat{\mm{G}}_{i,i}^\herm + \frac{1}{\rho_\mathrm{UL}} \mm{I}_N \big)^{-1} \nonumber \\
& \times \mm{\Psi}_{B, i} {\vv{h}}^*_{i,i,k} \Big) d_{i,k}.
\label{Eqn_IQU_UsefulSig1}
\end{align}
Thus, in the large system limit for $N \rightarrow \infty$, the useful signal power of the $k$th UT received at the $i$th BS converges to
\vspace*{-3mm}
\begin{align}
S^{\mathrm{IQU}^\circ}_{i,k} =& \lim_{N\rightarrow\infty} \rho_\mathrm{UL} \bigg( \Big| \frac{ \xi_{A,i,k} \hat{\vv{g}}_{i,i,k}^\herm}{N} \big( \frac{\hat{\mm{G}}_{i,i} \hat{\mm{G}}_{i,i}^\herm}{N} + \frac{1}{N\rho_\mathrm{UL}} \mm{I}_N \big)^{-1} \mm{\Psi}_{A,i} {\vv{h}}_{i,i,k} \Big|^2 + \Big| \frac{\xi^*_{B,i,k} \hat{\vv{g}}_{i,i,k}^\herm}{N} \big( \frac{\hat{\mm{G}}_{i,i} \hat{\mm{G}}_{i,i}^\herm}{N} + \nonumber \\
& \frac{1}{N \rho_\mathrm{UL}} \mm{I}_N \big)^{-1} \mm{\Psi}_{B, i} {\vv{h}}^*_{i,i,k} \Big|^2\bigg).
\label{Eqn_IQU_UsefulPow}
\end{align}
Applying the matrix inversion lemma \cite{Horn2013}, Lemmas \ref{Lemma_RMT1}, \ref{Lemma_Perturb}, and \ref{Lemma_RMT2}, the first term on the right hand side of (\ref{Eqn_IQU_UsefulPow}) can be rewritten as
\vspace*{-3mm}
\begin{align}
& \bigg| \frac{\xi_{A,i,k} \hat{\vv{g}}_{i,i,k}^\herm}{N} \big( \frac{\hat{\mm{G}}_{i,i} \hat{\mm{G}}_{i,i}^\herm}{N} + \frac{1}{N\rho_\mathrm{UL}} \mm{I}_N \big)^{-1} \mm{\Psi}_{A,i} {\vv{h}}_{i,i,k} \bigg|^2 \hspace*{-1mm}=\hspace*{-1mm} \bigg| \frac{\frac{\xi_{A,i,k} \hat{\vv{g}}_{i,i,k}^\herm}{N} {\mm{\Lambda}_{i,k}}^{-1} \mm{\Psi}_{A,i} {\vv{h}}_{i,i,k}}{1+ \frac{1}{N}\hat{\vv{g}}_{i,i,k}^\herm {\mm{\Lambda}_{i,k}}^{-1} \hat{\vv{g}}_{i,i,k} } \bigg|^2 \hspace*{-2mm} \xrightarrow[N\rightarrow\infty]{\mathrm{a.s.}} \hspace*{-2mm} \bigg| \frac{\xi_{A,i,k} \lambda^{(A)}_{i,i,k}}{1+\delta_{i,k}} \bigg|^2, 
\label{Eqn_IQU_UsefulPow1}
\end{align}
where ${\mm{\Lambda}_{i,k}}$ is defined as
\vspace*{-6mm}
\begin{align}
{\mm{\Lambda}_{i,k}} \triangleq \frac{1}{N}\hat{\mm{G}}_{i,i} \hat{\mm{G}}_{i,i}^\herm -\frac{1}{N}\hat{\vv{g}}_{i,i,k} \hat{\vv{g}}_{i,i,k}^\herm + \frac{1}{N \rho_\mathrm{UL}} \mm{I}_N,
\end{align}
$\delta_{i,k}$ is given in (\ref{Eqn_IQU_delta}), and $\lambda_{i,i,k}^{(A)}$ is given in (\ref{Eqn_IQU_lambda_A}) for $l=i$.
Similarly, the second term on the right hand side of (\ref{Eqn_IQU_UsefulPow}) can be expressed as
\vspace*{-3mm}
\begin{align}
\left| \frac{1}{N} \xi^*_{B,i,k} \hat{\vv{g}}_{i,i,k}^\herm \left( \frac{1}{N} \hat{\mm{G}}_{i,i} \hat{\mm{G}}_{i,i}^\herm + \frac{1}{N \rho_\mathrm{UL}} \mm{I}_N \right)^{-1} \mm{\Psi}_{B, i} {\vv{h}}^*_{i,i,k} \right|^2 \xrightarrow[N\rightarrow\infty]{\mathrm{a.s.}} \left| \frac{\xi^*_{B,i,k} \lambda^{(B)}_{i,i,k}}{1+\delta_{i,k}} \right|^2,
\label{Eqn_IQU_InterfAsy2}
\end{align}
where ${\lambda}^{(B)}_{i,i,k}$ is given in (\ref{Eqn_IQU_lambda_B}) for $l=i$.
%
Next, we derive an analytical expression for the asymptotic value of the interference power for $N \rightarrow \infty$. According to (\ref{Eqn_IQU_EstDataWithDet}), the interference part of the received signal of the $k$th UT at the $i$th BS is given by
\vspace*{-3mm}
\begin{align}
& \sqrt{\rho_\mathrm{UL}} \mathop{\sum_{l=1}^{L}\sum_{q=1}^{K}}_{(l, q) \neq (i, k)} \Bigg( \frac{\xi_{A,lq} \hat{\vv{g}}_{i,i,k}^\herm}{N} \bigg( \frac{\hat{\mm{G}}_{i,i} \hat{\mm{G}}_{i,i}^\herm}{N} \hspace*{-1mm} + \hspace*{-1mm} \frac{1}{N \rho_\mathrm{UL}} \mm{I}_N \bigg)^{-1} \mm{\Psi}_{A,i} {\vv{h}}_{i,l,q} \hspace*{-1mm} + \hspace*{-1mm} \frac{ \xi^*_{B,l,q} \hat{\vv{g}}_{i,i,k}^\herm}{N} \bigg( \frac{ \hat{\mm{G}}_{i,i} \hat{\mm{G}}_{i,i}^\herm}{N} \hspace*{-1mm} + \hspace*{-1mm} \frac{1}{N \rho_\mathrm{UL}} \mm{I}_N \bigg)^{-1} \nonumber \\
& \times \mm{\Psi}_{B, i} {\vv{h}}^*_{i,l,q} \Bigg) d_{l,q} \hspace*{-1mm} + \hspace*{-1mm} \sqrt{\rho_\mathrm{UL}} \sum_{l=1}^{L} \sum_{q=1}^{K} \Bigg( \frac{\xi_{B,l,q} \hat{\vv{g}}_{i,i,k}^\herm}{N} \bigg( \frac{\hat{\mm{G}}_{i,i} \hat{\mm{G}}_{i,i}^\herm}{N} +\frac{1}{N\rho_\mathrm{UL}} \mm{I}_N \bigg)^{-1} \mm{\Psi}_{A,i} {\vv{h}}_{i,l,q} + \frac{\xi^*_{A,l,q} \hat{\vv{g}}_{i,i,k}^\herm}{N} \nonumber \\
& \times \bigg( \frac{\hat{\mm{G}}_{i,i} \hat{\mm{G}}_{i,i}^\herm}{N} + \frac{1}{N \rho_\mathrm{UL}} \mm{I}_N \bigg)^{-1} \mm{\Psi}_{B, i} {\vv{h}}^*_{i,l,q} \Bigg) d^*_{l,q}.
\label{Eqn_IQU_InterfSig}
\end{align}
In the asymptotic regime, when $N \rightarrow \infty$, the interference power in (\ref{Eqn_IQU_InterfSig}) converges to 
\vspace*{-4mm}
\begin{align}
& I^{\mathrm{IQU}^\circ}_{i,k} = \lim_{N \rightarrow \infty} \rho_\mathrm{UL} \mathop{\sum_{l=1}^{L}\sum_{q=1}^{K}}_{(l, q) \neq (i, k)} \Bigg( \Big| \frac{\xi_{A,l,q} \hat{\vv{g}}_{i,i,k}^\herm}{N} \big( \frac{ \hat{\mm{G}}_{i,i} \hat{\mm{G}}_{i,i}^\herm }{N} \hspace*{-1mm}+\hspace*{-1mm} \frac{\mm{I}_N}{N\rho_\mathrm{UL}}  \big)^{-1}  \mm{\Psi}_{A,i} {\vv{h}}_{i,l,q} \Big|^2 \hspace*{-1mm}+\hspace*{-1mm} \Big| \frac{ \xi^*_{B,l,q} \hat{\vv{g}}_{i,i,k}^\herm}{N} \big(  \frac{\hat{\mm{G}}_{i,i} \hat{\mm{G}}_{i,i}^\herm}{N} \nonumber \\
& + \frac{\mm{I}_N}{N \rho_\mathrm{UL}}  \big)^{-1} \mm{\Psi}_{B, i} {\vv{h}}^*_{i,l,q} \Big|^2 \hspace*{-1mm}+\hspace*{-1mm} \Big| \frac{ \xi_{B,l,q} \hat{\vv{g}}_{i,i,k}^\herm }{N } \big( \frac{\hat{\mm{G}}_{i,i} \hat{\mm{G}}_{i,i}^\herm}{N} \hspace*{-1mm}+\hspace*{-1mm} \frac{\mm{I}_N}{N\rho_\mathrm{UL}} \big)^{-1} \mm{\Psi}_{A,i} {\vv{h}}_{i,l,q} \Big|^2 \hspace*{-1mm}+\hspace*{-1mm} \Big| \frac{ \xi^*_{A,l,q} \hat{\vv{g}}_{i,i,k}^\herm}{N} \big( \frac{ \hat{\mm{G}}_{i,i} \hat{\mm{G}}_{i,i}^\herm}{N} \nonumber \\
& \hspace*{-1mm}+\hspace*{-1mm} \frac{\mm{I}_N}{N \rho_\mathrm{UL}} \big)^{-1} \mm{\Psi}_{B, i} {\vv{h}}^*_{i,l,q} \Big|^2 \Bigg) \hspace*{-1mm}+\hspace*{-1mm} \rho_\mathrm{UL} \bigg( \Big| \frac{ \xi_{B,i,k} \hat{\vv{g}}_{i,i,k}^\herm}{N} \big( \frac{\hat{\mm{G}}_{i,i} \hat{\mm{G}}_{i,i}^\herm}{N} \hspace*{-1mm}+\hspace*{-1mm} \frac{\mm{I}_N}{N \rho_\mathrm{UL}} \big)^{-1} \mm{\Psi}_{A,i} {\vv{h}}_{i,i,k} \Big|^2 \hspace*{-1mm}+\hspace*{-1mm} \Big| \frac{ \xi^*_{A,i,k} \hat{\vv{g}}_{i,i,k}^\herm}{N} \nonumber \\
& \times \big( \frac{ \hat{\mm{G}}_{i,i} \hat{\mm{G}}_{i,i}^\herm }{N} \hspace*{-1mm}+\hspace*{-1mm} \frac{\mm{I}_N}{N \rho_\mathrm{UL}} \big)^{-1} \mm{\Psi}_{B, i} {\vv{h}}^*_{i,i,k} \Big|^2 \bigg).
\label{Eqn_IQU_InterfPow}
\end{align}
The first term on the right hand side of (\ref{Eqn_IQU_InterfPow}) can be rewritten as \cite{Hoydis2013}
\vspace*{-3mm}
\begin{align}
& \mathop{\sum_{l=1}^{L} \hspace*{-1mm} \sum_{q=1}^{K}}_{(l, q) \neq (i, k)} \hspace*{-1mm} \hspace*{-1mm} \Big| \frac{ \xi_{A,l,q} \hat{\vv{g}}_{i,i,k}^\herm}{N} \big( \frac{\hat{\mm{G}}_{i,i} \hat{\mm{G}}_{i,i}^\herm}{N} \hspace*{-1mm} + \hspace*{-1mm} \frac{\mm{I}_N}{N \rho_\mathrm{UL}}  \big)^{-1} \hspace*{-1mm} \mm{\Psi}_{A,i} {\vv{h}}_{i,l,q} \Big|^2 \hspace*{-3mm} = \hspace*{-2mm} \mathop{\sum_{l=1}^{L}}_{l \neq i} \hspace*{-1mm} \Big| \frac{ \xi_{A,l,k} \hat{\vv{g}}_{i,i,k}^\herm}{N} \big( \frac{\hat{\mm{G}}_{i,i} \hat{\mm{G}}_{i,i}^\herm}{N} \hspace*{-1mm} + \hspace*{-1mm} \frac{\mm{I}_N}{N \rho_\mathrm{UL}} \big)^{-1} \hspace*{-1mm} \mm{\Psi}_{A,i} {\vv{h}}_{i,l,k} \Big|^2 \nonumber \\
& + \mathop{\sum_{l=1}^{L}\sum_{q=1}^{K}}_{q \neq k} \Big| \frac{\xi_{A,l,q} \hat{\vv{g}}_{i,i,k}^\herm}{N} \big( \frac{\hat{\mm{G}}_{i,i} \hat{\mm{G}}_{i,i}^\herm}{N} + \frac{\mm{I}_N}{N \rho_\mathrm{UL}} \big)^{-1} \mm{\Psi}_{A,i} {\vv{h}}_{i,l,q} \Big|^2. 
\label{Eqn_IQU_Interf_Term1}
\vspace*{-4mm}
\end{align}
Applying the matrix inversion Lemma \cite{Horn2013} and Lemmas \ref{Lemma_RMT1}, \ref{Lemma_Perturb}, and \ref{Lemma_RMT2} yields
\vspace*{-2mm}
\begin{align}
& \frac{ \xi_{A,l,k} \hat{\vv{g}}_{i,i,k}^\herm}{N} \hspace*{-0.5mm} \big( \frac{ \hat{\mm{G}}_{i,i} \hat{\mm{G}}_{i,i}^\herm }{N} + \frac{\mm{I}_N}{N \rho_\mathrm{UL}} \big)^{-1} \hspace*{-0.5mm} \mm{\Psi}_{A,i} {\vv{h}}_{i,l,k} \hspace*{-1mm} \xrightarrow[N\rightarrow\infty]{\mathrm{a.s.}} \hspace*{-1mm} \frac{ \xi_{A,l,k} \left( \xi_{A,l,k}^* + \xi_{B,l,k}^* \right) \mathrm{tr} \left( \mm{\Psi}_{A,i} \mm{R}_{i,l,k} \mm{\Psi}_{A,i}^\herm \mm{\Omega}_{i,k} \mm{\Gamma}_{i} \right) }{N\left(1+\delta_{i,k}\right)} \nonumber \\
& =\frac{\xi_{A,l,k} \lambda^{(A)}_{i,l,k}}{1+\delta_{i,k}},
\vspace*{-4mm}
\end{align}
where $\delta_{i,k}$ and $\lambda^{(A)}_{i,l,k}$ are given in (\ref{Eqn_IQU_delta}) and (\ref{Eqn_IQU_lambda_A}), respectively. On the other hand, for the second term on the right hand side of (\ref{Eqn_IQU_Interf_Term1}), we define $\mm{\Theta}_{i,i,k} \triangleq \mathbb{E} \left\lbrace \hat{\vv{g}}_{i,i,k} \hat{\vv{g}}_{i,i,k}^\herm \right\rbrace$, apply the matrix inversion Lemma \cite{Horn2013} and Lemmas \ref{Lemma_RMT1}, \ref{Lemma_Perturb}, \ref{Lemma_RMT2}, and \ref{Lemma_RMT3}, and obtain \cite{Hoydis2013}
\vspace*{-3mm}
\begin{align}
&\left| \frac{1}{N} \xi_{A,l,q} \hat{\vv{g}}_{i,i,k}^\herm \left( \frac{\hat{\mm{G}}_{i,i} \hat{\mm{G}}_{i,i}^\herm }{N} + \frac{\mm{I}_N}{N\rho_\mathrm{UL}}  \right)^{-1} \mm{\Psi}_{A,i} {\vv{h}}_{i,l,q}\right|^2=\frac{\left|\xi_{A,l,q}\right|^2 {\vv{h}}_{i,l,q}^\herm \mm{\Psi}_{A,i}^\herm \mm{\Lambda}^{-1}_{i,k} \mm{\Theta}_{i,i,k} \mm{\Lambda}^{-1}_{i,k} \mm{\Psi}_{A,i} {\vv{h}}_{i,l,q} }{N^2\left(1+\frac{1}{N}\hat{\vv{g}}_{i,i,k}^\herm \mm{\Lambda}^{-1}_{i,k} \hat{\vv{g}}_{i,i,k}\right)^2}=\nonumber \\
& \frac{\left|\xi_{A,l,q}\right|^2}{\left(1+\delta_{i,k}\right)^2} \Bigg(\hspace*{-1.5mm} \frac{ {\vv{h}}_{i,l,q}^\herm \mm{\Psi}_{A,i}^\herm \mm{\Lambda}^{-1}_{i,k,q} \mm{\Theta}_{i,i,k} \mm{\Lambda}^{-1}_{i,k,q} \mm{\Psi}_{A,i} {\vv{h}}_{i,l,q} }{N^2} \hspace*{-1mm}+\hspace*{-1mm} \frac{ \left| {\vv{h}}_{i,l,q}^\herm \mm{\Psi}_{A,i}^\herm \mm{\Lambda}^{-1}_{i,k,q} \hat{\vv{g}}_{i,i,q} \right|^2 \hat{\vv{g}}_{i,i,q}^\herm \mm{\Lambda}^{-1}_{i,k,q} \mm{\Theta}_{i,i,k} \mm{\Lambda}^{-1}_{i,k,q} \hat{\vv{g}}_{i,i,q} }{N^4\left(1+\delta_{i,q}\right)^2} \nonumber \\ 
&- 2 \Re{\left\lbrace \frac{ \left(\hat{\vv{g}}_{i,i,q}^\herm \mm{\Lambda}^{-1}_{i,k,q} \mm{\Psi}_{A,i} {\vv{h}}_{i,l,q} \right)\left( {\vv{h}}_{i,l,q}^\herm \mm{\Psi}_{A,i}^\herm \mm{\Lambda}^{-1}_{i,k,q} \mm{\Theta}_{i,i,k} \mm{\Lambda}^{-1}_{i,k,q} \hat{\vv{g}}_{i,i,q} \right) }{N^3\left(1+\delta_{i,q}\right)}  \right\rbrace } \Bigg)\xrightarrow[N\rightarrow\infty]{\mathrm{a.s.}} \frac{\left|\xi_{A,l,q}\right|^2}{\left(1+\delta_{i,k}\right)^2} \Bigg( \zeta^{(A)}_{i,l,q} + \nonumber \\
& \frac{\left|\lambda^{(A)}_{i,l,q}\right|^2\kappa_{i,i,q}}{\left(1+\delta_{i,q}\right)^2} -2\Re{\left\lbrace \frac{\lambda^{(A)}_{i,l,q} \phi^{(A)}_{i,i,q}}{1+\delta_{i,q}} \right\rbrace } \Bigg)= \frac{\left|\xi_{A,l,q}\right|^2\varrho^{(A)}_{i,l,q}}{\left(1+\delta_{i,k}\right)^2}, 
\label{Eqn_IQU_InterPart2}
\vspace*{-4mm}
\end{align}
where $\varrho^{(A)}_{i,l,q}$ is given in (\ref{Eqn_IQU_Theo1_RhoA}) and ${\mm{\Lambda}_{i,k,q}}$ is defined as
\vspace*{-4mm}
\begin{align}
{\mm{\Lambda}_{i,k,q}} = \frac{1}{N} \hat{\mm{G}}_{i,i} \hat{\mm{G}}_{i,i}^\herm - \frac{1}{N} \hat{\vv{g}}_{i,i,k} \hat{\vv{g}}_{i,i,k}^\herm - \frac{1}{N} \hat{\vv{g}}_{i,i,q} \hat{\vv{g}}_{i,i,q}^\herm + \frac{1}{N \rho_\mathrm{UL}} \mm{I}_N,
\vspace*{-4mm}
\end{align}
and $\delta_{i,q}$ and $\lambda^{(A)}_{i,l,q}$ are obtained by replacing $k$ with $q$ in (\ref{Eqn_IQU_delta}) and (\ref{Eqn_IQU_lambda_A}), respectively. In (\ref{Eqn_IQU_InterPart2}), $\zeta^{(A)}_{i,l,q}$, $\kappa_{i,i,q}$, and $\phi^{(A)}_{i,i,q}$ are defined as
\vspace*{-3mm}
\begin{align}
& \frac{1}{N^2} {\vv{h}}_{i,l,q}^\herm \mm{\Psi}_{A,i}^\herm \mm{\Lambda}^{-1}_{i,k,q} \mm{\Theta}_{i,i,k} \mm{\Lambda}^{-1}_{i,k,q} \mm{\Psi}_{A,i} {\vv{h}}_{i,l,q} \xrightarrow[N\rightarrow\infty]{\mathrm{a.s.}} \frac{1}{N^2} \mathrm{tr} \left( \mm{R}_{i,l,q} \mm{\Psi}_{A,i}^\herm \mm{\Gamma}_{i,k}^\prime \mm{\Psi}_{A,i} \right) \triangleq \zeta^{(A)}_{i,l,q} \\
& \frac{1}{N^2} \hat{\vv{g}}_{i,i,q}^\herm \mm{\Lambda}^{-1}_{i,k,q} \mm{\Theta}_{i,i,k} \mm{\Lambda}^{-1}_{i,k,q} \hat{\vv{g}}_{i,i,q} \xrightarrow[N\rightarrow\infty]{\mathrm{a.s.}} \frac{1}{N^2} \mathrm{tr} \left( \mm{\Theta}_{i,i,q}  \mm{\Gamma}_{i,k}^\prime \right) \triangleq \kappa_{i,i,q}\\
& \frac{1}{N^2} {\vv{h}}_{i,l,q}^\herm \mm{\Psi}_{A,i}^\herm \mm{\Lambda}^{-1}_{i,k,q} \mm{\Theta}_{i,i,k} \mm{\Lambda}^{-1}_{i,k,q} \hat{\vv{g}}_{i,i,q} \xrightarrow[N\rightarrow\infty]{\mathrm{a.s.}} \frac{\xi_{A,l,q} }{N^2} \mathrm{tr} \left( \mm{R}_{i,l,q} \mm{\Psi}_{A, i}^\herm \mm{\Gamma}^\prime_{i,k} \mm{\Omega}_{i,k} \mm{\Psi}_{A, i} \right) \triangleq \phi^{(A)}_{i,i,q},
\label{Eqn_IQU_Interf1_rho_A}
\end{align}
where $\mm{\Gamma}^\prime_{i,k}$ is given by $\mm{T}^\prime$ in Lemma \ref{Lemma_RMT3} with $\mm{A}$, $\mm{B}$, $\mm{T}$, $\mm{C}$, and $\mm{\Delta}_k$ being set to $\mm{\Omega}_{i,k} \mm{\Psi}_{A, i} \mm{R}_{i,l,q} \mm{\Psi}_{A, i}^\herm$, $\mm{0}$, $\mm{\Gamma}_{i}$, $\mm{\Theta}_{i,i,k}$, and $\mm{\Theta}_{i,i,k}/N$, respectively, where $\mm{0}$ is an all-zero $N \times N$ matrix. Considering (\ref{Eqn_IQU_Interf_Term1})-(\ref{Eqn_IQU_Interf1_rho_A}), and performing similar mathematical operations for the remaining terms in (\ref{Eqn_IQU_InterfPow}), we obtain the expression in (\ref{Eqn_IQU_Theo2InterfPow}) for the interference power of the $k$th UT received at the $i$th BS.
%
Furthermore, applying the matrix inversion lemma \cite{Horn2013}, the asymptotic noise power at the output of the IQU-MMSE detector corresponding to the $k$th UT in the $i$th cell is obtained as
\vspace*{-2mm}
\begin{align}
Z^{\mathrm{IQU}^\circ}_{i,k}= \lim_{N \rightarrow \infty} \frac{ \hat{\vv{g}}_{i,i,k}^\herm \mm{\Lambda}_{i,k}^{-1} \mm{\Psi}_{A, i} \mm{\Psi}_{A, i}^\herm \mm{\Lambda}_{i,k}^{-1} \hat{\vv{g}}_{i,i,k}}{N^2\left( 1+\frac{1}{N} \hat{\vv{g}}_{i,i,k}^\herm \mm{\Lambda}^{-1}_{i,k} \hat{\vv{g}}_{i,i,k}\right)^2} + \frac{ \hat{\vv{g}}_{i,i,k}^\herm \mm{\Lambda}_{i,k}^{-1} \mm{\Psi}_{B, i} \mm{\Psi}_{B, i}^\herm \mm{\Lambda}_{i,k}^{-1} \hat{\vv{g}}_{i,i,k}}{N^2\left(1+ \frac{1}{N} \hat{\vv{g}}_{i,i,k}^\herm \mm{\Lambda}^{-1}_{i,k} \hat{\vv{g}}_{i,i,k}\right)^2}.
\label{Eqn_IQA-TheoProof_Noise}
\end{align}
Then applying Lemmas 1-4, the asymptotic expression for the noise power at the output of the IQU-MMSE detector in (\ref{Eqn_IQU_Theo1_Noise}) is obtained from (\ref{Eqn_IQA-TheoProof_Noise}). This completes the proof.
%
%
\vspace*{-4mm}
\section*{Appendix C - Proof of Corollary 1}\label{Coro1proof}
Applying the matrix inversion lemma [22], and Lemmas 1, 2, and 3, the useful signal power of the $k$th UT for IQU-MMSE detector in a single-cell uplink massive MIMO system with i.i.d. channels, IQI only at the BS and perfect CSI (i.e., with $\hat{\vv{g}}_{i,i,k}=\mm{\Psi}_A \vv{h}_{i,i,k}$) is obtained as
\vspace*{-3mm}
\begin{align}
S^{\mathrm{IQU}^\circ}_{i, k} = \lim_{N \rightarrow \infty} \rho_\mathrm{UL} \Big| \frac{1}{N} \vv{h}_{i,i,k}^\herm \mm{\Psi}_A^\herm \mm{\Lambda}^{-1}_i \mm{\Psi}_A \vv{h}_{i,i,k} \Big|^2 = \lim_{N \rightarrow \infty} \frac{\rho_\mathrm{UL} \Big| \frac{1}{N} \vv{h}_{i,i,k}^\herm \mm{\Psi}_A^\herm \mm{\Lambda}^{-1}_{i, k} \mm{\Psi}_A \vv{h}_{i,i,k} \Big|^2}{\left( 1+ \frac{1}{N} \vv{h}_{i,i,k}^\herm \mm{\Psi}_A^\herm \mm{\Lambda}^{-1}_{i, k} \mm{\Psi}_A \vv{h}_{i,i,k}\right)^2} = \frac{ \rho_\mathrm{UL} \delta^2}{\left(1+ \delta \right)^2}, \nonumber
\end{align}
where ${\mm{\Lambda}_{i}} \triangleq \frac{1}{N} {\mm{G}}_{i,i} {\mm{G}}_{i,i}^\herm + \frac{1}{N \rho_\mathrm{UL}} \mm{I}_N$, ${\mm{\Lambda}_{i, k}} \triangleq {\mm{\Lambda}_{i}} - \frac{1}{N} {\vv{g}}_{i,i,k} {\vv{g}}_{i,i,k}^\herm$, and $\delta$ is defined as
\vspace*{-4mm}
\begin{align}
\delta = \lim_{N \rightarrow \infty} \frac{1}{N} \vv{h}_{i,i,k}^\herm \mm{\Psi}_A^\herm \mm{\Lambda}^{-1}_{i, k} \mm{\Psi}_A \vv{h}_{i,i,k} = \lim_{N \rightarrow \infty} \frac{1}{N} \mathrm{tr} \left( \mm{\Psi}_A \mm{\Psi}^\herm_A \mm{\Gamma} \right),
\label{Eqn_Coro0delta}
\end{align}
and $\mm{\Gamma}$ is given by
\vspace*{-5mm}
\begin{align}
\mm{\Gamma} = \left( \frac{\beta \mm{\Psi}_A \mm{\Psi}^\herm_A }{1+\delta} + \frac{1}{N \rho_\mathrm{UL}} \mm{I}_{N} \right)^{-1}.
\label{Eqn_Coro0Gamma}
\end{align}
Assuming $\beta \ll 1$, $\delta \gg 1$, and applying Taylor series expansion, (\ref{Eqn_Coro0Gamma}) can be rewritten as
\vspace*{-3mm}
\begin{align}
\mm{\Gamma} \approx N \rho_\mathrm{UL} \left( \mm{I}_{N} - \frac{ N \rho_\mathrm{UL} \beta \mm{\Psi}_A \mm{\Psi}^\herm_A }{1+ \delta} \right).
\label{Eqn_Coro0Gamma2}
\end{align}
Substituting (\ref{Eqn_Coro0delta}) into (\ref{Eqn_Coro0Gamma2}) and considering $\beta \ll 1$ and $ \delta \gg 1$, the following quadratic equation is obtained
\vspace*{-4mm}
\begin{align}
\delta^2 + \left(1- N \rho_\mathrm{UL} \mu \right) \delta + \left( N^2 \rho_\mathrm{UL}^2 \beta \mu^\prime - N \rho_\mathrm{UL} \mu \right) = 0, 
\label{Eqn_IQU_delta_eqn}
\end{align}
where $\mu=\mathrm{tr} \left( \mm{\Psi}_A \mm{\Psi}^\herm_A \right)/N$ and $\mu^\prime=\mathrm{tr} \left( \left( \mm{\Psi}_A \mm{\Psi}^\herm_A \right)^2 \right)/N$. Solving (\ref{Eqn_IQU_delta_eqn}) and assuming $\beta \rightarrow 0$ leads to
\vspace*{-4mm}
\begin{align}
\delta = \frac{N \rho_\mathrm{UL} }{2} \left( \mu + \sqrt{ \mu^2 + 4 \beta \mu^\prime } \right) \xrightarrow[\beta \rightarrow 0]{} N \rho_\mathrm{UL} \mu.
\label{Eqn_Coro0mu}
\end{align}
Now, substituting (\ref{Eqn_Coro0mu}) into (\ref{Eqn_Coro0Gamma2}) and considering $\beta \ll 1$ yields
\vspace*{-4mm}
\begin{align}
\mm{\Gamma} \xrightarrow[\beta \rightarrow 0]{} N \rho_\mathrm{UL} \mm{I}_N.
\vspace*{-4mm}
\label{Eqn_Coro1_Gamma}
\end{align}
Furthermore, the interference power of the $k$th UT in the considered single-cell uplink massive MIMO system can be expressed as
\vspace*{-4mm}
\begin{align}
I^{\mathrm{IQU}^\circ}_{i, k} = \sum_{q=1, q \neq k}^{K} \rho_\mathrm{UL} \Big| \frac{1}{N} \vv{h}_{i,i,k}^\herm \mm{\Psi}^\herm_A \mm{\Lambda}^{-1}_i \mm{\Psi}_A \vv{h}_{i,i,q} \Big|^2 + \sum_{q=1}^{K} \rho_\mathrm{UL} \Big| \frac{1}{N} \vv{h}_{i,i,k}^\herm \mm{\Psi}^\herm_A \mm{\Lambda}^{-1}_i \mm{\Psi}_B \vv{h}^*_{i,i,q} \Big|^2.
\vspace*{-4mm}
\label{Eqn_IQU_Coro0_Int}
\end{align}
Using the matrix inversion lemma \cite{Horn2013}, and Lemmas \ref{Lemma_RMT1}, \ref{Lemma_Perturb}, and \ref{Lemma_RMT2}, the first term on the right hand side of (\ref{Eqn_IQU_Coro0_Int}) can be expressed as
\vspace*{-4mm}
\begin{align}
&\sum_{q=1, q \neq k}^{K} \rho_\mathrm{UL} \Big| \frac{1}{N} \vv{h}_{i,i,k}^\herm \mm{\Psi}^\herm_A \mm{\Lambda}^{-1}_i \mm{\Psi}_A \vv{h}_{i,i,q} \Big|^2 =\sum_{q=1, q \neq k}^{K} \frac{1}{N^2} \rho_\mathrm{UL} \vv{h}_{i,i,k}^\herm \mm{\Psi}^\herm_A \mm{\Lambda}^{-1}_i \mm{\Psi}_A \vv{h}_{i,i,q} \vv{h}_{i,i,q}^\herm \mm{\Psi}_A^\herm \mm{\Lambda}^{-1}_i \mm{\Psi}_A \vv{h}_{i,i,k} \nonumber \\
& \xrightarrow[N\rightarrow\infty]{\mathrm{a.s.}} \hspace*{-1mm}  \frac{\beta \rho_\mathrm{UL} \delta^\prime}{\left(1+\delta\right)^4}.
\end{align}
Here, $\delta^\prime$ is given by
\vspace*{-6mm}
\begin{align}
\delta^\prime = \frac{1}{N} \mathrm{tr} \left( \mm{\Psi}_A \mm{\Psi}^\herm_A \mm{\Lambda}^{-1}_i \mm{\Psi}_A \mm{\Psi}^\herm_A \mm{\Lambda}^{-1}_i \right) = \frac{ \frac{1}{N} \mathrm{tr} \left( \left( \mm{\Psi}_A \mm{\Psi}^\herm_A \mm{\Gamma} \right)^2 \right) }{1- \frac{\beta}{\left(1+ \delta \right)^2}  \frac{1}{N} \mathrm{tr} \left( \left( \mm{\Psi}_A \mm{\Psi}^\herm_A \mm{\Gamma} \right)^2 \right)} \xrightarrow[\beta\rightarrow 0]{} \mu^\prime N^2 \rho^2_\mathrm{UL},
\end{align}
where we used Lemma 4 and (\ref{Eqn_Coro1_Gamma}). Following a similar procedure, and considering $\delta \gg 1$, the second term on the right hand side of (\ref{Eqn_IQU_Coro0_Int}) can be approximated as
\vspace*{-4mm}
\begin{align}
&\sum_{q=1}^{K} \rho_\mathrm{UL} \Big| \frac{1}{N} \vv{h}_{i,i,k}^\herm \mm{\Psi}^\herm_A \mm{\Lambda}^{-1}_i \mm{\Psi}_B \vv{h}^*_{i,i,q} \Big|^2 =\sum_{q=1}^{K} \frac{1}{N^2} \rho_\mathrm{UL} \vv{h}_{i,i,q}^\transp \mm{\Psi}^\herm_B \mm{\Lambda}^{-1}_{i,q} \mm{\Psi}_A \vv{h}_{i,i,k} \vv{h}_{i,i,k}^\herm \mm{\Psi}_A^\herm \mm{\Lambda}^{-1}_{i,q} \mm{\Psi}_B \vv{h}^*_{i,i,q}  \nonumber \\
& \xrightarrow[N\rightarrow\infty]{\mathrm{a.s.}} \hspace*{-1mm} \frac{\beta \rho_\mathrm{UL} \delta^{\prime\prime}}{\left(1+\delta\right)^2},
\end{align}
where $\delta^{\prime\prime}$ is given by
\vspace*{-4mm}
\begin{align}
\delta^{\prime\prime} = \lim_{N \rightarrow \infty} \frac{1}{N} \mathrm{tr} \left( \mm{\Psi}_A \mm{\Psi}^\herm_A \mm{\Lambda}^{-1}_{i,q} \mm{\Psi}_B \mm{\Psi}^\herm_B \mm{\Lambda}^{-1}_{i,q} \right) = \frac{ \frac{1}{N} \mathrm{tr} \left( \left( \mm{\Psi}_A \mm{\Psi}^\herm_A \mm{\Gamma} \mm{\Psi}_B \mm{\Psi}^\herm_B \mm{\Gamma}\right) \right) }{1- \frac{\beta}{\left(1+ \delta \right)^2}  \frac{1}{N} \mathrm{tr} \left( \left( \mm{\Psi}_A \mm{\Psi}^\herm_A \mm{\Gamma} \right)^2 \right)} \xrightarrow[\beta\rightarrow 0]{} \mu^{\prime\prime} N^2 \rho_\mathrm{UL}^2,
\end{align}
with $\mu^{\prime\prime}=\mathrm{tr} \left( \mm{\Psi}_A \mm{\Psi}^\herm_A \mm{\Psi}_B \mm{\Psi}^\herm_B \right) /N$ and we used Lemma 4 and (\ref{Eqn_Coro1_Gamma}). Moreover, considering (\ref{Eqn_IQU_EstDataWithDet}) for $L=1$, perfect CSI, and no IQI at the UTs, applying the matrix inversion lemma \cite{Horn2013} and Lemmas \ref{Lemma_RMT1}, \ref{Lemma_Perturb}, and \ref{Lemma_RMT2}, the received noise power corresponding to the $k$th UT is given by
\vspace*{-4mm}
\begin{align}
Z^{\mathrm{IQU}^\circ}_{k} \xrightarrow[N\rightarrow\infty]{\mathrm{a.s.}} \frac{\delta^\prime + \delta^{\prime\prime}}{N\left(1+\delta\right)^2}.
\label{Eqn_IQA_Coro0_Noise}
\end{align}
Furthermore, considering the definition of $\mm{\Psi}_A$ and $\mm{\Psi}_B$ in Section \ref{Subsec_SystemModel_IQU}, $\epsilon_{i, n}, \theta_{i, n} \ll 1$, and performing simple algebraic operations yields $\mu=1+\sum_{n=1}^{N} \left(\epsilon_{n}^2-1\right) \sin^2 \frac{\theta_n}{2}$, $\mu^\prime= \sum_{n=1}^{N} \big( 1 + \big( \epsilon^2_{n} -1\big) \sin^2 \frac{\theta_{n}}{2} \big)^2 /N $, $\mu^{\prime\prime}= \sum_{n=1}^{N} \big( \frac{1}{^4} \ {\sin^2 \theta} $ $  + \epsilon_n^2 \big)/N$. This completes the proof.
%
\vspace*{-4mm}
\section*{Appendix D - Proof of Corollary 3}\label{Coro2proof}
Applying the matrix inversion lemma [22], and Lemmas 1, 2, and 3, the useful signal power of the $k$th UT for the IQU-MMSE detector in a single-cell uplink massive MIMO system with i.i.d. channels, IQI only at the UT, and perfect CSI (i.e., with $\hat{\vv{g}}_{i,i,k} = \xi_A \vv{h}_{i,i,k}$) is obtained as
\vspace*{-3mm}
\begin{align}
S^{\mathrm{IQU}^\circ}_{i, k} = \lim_{N \rightarrow \infty} \rho_\mathrm{UL} \Big| \frac{1}{N} \vv{h}_{i,i,k}^\herm \xi_A^* \mm{\Lambda}^{-1}_i \xi_A \vv{h}_{i,i,k} \Big|^2 = \lim_{N \rightarrow \infty} \frac{\rho_\mathrm{UL} \Big| \frac{1}{N} \vv{h}_{i,i,k}^\herm \xi_A^* \mm{\Lambda}^{-1}_{i, k} \xi_A \vv{h}_{i,i,k} \Big|^2}{\left( 1+ \frac{1}{N} \vv{h}_{i,i,k}^\herm \xi_A^* \mm{\Lambda}^{-1}_{i, k} \xi_A \vv{h}_{i,i,k}\right)^2} = \frac{ \rho_\mathrm{UL} \delta^2}{\left(1+ \delta \right)^2},
\label{Eqn_CoroIQU_SigPow}
\end{align}
where ${\mm{\Lambda}_{i}} \triangleq \frac{1}{N} {\mm{G}}_{i,i} {\mm{G}}_{i,i}^\herm + \frac{1}{N \rho_\mathrm{UL}} \mm{I}_N$, ${\mm{\Lambda}_{i, k}} \triangleq {\mm{\Lambda}_{i}} - \frac{1}{N} {\vv{g}}_{i,i,k} {\vv{g}}_{i,i,k}^\herm$ and $\delta$ is defined as
\vspace*{-5mm}
\begin{align}
\delta = \lim_{N \rightarrow \infty} \frac{1}{N} \vv{h}_{i,i,k}^\herm \xi_A^* \mm{\Lambda}^{-1}_i \xi_A \vv{h}_{i,i,k} = \lim_{N \rightarrow \infty} \frac{\big| \xi_A \big|^2}{N} \mathrm{tr} \left( \mm{\Gamma} \right),
\label{Eqn_CoroIQU_UT_delta}
\end{align}
and $\mm{\Gamma}$ is given by
\vspace*{-5mm}
\begin{align}
\mm{\Gamma} = \left( \frac{\beta \big| \xi_A \big|^2 }{1+\delta} + \frac{1}{N \rho_\mathrm{UL}}  \right)^{-1} \mm{I}_{N}.
\label{Eqn_CoroIQU_UT_Gamma}
\end{align}
Applying similar techniques as in the proof of Corollary 1, we obtain
\vspace*{-4mm}
\begin{align}
\delta & \xrightarrow[\beta \rightarrow 0]{} \big| \xi_A \big|^2 N \rho_\mathrm{UL}, \\
\mm{\Gamma} & \xrightarrow[\beta \rightarrow 0]{} N \rho_\mathrm{UL} \mm{I}_N.
\vspace*{-4mm}
\label{Eqn_CoroIQU_UT_Gamma}
\end{align}
Furthermore, the interference power of the $k$th UT in the considered single-cell uplink massive MIMO system can be expressed as
\vspace*{-2mm}
\begin{align}
I^{\mathrm{IQU}^\circ}_{i, k} = \sum_{q=1, q \neq k}^{K} \rho_\mathrm{UL} \Big| \frac{1}{N} \vv{h}_{i,i,k}^\herm \xi_A^* \mm{\Lambda}^{-1}_i \xi_A \vv{h}_{i,i,q} \Big|^2 + \sum_{q=1}^{K} \rho_\mathrm{UL} \Big| \frac{1}{N} \vv{h}_{i,i,k}^\herm \xi_A^* \mm{\Lambda}^{-1}_i \xi_B \vv{h}^*_{i,i,q} \Big|^2.
\vspace*{-4mm}
\label{Eqn_CoroIQU_UT_Int}
\end{align}
Using the matrix inversion lemma \cite{Horn2013} and Lemmas \ref{Lemma_RMT1}, \ref{Lemma_Perturb}, and \ref{Lemma_RMT2}, the first term on the right hand side of (\ref{Eqn_CoroIQU_UT_Int}) can be expressed as
\vspace*{-2mm}
\begin{align}
&\sum_{q=1, q \neq k}^{K} \rho_\mathrm{UL} \Big| \frac{1}{N} \vv{h}_{i,i,k}^\herm \xi_A^* \mm{\Lambda}^{-1}_i \xi_A \vv{h}_{i,i,q} \Big|^2 =\sum_{q=1, q \neq k}^{K} \frac{1}{N^2} \rho_\mathrm{UL} \vv{h}_{i,i,k}^\herm \xi_A^* \mm{\Lambda}^{-1}_i \xi_A \vv{h}_{i,i,q} \vv{h}_{i,i,q}^\herm \xi_A^* \mm{\Lambda}^{-1}_i \xi_A \vv{h}_{i,i,k} \nonumber \\
& \xrightarrow[N\rightarrow\infty]{\mathrm{a.s.}} \hspace*{-1mm}  \frac{\beta \rho_\mathrm{UL} \delta^\prime}{\left(1+\delta\right)^4}.
\label{Eqn_CoroIQU_UT_Interf0}
\end{align}
Here, $\delta^\prime$ is given by
\vspace*{-2mm}
\begin{align}
\delta^\prime = \frac{1}{N} \mathrm{tr} \left( \big| \xi_A \big|^4 \mm{\Lambda}^{-2}_i \right) = \frac{ \frac{1}{N} \mathrm{tr} \left(  \big| \xi_A \big|^4 \mm{\Gamma}^2 \right) }{1- \frac{\beta}{\left(1+ \delta \right)^2}  \frac{1}{N} \mathrm{tr} \left( \big| \xi_A \big|^4 \mm{\Gamma}^2 \right)} \xrightarrow[\beta\rightarrow 0]{} \big| \xi_A \big|^4 N^2 \rho^2_\mathrm{UL},
\end{align}
where we used Lemma 4 and (\ref{Eqn_CoroIQU_UT_Gamma}). Following a similar procedure, and considering $\delta \gg 1$, the second term on the right hand side of (\ref{Eqn_CoroIQU_UT_Int}) can be approximated as
\vspace*{-2mm}
\begin{align}
&\sum_{q=1}^{K} \rho_\mathrm{UL} \Big| \frac{1}{N} \vv{h}_{i,i,k}^\herm \xi_A^* \mm{\Lambda}^{-1}_i \xi_B \vv{h}^*_{i,i,q} \Big|^2 =\sum_{q=1, q \neq k}^{K} \frac{1}{N^2} \rho_\mathrm{UL} \vv{h}_{i,i,q}^\transp \xi_B^* \mm{\Lambda}^{-1}_{i} \xi_A \vv{h}_{i,i,k} \vv{h}_{i,i,k}^\herm \xi_A^* \mm{\Lambda}^{-1}_{i} \xi_B \vv{h}^*_{i,i,q}  \nonumber \\
& + \rho_\mathrm{UL} \Big| \frac{1}{N} \vv{h}_{i,i,k}^\herm \xi_A^* \mm{\Lambda}^{-1}_i \xi_B \vv{h}^*_{i,i,k} \Big|^2 \xrightarrow[N\rightarrow\infty]{\mathrm{a.s.}} \hspace*{-1mm} \frac{ \rho_\mathrm{UL} \delta^2}{\left(1+\delta\right)^2} \left| \frac{\xi_B}{\xi_A} \right|^2.
\label{Eqn_CoroIQU_UT_Interf}
\end{align}
Moreover, considering (\ref{Eqn_IQU_EstDataWithDet}) for $L=1$, perfect CSI, and no IQI at the BS and applying the matrix inversion lemma \cite{Horn2013} and Lemmas \ref{Lemma_RMT1}, \ref{Lemma_Perturb}, and \ref{Lemma_RMT2}, the received noise power of the $k$th UT is given by
\vspace*{-4mm}
\begin{align}
Z^{\mathrm{IQU}^\circ}_{k} = \left| \frac{1}{N} \vv{h}_{i,i,k}^\herm \xi_A^* \mm{\Lambda}^{-1}_i \vv{z}_i \right|^2 \xrightarrow[N\rightarrow\infty]{\mathrm{a.s.}}   \frac{\left| \xi_A \right|^2}{N\left(1+\delta\right)^2} \frac{1}{N} \mathrm{tr} \left( \mm{\Lambda}^{-2}_i \right) \xrightarrow[N\rightarrow\infty]{\mathrm{a.s.}} \frac{\rho_\mathrm{UL}^2 \left| \xi_A \right|^2 }{N \left(1+\delta\right)^2} \approx \frac{1}{N \left| \xi_A \right|^2 }.
\label{Eqn_CoroIQU_UT_Noise}
\end{align}
Considering (\ref{Eqn_CoroIQU_SigPow}) (\ref{Eqn_CoroIQU_UT_Int}), (\ref{Eqn_CoroIQU_UT_Interf0}) (\ref{Eqn_CoroIQU_UT_Interf}), (\ref{Eqn_CoroIQU_UT_Noise}), and performing straightforward algebraic simplifications yields (\ref{Eqn_Coro3_IQU}). This completes the proof.
%
\vspace*{-4mm}
\section*{Appendix E - Proof of Theorem 2}\label{Theo2proof}
According to (\ref{IQA_DetectedData}), for the IQA-WLMMSE detector, the useful signal power corresponding to the $k$th UT in the $i$th cell is given by
\vspace*{-3mm}
\begin{align}
S^{\mathrm{IQA}}_{i,k} = & 0.5 {\rho_\mathrm{UL}} \Bigg( \Bigg| \frac{\hat{\tilde{\vv{g}}}_{i,i,k}^\transp}{2N} \hspace*{-1mm} \Bigg( \frac{\hat{\tilde{\mm{G}}}_{i,i} \hat{\tilde{\mm{G}}}_{i,i}^\transp}{2N} + \frac{1}{2N \rho_\mathrm{UL}} \mm{I}_{2N} \Bigg)^{-1} \hspace*{-4mm} {\tilde{\vv{g}}}_{i,i,k} \Bigg|^2 + \Bigg| \frac{ \hat{\tilde{\vv{g}}}_{i,i,k+K}^\transp}{2N} \hspace*{-1mm} \Bigg( \frac{\hat{\tilde{\mm{G}}}_{i,i} \hat{\tilde{\mm{G}}}_{i,i}^\transp}{2N} + \frac{1}{2N\rho_\mathrm{UL}} \mm{I}_{2N} \Bigg)^{-1} \nonumber \\
& \times {\tilde{\vv{g}}}_{i,i,k+K} \Bigg|^2 \Bigg),
\label{Eqn_IQA_SigPow}
\end{align}
where the first and second terms on the right hand side represent the power of the real and imaginary parts of the useful signal, respectively. Applying the matrix inversion lemma \cite{Horn2013}, we obtain the following expression for the first term on the right hand side of (\ref{Eqn_IQA_SigPow})
\vspace*{-4mm}
\begin{align}
\left| \frac{\hat{\tilde{\vv{g}}}_{i,i,k}^\transp}{2N} \left( \frac{\hat{\tilde{\mm{G}}}_{i,i} \hat{\tilde{\mm{G}}}_{i,i}^\transp}{2N} + \frac{1}{2N\rho_\mathrm{UL}} \mm{I}_{2N} \right)^{-1} {\tilde{\vv{g}}}_{i,i,k} \right|^2 = \left| \frac{ \frac{1}{2N}\hat{\tilde{\vv{g}}}_{i,i,k}^\transp {\tilde{\mm{\Lambda}}}_{i,k}^{-1} {\tilde{\vv{g}}}_{i,i,k}}{1+\frac{1}{2N}\hat{\tilde{\vv{g}}}_{i,i,k}^\transp {\tilde{\mm{\Lambda}}}_{i,k}^{-1} \hat{\tilde{\vv{g}}}_{i,i,k} }  \right|^2,
\label{Eqn_IQA_SigPow1}
\end{align}
where ${\tilde{\mm{\Lambda}}}_{i,k}$ is given by
\vspace*{-6mm}
\begin{align}
{\tilde{\mm{\Lambda}}}_{i,k} = \frac{1}{2N} \hat{\tilde{\mm{G}}}_{i,i} \hat{\tilde{\mm{G}}}_{i,i}^\transp - \frac{1}{2N} \hat{\tilde{\vv{g}}}_{i,i,k} \hat{\tilde{\vv{g}}}_{i,i,k}^\transp + \frac{1}{2N \rho_\mathrm{UL}} \mm{I}_{2N}.
\end{align}
Considering (\ref{Eqn_IQA_ChEst_Omega}) and applying Lemmas \ref{Lemma_RMT1}, \ref{Lemma_Perturb}, and \ref{Lemma_RMT2} yields
\vspace*{-2mm}
\begin{align}
\left| \frac{\frac{1}{2N} \hat{\tilde{\vv{g}}}_{i,i,k}^\transp {\tilde{\mm{\Lambda}}}^{-1}_{i,k} {\tilde{\vv{g}}}_{i,i,k}}{1+\frac{1}{2N}\hat{\tilde{\vv{g}}}_{i,i,k}^\transp {\tilde{\mm{\Lambda}}}^{-1}_{i,k} \hat{\tilde{\vv{g}}}_{i,i,k} }  \right|^2 \xrightarrow[N\rightarrow\infty]{\mathrm{a.s.}} \frac{  \left|\tilde{\chi}_{i,k} \right|^2 }{\left( 1+ {\tilde{\delta}}_{i,k} \right)^2}, 
\label{Eqn_IQA_UsefulPowAsy1}
\end{align}
where $\tilde{\chi}_{i,k}$ and ${\tilde{\delta}}_{i,k}$ are defined as
\vspace*{-3mm}
\begin{align}
{\tilde{\delta}}_{i,k} &\triangleq \frac{1}{2N} \mathrm{tr} \left(  \tilde{\mm{\Phi}}^{(1)}_{i,i,k} \tilde{\mm{\Gamma}}_{i} \right) = \lim_{N \rightarrow \infty} \frac{1}{2N} \hat{\tilde{\vv{g}}}_{i,i,k}^\transp {\tilde{\mm{\Lambda}}}^{-1}_{i,i,k} \hat{\tilde{\vv{g}}}_{i,i,k}
\label{Eqn_IQA_delta_i_k} \\
\tilde{\chi}_{i,k} & \triangleq \frac{1}{2N} \mathrm{tr} \left( \tilde{\mm{\Phi}}^{(2)}_{i,i,k} \tilde{\mm{\Gamma}}_{i} \right) = \lim_{N \rightarrow \infty} \frac{1}{2N} \hat{\tilde{\vv{g}}}_{i,i,k}^\transp {\tilde{\mm{\Lambda}}}^{-1}_{i,i,k} {\tilde{\vv{g}}}_{i,i,k} \label{Eqn_IQA_chi_i_k}.
\end{align}
Moreover, $\tilde{\mm{\Phi}}^{(1)}_{i,i,k}$ and $\tilde{\mm{\Phi}}^{(2)}_{i,l,k}$ are given by
\vspace*{-4mm}
\begin{align}
\tilde{\mm{\Phi}}^{(1)}_{i,i,k} &=  \tilde{\mm{\Omega}}_{i,i,k} \left( 0.5 \sum_{l=1}^{L} \left({\left[ \mm{\Xi}_{l,k} \right]}_{1,1}^2 +{\left[ \mm{\Xi}_{l,k} \right]}_{2,1}^2 \right)\tilde{\mm{\Psi}}_{i} \tilde{\mm{R}}_{i,l,k} \tilde{\mm{\Psi}}_{i}^\transp + \frac{1}{2\rho_\mathrm{TR}} \tilde{\mm{\Psi}}_{i} \tilde{\mm{\Psi}}_{i}^\transp \ \right)  \tilde{\mm{\Omega}}_{i,i,k}^\transp \\
\tilde{\mm{\Phi}}^{(2)}_{i,l,k} &= 0.5 \left({\left[ \mm{\Xi}_{l,k} \right]}_{1,1}^2 +{\left[ \mm{\Xi}_{l,k} \right]}_{2,1}^2 \right) \tilde{\mm{\Omega}}_{i,l,k} \tilde{\mm{\Psi}}_{i} \tilde{\mm{R}}_{i,l,k} \tilde{\mm{\Psi}}_{i}^\transp, \label{Eqn_IQA_Phi2_00}
\end{align}
and $\tilde{\mm{\Gamma}}_{i}$ is determined as
\vspace*{-4mm}
\begin{align}
\tilde{\mm{\Gamma}}_{i} = \left( \frac{1}{2N} \sum_{k=1}^{K} \left( \frac{\tilde{\mm{\Phi}}^{(1)}_{i,i,k}}{1+\tilde{\delta}_{i,k}} + \frac{ \tilde{\mm{\Phi}}^{(1)}_{i,i,k+K}}{1+\tilde{\delta}_{i, k+K} }\right) + \frac{1}{2N\rho_\mathrm{UL}} \mm{I}_{2N} \right)^{-1},
\end{align}
where $\tilde{\mm{\Phi}}^{(1)}_{i,i,k+K}$ is defined as
\vspace*{-4mm}
\begin{align}
\tilde{\mm{\Phi}}^{(1)}_{i,i,k+K} &=  \tilde{\mm{\Omega}}_{i,i,k} \left( 0.5 \sum_{l=1}^{L} \left({\left[ \mm{\Xi}_{l,k} \right]}_{1,2}^2 +{\left[ \mm{\Xi}_{l,k} \right]}_{2,2}^2  \right)\tilde{\mm{\Psi}}_{i} \tilde{\mm{R}}_{i,l,k} \tilde{\mm{\Psi}}_{i}^\transp + \frac{1}{2\rho_\mathrm{TR}} \tilde{\mm{\Psi}}_{i} \tilde{\mm{\Psi}}_{i}^\transp \ \right)  \tilde{\mm{\Omega}}_{i,i,k}^\transp,
\end{align}
and $\tilde{\delta}_{i,k}$ and $\tilde{\delta}_{i,k+K}$ are the solutions to the following fixed-point equations
\vspace*{-4mm}
\begin{align}
\tilde{\delta}_{i,k} =& \frac{1}{2N} \mathrm{tr} \left( \tilde{\mm{\Phi}}^{(1)}_{i,i,k} \left( \frac{1}{2N} \sum_{q=1}^{K} \left( \frac{\tilde{\mm{\Phi}}^{(1)}_{i,i,q}}{1+\tilde{\delta}_{i,q}} + \frac{ \tilde{\mm{\Phi}}^{(1)}_{i,i,q+K}}{1+\tilde{\delta}_{i,q+K} }\right) + \frac{1}{2N\rho_\mathrm{UL}} \mm{I}_{2N} \right)^{-1} \right) \label{Eqn_IQA_delta0} \\
\tilde{\delta}_{i,k+K} =& \frac{1}{2N} \mathrm{tr} \left( \tilde{\mm{\Phi}}^{(1)}_{i,i,k+K} \left( \frac{1}{2N} \sum_{q=1}^{K} \left( \frac{\tilde{\mm{\Phi}}^{(1)}_{i,i,q}}{1+\tilde{\delta}_{i,q}} + \frac{ \tilde{\mm{\Phi}}^{(1)}_{i,i,q+K}}{1+\tilde{\delta}_{i,q+K} }\right) + \frac{1}{2N\rho_\mathrm{UL}} \mm{I}_{2N} \right)^{-1} \right).
\label{Eqn_IQA_delta1}
\end{align}
Similarly, the second term on the right hand side of (\ref{Eqn_IQA_SigPow}) can be expressed as
\vspace*{-4mm}
\begin{align}
& \left| \frac{1}{2N} \hat{\tilde{\vv{g}}}_{i,i,k+K}^\transp \left( \frac{1}{2N} \hat{\tilde{\mm{G}}}_{i,i} \hat{\tilde{\mm{G}}}_{i,i}^\transp + \frac{1}{2N\rho_\mathrm{UL}} \mm{I}_{2N} \right)^{-1} {\tilde{\vv{g}}}_{i,i,k+K} \right|^2 = \left| \frac{\frac{1}{2N}\hat{\tilde{\vv{g}}}_{i,i,k+K}^\transp {\tilde{\mm{\Lambda}}}^{-1}_{i,i,k+K} {\tilde{\vv{g}}}_{i,i,k+K}}{1+\frac{1}{2N}\hat{\tilde{\vv{g}}}_{i,i,k+K}^\transp {\tilde{\mm{\Lambda}}}^{-1}_{i,i,k+K} \hat{\tilde{\vv{g}}}_{i,i,k+K} }  \right|^2 \nonumber \\
& \xrightarrow[N\rightarrow\infty]{\mathrm{a.s.}} \frac{  \left|\tilde{\chi}_{i,k+K} \right|^2 }{\left( 1+ {\tilde{\delta}}_{i,k+K} \right)^2},
\label{Eqn_IQA_UsefulPowAsy2}
\end{align}
where ${\tilde{\delta}}_{i,k+K}$ is given in (\ref{Eqn_IQA_delta1}) and $\tilde{\chi}_{i,k+K}$ is defined as
\vspace*{-4mm}
\begin{align}
\tilde{\chi}_{i,k+K} = \frac{1}{2N} \mathrm{tr} \left( \tilde{\mm{\Phi}}^{(2)}_{i,i,k+K} \tilde{\mm{\Gamma}}_{i} \right)
\end{align}
with
\vspace*{-4mm}
\begin{align}
\tilde{\mm{\Phi}}^{(2)}_{i,i,k+K} &= 0.5 \left( {\left[ \mm{\Xi}_{l,k} \right]}_{1,2}^2 +{\left[ \mm{\Xi}_{l,k} \right]}_{2,2}^2 \right) \tilde{\mm{\Omega}}_{i,i,k} \tilde{\mm{\Psi}}_{i} \tilde{\mm{R}}_{i,i,k} \tilde{\mm{\Psi}}_{i}^\transp.
\end{align}
Considering (\ref{Eqn_IQA_SigPow}), (\ref{Eqn_IQA_UsefulPowAsy1}), and (\ref{Eqn_IQA_UsefulPowAsy2}) the asymptotic useful signal power can be expressed as
\vspace*{-4mm}
\begin{align}
S^{\mathrm{IQA}^\circ}_{i,k} = \lim_{N \rightarrow \infty} S^{\mathrm{IQA}}_{i,k} = 0.5 \rho_\mathrm{UL} \left( \frac{  \left|\tilde{\chi}_{i,k} \right|^2 }{\left( 1+ {\tilde{\delta}}_{i,k} \right)^2} + \frac{  \left|\tilde{\chi}_{i,k+K} \right|^2 }{\left( 1+ {\tilde{\delta}}_{i,k+K} \right)^2} \right).
\end{align}
%
According to (\ref{IQA_DetectedData}), the interference power of the signal corresponding to the $k$th UT observed at the output of the IQA-WLMMSE detector of the $i$th BS is given by
\vspace*{-4mm}
\begin{align}
&I^{\mathrm{IQA}}_{ik} = \frac{0.5\rho_\mathrm{UL}}{4N^2} \mathop{\sum_{l=1}^{L}\sum_{q=1}^{K}}_{(l, q) \neq (i, k)} \Bigg( \left| \hat{\tilde{\vv{g}}}_{i,i,k}^\transp {\tilde{\mm{\Lambda}}}_{i}^{-1} \tilde{\vv{g}}_{i,l,q} \right|^2 + \bigg| \hat{\tilde{\vv{g}}}_{i,i,k}^\transp {\tilde{\mm{\Lambda}}}_{i}^{-1} \tilde{\vv{g}}_{i,l,q+K} \bigg|^2 + \left| \hat{\tilde{\vv{g}}}_{i,i,k+K}^\transp {\tilde{\mm{\Lambda}}}_{i}^{-1} \tilde{\vv{g}}_{i,l,q} \right|^2 + \nonumber \\
& \left| \hat{\tilde{\vv{g}}}_{i,i,k+K}^\transp {\tilde{\mm{\Lambda}}}_{i}^{-1} \tilde{\vv{g}}_{i,l,q+K} \right|^2\Bigg),
\label{Eqn_IQA_InterfPow1}
\end{align}
where ${\tilde{\mm{\Lambda}}}_{i} = \frac{1}{2N} \hat{\tilde{\mm{G}}}_{i,i} \hat{\tilde{\mm{G}}}_{i,i}^\transp + \frac{1}{2N\rho_\mathrm{UL}} \mm{I}_{2N}$.
The first term on the right hand side of (\ref{Eqn_IQA_InterfPow1}) can be reformulated as
\begin{align}
& \mathop{\sum_{l=1}^{L}\sum_{q=1}^{K}}_{(l, q) \neq (i, k)} \left| \frac{1}{2N} \hat{\tilde{\vv{g}}}_{i,i,k}^\transp {\tilde{\mm{\Lambda}}}_{i}^{-1} \tilde{\vv{g}}_{i,l,q} \right|^2 = \mathop{\sum_{l=1}^{L}}_{l \neq i} \left| \frac{1}{2N} \hat{\tilde{\vv{g}}}_{i,i,k}^\transp {\tilde{\mm{\Lambda}}}_{i}^{-1} \tilde{\vv{g}}_{i,l,k} \right|^2 + \mathop{\sum_{l=1}^{L}\sum_{q=1}^{K}}_{q \neq k} \left| \frac{1}{2N} \hat{\tilde{\vv{g}}}_{i,i,k}^\transp {\tilde{\mm{\Lambda}}}_{i}^{-1} \tilde{\vv{g}}_{i,l,q} \right|^2.
\label{Eqn_IQA_InterfMcellTerm}
\end{align}
Applying the matrix inversion lemma \cite{Horn2013}, Lemmas 1-3, and considering (\ref{Eqn_IQA_delta_i_k}) yields
\vspace*{-3mm}
\begin{align}
\frac{1}{2N} \hat{\tilde{\vv{g}}}_{i,i,k}^\transp {\tilde{\mm{\Lambda}}}_{i}^{-1} \tilde{\vv{g}}_{i,l,k} = \frac{\frac{1}{2N}\hat{\tilde{\vv{g}}}_{i,i,k}^\transp {\tilde{\mm{\Lambda}}}^{-1}_{i,k} {\tilde{\vv{g}}}_{i,l,k}}{1+\frac{1}{2N}\hat{\tilde{\vv{g}}}_{i,i,k}^\transp {\tilde{\mm{\Lambda}}}^{-1}_{i,k} \hat{\tilde{\vv{g}}}_{i,i,k} }  \xrightarrow[N\rightarrow\infty]{\mathrm{a.s.}} &\frac{ \mathrm{tr} \left( \tilde{\mm{\Phi}}^{(2)}_{i, l, k} \mm{\Gamma}_{i} \right) }{2N\left(1+\tilde{\delta}_{i,k}\right)} =\frac{\tilde{\lambda}_{i,l,k}}{1+\tilde{\delta}_{i,k}},
\end{align}
where $\tilde{\lambda}_{i,l,k}$ is defined as
\vspace*{-6mm}
\begin{align}
\tilde{\lambda}_{i,l,k} \triangleq \frac{1}{2N} \mathrm{tr} \left( \tilde{\mm{\Phi}}^{(2)}_{i, l, k} \mm{\Gamma}_{i} \right),
\label{Eqn_lambda_IQA}
\end{align}
and $\tilde{\mm{\Phi}}^{(2)}_{i, l, k}$ is given by (\ref{Eqn_IQA_Phi2_00}).
On the other hand, for the second term on the right hand side of (\ref{Eqn_IQA_InterfMcellTerm}), using matrix the inversion lemma \cite{Horn2013} and Lemmas 1-3 yields \cite{Hoydis2013}
\vspace*{-3mm}
\begin{align}
&\left| \frac{1}{2N} \hat{\tilde{\vv{g}}}_{i,i,k}^\transp {\tilde{\mm{\Lambda}}}_{i}^{-1} \tilde{\vv{g}}_{i,l,q} \right|^2=\left| \frac{\frac{1}{2N} \hat{\tilde{\vv{g}}}_{i,i,k}^\transp \tilde{\mm{\Lambda}}^{-1}_{i,k} \tilde{\vv{g}}_{i,l,q} }{1+\frac{1}{2N}\hat{\tilde{\vv{g}}}^\transp_{i,i,k} \tilde{\mm{\Lambda}}^{-1}_{i,k} \hat{\tilde{\vv{g}}}_{i,i,k} } \right|^2 = \frac{  \tilde{\vv{g}}^\transp_{i,l,q}  \tilde{\mm{\Lambda}}^{-1}_{i,k} \tilde{\mm{\Phi}}^{(1)}_{i,i,k} \tilde{\mm{\Lambda}}^{-1}_{i,k} \tilde{\vv{g}}_{i,l,q} }{4N^2\left(1+\tilde{\delta}_{i,k}\right)^2}=\nonumber \\
& \frac{1}{\left(1+\tilde{\delta}_{i,k}\right)^2} \Bigg( \frac{1}{4N^2} \tilde{\vv{g}}_{i,l,q}^\transp \tilde{\mm{\Lambda}}^{-1}_{i,k,q} \tilde{\mm{\Phi}}^{(1)}_{i,i,k} \tilde{\mm{\Lambda}}^{-1}_{i,k,q} \tilde{\vv{g}}_{i,l,q} + \frac{ \left| \frac{1}{2N} \tilde{\vv{g}}_{i,l,q}^\transp \tilde{\mm{\Lambda}}^{-1}_{i,k,q} \hat{ \tilde{\vv{g}}}_{i,i,q} \right|^2 \hat{ \tilde{\vv{g}}}_{i,i,q}^\transp \tilde{\mm{\Lambda}}^{-1}_{i,k,q} \tilde{\mm{\Phi}}^{(1)}_{i,i,k} \tilde{\mm{\Lambda}}^{-1}_{i,k,q} \hat{ \tilde{\vv{g}}}_{i,i,q} }{4N^2\left(1+ \tilde{\delta}_{i,k}\right)^2} \nonumber \\ 
&- 2 \Re{\left\lbrace \frac{ \left( \frac{1}{2N} \hat{ \tilde{\vv{g}}}_{i,i,q}^\transp \tilde{\mm{\Lambda}}^{-1}_{i,k,q} \tilde{\vv{g}}_{i,l,q} \right) \left( \frac{1}{4N^2} \tilde{\vv{g}}_{i,l,q}^\transp  \tilde{\mm{\Lambda}}^{-1}_{i,k,q} \tilde{\mm{\Phi}}^{(1)}_{i,i,k} \tilde{\mm{\Lambda}}^{-1}_{i,k,q} \hat{ \tilde{\vv{g}}}_{i,i,q} \right) }{\left(1+\tilde{\delta}_{i,q}\right)}  \right\rbrace } \Bigg)\xrightarrow[N\rightarrow\infty]{\mathrm{a.s.}} \frac{1}{\left(1+\tilde{\delta}_{i,k}\right)^2} \Bigg( \tilde{\zeta}_{i,l,q} + \nonumber \\
& \frac{\left|\tilde{\lambda}_{i,l,q}\right|^2 \tilde{\kappa}_{i,i,q}}{\left(1+\tilde{\delta}_{i,k}\right)^2} -2\Re{\left\lbrace \frac{\tilde{\lambda}_{i,l,q} \tilde{\phi}_{i,i,q}}{1+\tilde{\delta}_{i,k}} \right\rbrace } \Bigg)= \frac{\tilde{\varrho}_{i,l,q}}{\left(1+\tilde{\delta}_{i,k}\right)^2}, 
\label{Eqn_IQA_InterPart2}
\end{align}
where $\tilde{\mm{\Lambda}}_{i,k,q} = \frac{1}{2N} \hat{\tilde{\mm{G}}}_{i,i} \hat{\tilde{\mm{G}}}_{i,i}^\transp -\frac{1}{2N}\hat{\tilde{\vv{g}}}_{i,i,k} \hat{\tilde{\vv{g}}}_{i,i,k}^\transp -\frac{1}{2N}\hat{\tilde{\vv{g}}}_{i,i,q} \hat{\tilde{\vv{g}}}_{i,i,q}^\transp + \frac{1}{2N\rho_\mathrm{UL}} \mm{I}_{2N}$, and $\tilde{\delta}_{i,q}$ and $\tilde{\lambda}_{i,l,q}$ are obtained by replacing $k$ with $q$ in (\ref{Eqn_IQA_delta0}) and (\ref{Eqn_lambda_IQA}), respectively. Moreover, $\tilde{\zeta}_{i,l,q}$, $\tilde{\kappa}_{i,i,q}$, and $\tilde{\phi}_{i,i,q}$ are defined as
\vspace*{-6mm}
\begin{align}
& \frac{1}{4N^2} \tilde{\vv{g}}_{i,l,q}^\transp \tilde{\mm{\Lambda}}^{-1}_{i,k,q} \tilde{\mm{\Phi}}^{(1)}_{i,i,k} \tilde{\mm{\Lambda}}^{-1}_{i,k,q} \tilde{\vv{g}}_{i,l,q} \xrightarrow[N\rightarrow\infty]{\mathrm{a.s.}} \frac{1}{4N^2} \mathrm{tr} \left( \tilde{\mm{R}}_{i,l,q} \tilde{\mm{\Gamma}}^\prime_{i,k} \right) = \tilde{\zeta}_{i,l,q} \label{Eqn_IQAzeta} \\
& \frac{1}{4N^2} \hat{ \tilde{\vv{g}}}_{i,i,q}^\transp \tilde{\mm{\Lambda}}^{-1}_{i,k,q} \tilde{\mm{\Phi}}^{(1)}_{i,i,k} \tilde{\mm{\Lambda}}^{-1}_{i,k,q} \hat{ \tilde{\vv{g}}}_{i,i,q} \xrightarrow[N\rightarrow\infty]{\mathrm{a.s.}} \frac{1}{4N^2} \mathrm{tr} \left( \tilde{\mm{\Phi}}^{(1)}_{i,i,k} \tilde{\mm{\Gamma}}^\prime_{i,k} \right) = \tilde{\kappa}_{i,i,q} \label{Eqn_IQAkappa}\\
& \frac{1}{4N^2} \tilde{\vv{g}}_{i,l,q}^\transp  \tilde{\mm{\Lambda}}^{-1}_{i,k,q} \tilde{\mm{\Phi}}^{(1)}_{i,i,k} \tilde{\mm{\Lambda}}^{-1}_{i,k,q} \hat{ \tilde{\vv{g}}}_{i,i,q} \xrightarrow[N\rightarrow\infty]{\mathrm{a.s.}} \frac{1}{4N^2} \mathrm{tr} \left( \tilde{\mm{R}}_{i,l,q} \tilde{\mm{\Gamma}}^\prime_{i,k} \tilde{\mm{\Omega}}_{i,k} \right) = \tilde{\phi}_{i,i,q},
\label{Eqn_IQAphi}
\end{align}
where $\tilde{\mm{\Gamma}}^\prime_{i,k}$ is given by $\mm{T}^\prime$ in Lemma \ref{Lemma_RMT3} with $N$ being replaced by $2N$ and $\mm{T}$, $\mm{C}$, and $\mm{\Delta}_k$ being equal to $\tilde{\mm{\Gamma}}_{i}$, $\tilde{\mm{\Phi}}^{(1)}_{i,i,k}$, and $\tilde{\mm{\Phi}}^{(1)}_{i,i,k}/(2N)$, respectively. Performing a similar procedure for the other terms in (\ref{Eqn_IQA_InterfMcellTerm}), the interference power as given in (\ref{Eqn_IQA_AsySigPow}) is obtained.
Next, we derive the asymptotic value of the noise power for the IQA-WLMMSE detector. According to (\ref{IQA_DetectedData}), the noise power corresponding to the $k$th UT in the $i$th cell at the output of the IQA-WLMMSE detector is given by
\vspace*{-6mm}
\begin{align}
Z^{\mathrm{IQA}}_{i,k} = \left| \frac{1}{2N} \hat{\tilde{\vv{g}}}_{i,i,k}^\transp \tilde{\mm{\Lambda}}^{-1}_{i} \tilde{\mm{\Psi}}_{i} \tilde{\vv{z}}_i \right|^2+  \left| \frac{1}{2N} \hat{\tilde{\vv{g}}}_{i,i,k+K}^\transp \tilde{\mm{\Lambda}}^{-1}_{i} \tilde{\mm{\Psi}}_{i} \tilde{\vv{z}}_i \right|^2.
\end{align}
Applying the \vspace*{-2mm}
\begin{align}
Z^{\mathrm{IQA}}_{i,k} = \frac{ 0.5\hat{\tilde{\vv{g}}}_{i,i,k}^\transp \tilde{\mm{\Lambda}}^{-1}_{i,k} \tilde{\mm{\Psi}}_{i} \tilde{\mm{\Psi}}_{i}^\transp \tilde{\mm{\Lambda}}^{-1}_{i,k} \hat{\tilde{\vv{g}}}_{i,i,k}}{4N^2\left(1+\frac{1}{2N}\hat{\tilde{\vv{g}}}_{i,i,k}^\transp \tilde{\mm{\Lambda}}^{-1}_{i,k}\hat{\tilde{\vv{g}}}_{i,i,k}\right)^2}  +  \frac{0.5\hat{\tilde{\vv{g}}}_{i,i,k+K}^\transp \tilde{\mm{\Lambda}}^{-1}_{i,k+K} \tilde{\mm{\Psi}}_{i} \tilde{\mm{\Psi}}_{i}^\transp \tilde{\mm{\Lambda}}^{-1}_{i,k+K} \hat{\tilde{\vv{g}}}_{i,i,k+K}}{4N^2\left(1+\frac{1}{2N}\hat{\tilde{\vv{g}}}_{i,i,k+K}^\transp \tilde{\mm{\Lambda}}^{-1}_{i,k+K}\hat{\tilde{\vv{g}}}_{i,i,k+K}\right)^2}.
\label{Theo2Proof_Noise}
\end{align}
Finally, applying Lemmas \ref{Lemma_RMT1}, \ref{Lemma_Perturb}, and \ref{Lemma_RMT2} to (\ref{Theo2Proof_Noise}) leads to (\ref{Eqn_IQA_AsyNoisePow}). This completes the proof.

%


%
\vspace*{-2mm}
\section*{Appendix F - Proof of Corollary 4}\label{Coro3proof}
Applying the matrix inversion lemma \cite{Horn2013}, and Lemmas 1-3 in (\ref{Eqn_IQA_SigPow}), the asymptotic power of the useful signal of the $k$th UT in a single-cell uplink massive MIMO system with perfect CSI and IQI only at the BS is given by
\vspace*{-3mm}
\begin{align}
S^{\mathrm{IQA}^\circ}_{i, k} \hspace*{-2mm} = \frac{0.5\rho_\mathrm{UL}\Big| \tilde{\vv{g}}_{i,i,k}^\transp \tilde{\mm{\Lambda}}^{-1}_{i,k} \tilde{\vv{g}}_{i,i,k} \Big|^2}{4N^2\left( 1+\frac{1}{2N}\tilde{\vv{g}}_{i,i,k}^\transp \tilde{\mm{\Lambda}}^{-1}_{i,k} \tilde{\vv{g}}_{i,i,k}\right)^2} + \frac{0.5 \rho_\mathrm{UL}\Big| \tilde{\vv{g}}_{i,i,k+K}^\transp \tilde{\mm{\Lambda}}^{-1}_{i,k+K} \tilde{\vv{g}}_{i,i,k+K} \Big|^2}{4N^2\left( 1+\frac{1}{2N}\tilde{\vv{g}}_{i,i,k+K}^\transp \tilde{\mm{\Lambda}}^{-1}_{i,k+K} \tilde{\vv{g}}_{i,i,k+K}\right)^2} = \frac{\rho_\mathrm{UL}\tilde{\delta}^2}{\left(1+\tilde{\delta} \right)^2},
\label{Eqn_IQA_Coro1_Sig}
\end{align}
where 
\vspace*{-6mm}
\begin{align}
\tilde{\delta} = \frac{1}{2N} \mathrm{tr} \left( \tilde{\mm{\Psi}} \tilde{\mm{\Psi}}^\transp \tilde{\mm{\Gamma}} \right),
\label{Eqn_Coro1delta}
\end{align}
and $\tilde{\mm{\Gamma}}$ is given by
\vspace*{-6mm}
\begin{align}
\tilde{\mm{\Gamma}} = \left( \frac{\beta \tilde{\mm{\Psi}} \tilde{\mm{\Psi}}^\transp }{1+\tilde{\delta}} + \frac{1}{2N\rho_\mathrm{UL}} \mm{I}_{2N} \right)^{-1}.
\label{Eqn_Coro1Gamma}
\end{align}
Assuming $\beta \ll 1$, $\tilde{\delta} \gg 1$, and applying Taylor series expansion, (\ref{Eqn_Coro1Gamma}) can be rewritten as
\vspace*{-3mm}
\begin{align}
\tilde{\mm{\Gamma}} \approx 2N \rho_\mathrm{UL} \left( \mm{I}_{2N} - \frac{ 2N \rho_\mathrm{UL} \beta \tilde{\mm{\Psi}} \tilde{\mm{\Psi}}^\transp }{1+\tilde{\delta} }  \right).
\label{Eqn_Coro1Gamma2}
\end{align}
Substituting (\ref{Eqn_Coro1delta}) into (\ref{Eqn_Coro1Gamma2}) and considering $\beta \ll 1$,  the following quadratic equation is obtained
\vspace*{-6mm}
\begin{align}
\tilde{\delta}^2 + \left(1-2N\rho_\mathrm{UL} \tilde{\mu} \right) \tilde{\delta} + \left( 4N^2\rho_\mathrm{UL}^2 \beta \check{\tilde{\mu}} - 2N \rho_\mathrm{UL} \tilde{\mu} \right) = 0, 
\label{Eqn_IQU_Coro2_delta_eq}
\end{align}
where $\tilde{\mu}=\mathrm{tr} \left( \tilde{\mm{\Psi}} \tilde{\mm{\Psi}}^\transp \right)/(2N)$ and $\check{\tilde{\mu}}=\mathrm{tr} \left( \left( \tilde{\mm{\Psi}} \tilde{\mm{\Psi}}^\transp \right)^2 \right)/(2N)$. Solving (\ref{Eqn_IQU_Coro2_delta_eq}) and applying further straightforward simplifications leads to
\vspace*{-4mm}
\begin{align}
\tilde{\delta} = \frac{2N \rho_\mathrm{UL} }{2} \left( \tilde{\mu} + \sqrt{ \tilde{\mu}^2 + 4 \beta \check{\tilde{\mu}} } \right) \xrightarrow[\beta \rightarrow 0]{} 2N \rho_\mathrm{UL} \tilde{\mu}.
\label{Eqn_Coro1mu}
\end{align}
Substituting (\ref{Eqn_Coro1mu}) into (\ref{Eqn_Coro1Gamma2}) and considering $\beta \ll 1$ yields
\vspace*{-4mm}
\begin{align}
\tilde{\mm{\Gamma}} = 2N \rho_\mathrm{UL} \mm{I}_{2N}.
\end{align}
Now, we evaluate the interference power of the $k$th UT in the considered single-cell uplink massive MIMO system, which can be expressed as
\vspace*{-3mm}
\begin{align}
& I^{\mathrm{IQA}^\circ}_{i, k} \hspace*{-2mm} = \hspace*{-3mm} \sum_{q=1, q \neq k}^{2K} \hspace*{-3mm} \frac{0.5 \rho_\mathrm{UL}}{4N^2} \left( \Big|  \tilde{\vv{g}}_{i,i,k}^\transp \tilde{\mm{\Lambda}}^{-1}_i \tilde{\vv{g}}_{i,i,q} \Big|^2 \hspace*{-2mm} + \hspace*{-1mm} \Big| \tilde{\vv{g}}_{i,i,k+K}^\transp \tilde{\mm{\Lambda}}^{-1}_i \tilde{\vv{g}}_{i,i,q} \Big|^2 \right) \hspace*{-1mm} = \frac{\rho_\mathrm{UL}}{8N^2} \hspace*{-3mm} \sum_{q=1, q \neq k}^{2K} \hspace*{-2mm} \tilde{\vv{g}}_{i,i,k}^\transp \tilde{\mm{\Lambda}}^{-1}_i \tilde{\vv{g}}_{i,i,q} \tilde{\vv{g}}_{i,i,q}^\transp \tilde{\mm{\Lambda}}^{-1}_i \tilde{\vv{g}}_{i,i,k} \nonumber\\
& + \frac{\rho_\mathrm{UL}}{8N^2} \sum_{q=1, q \neq k}^{2K} \tilde{\vv{g}}_{i,i,k+K}^\transp \tilde{\mm{\Lambda}}^{-1}_i \tilde{\vv{g}}_{i,i,q} \tilde{\vv{g}}_{i,i,q}^\transp \tilde{\mm{\Lambda}}^{-1}_i \tilde{\vv{g}}_{i,i,k+K} \xrightarrow[N \rightarrow \infty]{\mathrm{a.s.}} \frac{\rho_\mathrm{UL} \beta \tilde{\delta}^\prime }{\left( 1+ \tilde{\delta} \right)^4},
\label{Eqn_IQA_Coro1_Int}
\end{align}
where we applied the matrix inversion lemma \cite{Horn2013} and Lemmas 1 and 2. Here, $\tilde{\delta}^\prime$ is given by
\vspace*{-3mm}
\begin{align}
\tilde{\delta}^\prime = \frac{1}{2N} \mathrm{tr} \left( \tilde{\mm{\Psi}} \tilde{\mm{\Psi}}^\transp \tilde{\mm{\Lambda}}^{-1}_i  \tilde{\mm{\Psi}} \tilde{\mm{\Psi}}^\transp \tilde{\mm{\Lambda}}^{-1}_i \right) = \frac{ \frac{1}{2N} \mathrm{tr} \left( \left( \tilde{\mm{\Psi}} \tilde{\mm{\Psi}}^\transp \tilde{\mm{\Gamma}} \right)^2 \right) }{1- \frac{\beta}{\left(1+ \tilde{\delta} \right)^2}  \frac{1}{2N} \mathrm{tr} \left( \left( \tilde{\mm{\Psi}} \tilde{\mm{\Psi}}^\transp \tilde{\mm{\Gamma}} \right)^2 \right)},
\end{align}
where we used Lemma 3. Assuming $\beta \ll 1$ leads to
\vspace*{-4mm}
\begin{align}
\tilde{\delta}^\prime = 4N^2 \rho^2 \check{\tilde{\mu}}.
\end{align}
Moreover, the received noise power corresponding to the $k$th UT is given by
\vspace*{-3mm}
\begin{align}
Z^{\mathrm{IQA}^\circ}_{i, k} =  & \Big| \frac{1}{2N} \tilde{\vv{g}}_{i,i,k}^\transp \tilde{\mm{\Lambda}}^{-1}_i \tilde{\mm{\Psi}} \tilde{\vv{z}}_{i} \Big|^2 + \Big| \frac{1}{2N} \tilde{\vv{g}}_{i,i,k+K}^\transp \tilde{\mm{\Lambda}}^{-1}_i \tilde{\mm{\Psi}} \tilde{\vv{z}}_{i} \Big|^2  = \frac{1}{4N^2} \tilde{\vv{g}}_{i,i,k}^\transp \tilde{\mm{\Lambda}}^{-1}_i \tilde{\mm{\Psi}} \tilde{\vv{z}}_{i} \tilde{\vv{z}}^\transp_{i} \tilde{\mm{\Psi}}^\transp \tilde{\mm{\Lambda}}^{-1}_i \tilde{\vv{g}}_{i,i,k} + \nonumber \\
& \frac{1}{4N^2} \tilde{\vv{g}}_{i,i,k+K}^\transp \tilde{\mm{\Lambda}}^{-1}_i \tilde{\mm{\Psi}} \tilde{\vv{z}}_{i} \tilde{\vv{z}}^\transp_{i} \tilde{\mm{\Psi}}^\transp \tilde{\mm{\Lambda}}^{-1}_i \tilde{\vv{g}}_{i,i,k+K} \xrightarrow[N \rightarrow \infty]{\mathrm{a.s.}} \frac{ \tilde{\delta}^\prime }{N\left(1+\tilde{\delta}\right)^2},
\label{Eqn_IQA_Coro1_Noise}
\end{align}
where we applied the matrix inversion lemma \cite{Horn2013} and Lemmas 1-4. Now, combining (\ref{Eqn_IQA_Coro1_Sig}), (\ref{Eqn_IQA_Coro1_Int}), (\ref{Eqn_IQA_Coro1_Noise}), and $\tilde{\delta} \gg 1$ yields
\vspace*{-5mm}
\begin{align}
\mathrm{SINR}^{\mathrm{IQA}^\circ}_{i, k} = \frac{ N \rho^2_\mathrm{UL} \tilde{\mu}^4 }{ \beta \check{\tilde{\mu}} N + \rho_\mathrm{UL} \check{\tilde{\mu}} \tilde{\mu}^2 }.
\end{align}
Next, we derive closed-form expressions for $\tilde{\mu}=\mathrm{tr}\left( \tilde{\mm{\Psi}} \tilde{\mm{\Psi}}^\transp \right)/2N$ and $\check{\tilde{\mu}}=\mathrm{tr} \left( \tilde{\mm{\Psi}} \tilde{\mm{\Psi}}^\transp \right)^2 /2N$. It can be easily shown that $\mm{\Pi} \mm{\Pi}^\transp=\mm{I}_{2N}$. Considering this and using (\ref{Eqn_IQA_MatrixA}), we obtain $\tilde{\mu}=\sum_{n=1}^N \norm{\mm{A}_{n}}^2_F/2N=1+\frac{1}{N}\sum_{n=1}^{N} \epsilon_{n}^2$. Similarly, $\check{\tilde{\mu}}$ is obtained as $\check{\tilde{\mu}}=\frac{1}{2N}\sum_{n=1}^{N} \left(1+\epsilon_{n}\right)^4 + \left(1-\epsilon_{n}\right)^4 + 2\left(1-\epsilon_{n}^2\right)^2 \sin^2 \theta_{n}$. Using  $\epsilon_{n} \ll 1$ and performing simple algebraic operations yields $\check{\tilde{\mu}} = 1+\frac{1}{2N} \sum_{n=1}^{N} \left(12-4\sin^2 \theta_{n} \right) \epsilon^2_{n} + 2 \sin^2 \theta_{n}$. This completes the proof.
%

\vspace*{-2mm}
\section*{Appendix G - Proof of Corollary 5}\label{Coro4proof}
From (\ref{Eqn_IQA_SigPow}), the matrix inversion lemma \cite{Horn2013}, and Lemmas 1-3, we obtain the asymptotic power of the useful signal of the $k$th UT in a single-cell uplink massive MIMO system with perfect CSI and IQI only at the UT as
\vspace*{-3mm}
\begin{align}
S^{\mathrm{IQA}^\circ}_{i, k} = & \frac{ \rho_\mathrm{UL}\Big| \frac{1}{2N} \tilde{\vv{g}}_{i,i,k}^\transp \tilde{\mm{\Lambda}}^{-1}_{i,k} \tilde{\vv{g}}_{i,i,k} \Big|^2}{2\left( 1+ \frac{1}{2N} \tilde{\vv{g}}_{i,i,k}^\transp \tilde{\mm{\Lambda}}^{-1}_{i,k} \tilde{\vv{g}}_{i,i,k}\right)^2} + \frac{\rho_\mathrm{UL}\Big| \frac{1}{2N} \tilde{\vv{g}}_{i,i,k+K}^\transp \tilde{\mm{\Lambda}}^{-1}_{i,k+K} \tilde{\vv{g}}_{i,i,k+K} \Big|^2}{2\left( 1+ \frac{1}{2N} \tilde{\vv{g}}_{i,i,k+K}^\transp \tilde{\mm{\Lambda}}^{-1}_{i,k+K} \tilde{\vv{g}}_{i,i,k+K}\right)^2} \xrightarrow[N \rightarrow \infty]{\mathrm{a.s.}} \frac{\rho_\mathrm{UL}\tilde{\delta}_k^2}{2\left(1+\tilde{\delta_k} \right)^2} \nonumber \\ 
& + \frac{\rho_\mathrm{UL}\tilde{\delta}_{k+K}^2}{2\left(1+\tilde{\delta}_{k+K} \right)^2},
\label{Eqn_IQA_Coro4_Sig}
\end{align}
where 
\vspace*{-3mm}
\begin{align}
\tilde{\delta}_k = \frac{ \left[ \mm{\Xi}_{i, k}\right]^2_{1,1} + \left[ \mm{\Xi}_{i, k}\right]^2_{2,1} }{4N} \mathrm{tr} \left( \tilde{\mm{\Gamma}} \right), \\
\tilde{\delta}_{k+K} = \frac{ \left[ \mm{\Xi}_{i, k}\right]^2_{1,2} + \left[ \mm{\Xi}_{i, k}\right]^2_{2,2} }{4N} \mathrm{tr} \left( \tilde{\mm{\Gamma}} \right).
\label{Eqn_Coro4delta}
\end{align}
Applying similar techniques as in the proof of Corollary 4, we obtain
\vspace*{-3mm}
\begin{align}
\tilde{\delta}_k \xrightarrow[\beta \rightarrow 0]{} N \rho_\mathrm{UL}\left( \left[ \mm{\Xi}_{i, k}\right]^2_{1,1} + \left[ \mm{\Xi}_{i, k}\right]^2_{2,1} \right), \label{Equ_IQA_Coro4_delta_k}\\
\tilde{\delta}_{k+K} \xrightarrow[\beta \rightarrow 0]{} N \rho_\mathrm{UL} \left( \left[ \mm{\Xi}_{i, k}\right]^2_{1,2} + \left[ \mm{\Xi}_{i, k}\right]^2_{2,2} \right), \label{Equ_IQA_Coro4_delta_kK}\\
\tilde{\mm{\Gamma}} \xrightarrow[\beta \rightarrow 0]{} 2N \rho_\mathrm{UL} \mm{I}_{2N}. \label{Equ_IQA_Coro4_Gamma}
\end{align}
Substituting (\ref{Equ_IQA_Coro4_delta_k}) and (\ref{Equ_IQA_Coro4_delta_kK}) into (\ref{Eqn_IQA_Coro4_Sig}) yields
\begin{align}
S^{\mathrm{IQA}^\circ}_{i, k} \hspace*{-2mm}=\hspace*{-1mm} 0.5 \rho_\mathrm{UL} \left( \left( \frac{N\left( \left[ \mm{\Xi}_{i, k}\right]^2_{1,1} \hspace*{-1mm}+\hspace*{-1mm} \left[ \mm{\Xi}_{i, k}\right]^2_{2,1} \right) \rho_\mathrm{UL}}{1\hspace*{-1mm}+\hspace*{-1mm}N\left( \left[ \mm{\Xi}_{i, k}\right]^2_{1,1} \hspace*{-1mm}+\hspace*{-1mm} \left[ \mm{\Xi}_{i, k}\right]^2_{2,1} \right) \rho_\mathrm{UL}}\right)^2 \hspace*{-2mm}+\hspace*{-1mm} \left( \frac{N\left( \left[ \mm{\Xi}_{i, k}\right]^2_{1,2} \hspace*{-1mm}+\hspace*{-1mm} \left[ \mm{\Xi}_{i, k}\right]^2_{2,2} \right) \rho_\mathrm{UL}}{1+N\left( \left[ \mm{\Xi}_{i, k}\right]^2_{1,2} \hspace*{-1mm}+\hspace*{-1mm} \left[ \mm{\Xi}_{i, k}\right]^2_{2,2} \right) \rho_\mathrm{UL}}\right)^2 \right).
\end{align}
Performing straightforward algebraic manipulations leads to
\begin{align}
S^{\mathrm{IQA}^\circ}_{i, k} = 0.5 \rho_\mathrm{UL} \left( \left( \frac{ N\rho_\mathrm{UL} \left( 1+2\check{\epsilon} \right) }{ 1 + N\rho_\mathrm{UL} \left( 1 + 2\check{\epsilon} \right) }\right)^2 + \left( \frac{ N\rho_\mathrm{UL} \left( 1-2\check{\epsilon} \right) }{ 1 + N\rho_\mathrm{UL} \left( 1 - 2\check{\epsilon} \right) }\right)^2 \right).
\end{align}
Applying similar techniques as in the proof of Corollary 4, the asymptotic interference power of the $k$th UT in the considered single-cell uplink massive MIMO system can be expressed as
\vspace*{-3mm}
\begin{align}
I^{\mathrm{IQA}^\circ}_{i, k} \hspace*{-2mm} = & \hspace*{-2mm} \sum_{q=1, q \neq k}^{2K} \hspace*{-2mm} \frac{\rho_\mathrm{UL}}{8N^2} \left( \Big| \tilde{\vv{g}}_{i,i,k}^\transp \tilde{\mm{\Lambda}}^{-1}_{i} \tilde{\vv{g}}_{i,i,q} \Big|^2 \hspace*{-2mm} + \hspace*{-1mm} \Big| \tilde{\vv{g}}_{i,i,k+K}^\transp \tilde{\mm{\Lambda}}^{-1}_{i} \tilde{\vv{g}}_{i,i,q} \Big|^2 \right) \hspace*{-1mm} =  \rho_\mathrm{UL} \hspace*{-3mm} \sum_{q=1, q \neq k}^{2K} \hspace*{-2mm} \frac{ \tilde{\vv{g}}_{i,i,k}^\transp \tilde{\mm{\Lambda}}^{-1}_{i, k} \tilde{\vv{g}}_{i,i,q} \tilde{\vv{g}}_{i,i,q}^\transp \tilde{\mm{\Lambda}}^{-1}_{i, k} \tilde{\vv{g}}_{i,i,k} }{8N^2\left( 1+ \tilde{\delta}_k \right)^2}\nonumber\\
& + \rho_\mathrm{UL} \sum_{q=1, q \neq k}^{2K} \frac{ \tilde{\vv{g}}_{i,i,k+K}^\transp \tilde{\mm{\Lambda}}^{-1}_{i, k+K} \tilde{\vv{g}}_{i,i,q} \tilde{\vv{g}}_{i,i,q}^\transp \tilde{\mm{\Lambda}}^{-1}_{i, k+K} \tilde{\vv{g}}_{i,i,k+K} }{ 8N^2\left( 1+ \tilde{\delta}_{k+K} \right)^2 } \xrightarrow[N \rightarrow \infty]{\mathrm{a.s.}} \frac{ \beta \rho_\mathrm{UL}}{\left(1+\tilde{\delta}_{k}\right) \left(1+\tilde{\delta}_{k+K}\right)} \nonumber \\ 
& + \frac{ \beta \rho_\mathrm{UL}}{2\left(1+\tilde{\delta}_{k}\right)^2 } + \frac{ \beta \rho_\mathrm{UL}}{2 \left(1+\tilde{\delta}_{k+K}\right)^2}.
\label{Eqn_IQA_Coro4_Int}
\end{align}
Similar to the proof of Corollary 4, the asymptotic noise power of the $k$th UT can be written as
\vspace*{-3mm}
\begin{align}
& Z^{\mathrm{IQA}^\circ}_{i, k} =  \Big| \frac{1}{2N}\tilde{\vv{g}}_{i,i,k}^\transp \tilde{\mm{\Lambda}}^{-1}_{i} \tilde{\vv{z}}_{i} \Big|^2 + \Big| \frac{1}{2N} \tilde{\vv{g}}_{i,i,k+K}^\transp \tilde{\mm{\Lambda}}^{-1}_{i} \tilde{\vv{z}}_{i} \Big|^2 = \frac{\tilde{\vv{g}}_{i,i,k}^\transp \tilde{\mm{\Lambda}}^{-1}_{i,k} \tilde{\vv{z}}_{i} \tilde{\vv{z}}^\transp_{i} \tilde{\mm{\Lambda}}^{-1}_{i,k} \tilde{\vv{g}}_{i,i,k} }{ 4N^2\left( 1+ \tilde{\delta}_k \right)^2 } + \nonumber \\
& \frac{\tilde{\vv{g}}_{i,i,k+K}^\transp \tilde{\mm{\Lambda}}^{-1}_{i,k} \tilde{\vv{z}}_{i} \tilde{\vv{z}}^\transp_{i} \tilde{\mm{\Lambda}}^{-1}_{i,k} \tilde{\vv{g}}_{i,i,k+K} }{ 4N^2\left( 1+ \tilde{\delta}_{k+K} \right)^2 } \hspace*{-1mm} \xrightarrow[N \rightarrow \infty]{\mathrm{a.s.}} \hspace*{-1mm} \frac{\left( \left[ \mm{\Xi}_{i, k}\right]^2_{1,1} + \left[ \mm{\Xi}_{i, k}\right]^2_{2,1} \right) \mathrm{tr} \left( \tilde{\mm{\Gamma}}_i^2 \right)}{16N^2 \left( 1+ \tilde{\delta}_k \right)^2 } \hspace*{-1mm} + \hspace*{-1mm} \frac{\left( \left[ \mm{\Xi}_{i, k}\right]^2_{1,2} + \left[ \mm{\Xi}_{i, k}\right]^2_{2,2} \right) \mathrm{tr} \left( \tilde{\mm{\Gamma}}_i^2 \right)}{16N^2 \left( 1+ \tilde{\delta}_{k+K} \right)^2 } \nonumber \\
& \xrightarrow[N \rightarrow \infty]{\mathrm{a.s.}} \frac{1}{2N\left( \left[ \mm{\Xi}_{i, k}\right]^2_{1,1} + \left[ \mm{\Xi}_{i, k}\right]^2_{2,1} \right) } + \frac{1}{2N\left( \left[ \mm{\Xi}_{i, k}\right]^2_{1,2} + \left[ \mm{\Xi}_{i, k}\right]^2_{2,2} \right)}.
\label{Eqn_IQA_Coro4_Noise}
\end{align}
Observing (\ref{Eqn_IQA_Coro4_Int}) and (\ref{Eqn_IQA_Coro4_Noise}) and considering $\beta \ll 1$ and $\tilde{\delta}_{k}, \tilde{\delta}_{k+K} \gg 1$ the interference power vanishes and the only remaining disturbance is the noise power. Taking this into account and performing straightforward algebraic operations leads to (\ref{Eqn_IQA_Coro_UT}). This completes the proof.

%
\vspace*{-3mm}
\bibliographystyle{IEEEtran}
\bibliography{Massive_MIMO}
\end{document}